\documentclass[11 pt, a4paper]{article}
\usepackage[hmargin=1.25in,vmargin=1.25in]{geometry}
\usepackage{amsmath,amssymb,amsthm,mathrsfs,amsfonts,mathtools,calculator,dsfont,bm} 
\usepackage[dvipsnames,svgnames, x11names]{xcolor}
\usepackage{natbib,bibentry,bibunits,lmodern}
\usepackage[bottom,hang,flushmargin]{footmisc} 
\usepackage{enumitem,pdfpages,selectp} 
\usepackage[hyphens]{url}
\usepackage[labelsep=period]{caption} 
\usepackage{subcaption}
\usepackage{environ,trimspaces,xspace}
\usepackage[titletoc]{appendix}
\usepackage{fancyhdr,setspace,titlesec,titling} 
\usepackage{siunitx,booktabs,longtable,threeparttable,multirow,makecell,float,rotating,tabularx} 
\usepackage{tikz,pgfplots} 
\pgfplotsset{compat=1.11}
\usetikzlibrary{arrows,intersections,plotmarks}
\usetikzlibrary{backgrounds,positioning,fit,petri}
\usetikzlibrary{mindmap,trees,calendar,external,shapes}
\usetikzlibrary{decorations.fractals,decorations.pathmorphing,shadows,shadings,patterns}
\usetikzlibrary{spy,calc} 
\usetikzlibrary{colorbrewer}
\usepgfplotslibrary{fillbetween,groupplots,colorbrewer}
\usepackage[bookmarksnumbered,bookmarksopen,psdextra,unicode,hyperindex,breaklinks,hypertexnames=false]{hyperref}
\usepackage[open,openlevel=2,numbered,atend]{bookmark} 

\makeatletter
\renewcommand\paragraph{\@startsection{paragraph}{4}{\z@}%
            {-2.5ex\@plus -1ex \@minus -.25ex}%
            {1.25ex \@plus .25ex}%
            {\normalfont\normalsize\bfseries}}
\makeatother
\hypersetup{colorlinks, linkcolor = {NavyBlue}, citecolor = {NavyBlue}, urlcolor ={NavyBlue}}

\newtheorem{thm}{Theorem}[section]

\newtheorem{lem}{Lemma}[section]
\newtheorem{pro}{Proposition}[section]

\newtheorem{ass}{Assumption}[section]
\theoremstyle{definition}

\newtheorem{rem}{Remark}[section]

\theoremstyle{definition}

\newtheorem*{ex*}{Example}

\newtheoremstyle{exctd}
{\topsep} {\topsep}%
{\upshape}
{}
{\bfseries\scshape}
{.}
{1em}
{\thmname{#1} \thmnumber{ #2}\thmnote{#3} (cont.)}
\theoremstyle{exctd}

\newcommand{\transpose}{\text{\scalebox{0.8}{$\intercal$}}}

\newcommand{\convp}{\xrightarrow{p}}

\pgfmathdeclarefunction{npdf}{3}{%
\pgfmathparse{1/(#3*sqrt(2*pi))*exp(-((#1-#2)^2)/(2*#3^2))}%
}
\tikzset{
    declare function={
        ncdf(\x,\m,\s)=1/(1 + exp(-0.07056*((\x-\m)/\s)^3 - 1.5976*(\x-\m)/\s));
    }
}

\begin{document}
\pdfbookmark[1]{Title}{title}
\title{A Unifying Framework for Testing Shape Restrictions\thanks{I thank Dora Gicheva for helping me make sense of her data set. Advanced computing resources provided by Texas A\&M High Performance Research Computing are gratefully acknowledged.}}
\author{Zheng Fang \\ Department of Economics \\ Texas A\&M University\\ zfang@tamu.edu}
\date{August 1, 2021}
\maketitle

\begin{abstract}
This paper makes the following original contributions. First, we develop a unifying framework for testing shape restrictions based on the Wald principle. The test has asymptotic uniform size control and is uniformly consistent. Second, we examine the applicability and usefulness of some prominent shape enforcing operators in implementing our framework. In particular, in stark contrast to its use in point and interval estimation, the rearrangement operator is inapplicable due to a lack of convexity. The greatest convex minorization and the least concave majorization are shown to enjoy the analytic properties required to employ our framework.  Third, we show that, despite that the projection operator may not be well-defined/behaved in general parameter spaces such as those defined by uniform norms, one may nonetheless employ a powerful distance-based test by applying our framework. Monte Carlo simulations confirm that our test works well. We further showcase the empirical relevance by investigating the relationship between weekly working hours and the annual wage growth in the high-end labor market.
\end{abstract}


\begin{center}
\end{center}

\newpage

\section{Introduction}

Shape restrictions are ubiquitous in economics, often arising as characterizations or implications of economic theory; see \citet{Matzkin1994Handbook}, \citet{MWG1995}, and \citet{ChetverikovSantosAzeem2018Shape} for surveys and textbook treatment, as well as \citet{CoupriePelusoTrannoy2010Power}, \citet{Ellison2011Strategic}, \citet{CardMasMorettiSaez2012Inequality}, \citet{Gicheva2013Working}, \citet{ChandraHeadTappata2014Border}, \citet{BoffaPiolattoPonzetto2016Political}, \citet{ScheuerWerning2017Superstar}, and \citet{ElliottKudrinWuthrich2021Detecting} for some recent concrete examples. Contrary to their extensive roles in economics and despite the sizable literature, the formal statistical analysis of shape restrictions appears to be relatively scant in empirical work. One possible explanation is that many existing inferential procedures are either problem-specific or lack computational tractability without sacrificing statistical power \citep{FangSeo2019Shape}.

In this paper, we develop a unifying framework for testing shape restrictions concerning a parameter of interest $\theta_0$, where the hypotheses are formulated as:
\begin{align}\label{Eqn: hypotheses}
\mathrm H_0: \phi(\theta_0)=0 \qquad \mathrm{vs.}\qquad  \mathrm H_1: \phi(\theta_0)>0~,
\end{align}
for $\phi$ some known map taking nonnegative values. In the spirit of the Wald test, we base our framework on the plug-in statistic $\phi(\hat\theta_n)$ for an unconstrained estimator $\hat\theta_n$ of $\theta_0$. Statistical properties of the resulting test then depend on the choice of the map $\phi$, which we shall call the Wald functional in what follows. As an example, for $\Lambda$ the family of all elements in the parameter space satisfying the shape restriction (hereafter), \citet{FangSeo2019Shape} opt for $\phi(\theta)=\|\theta-\Pi_\Lambda(\theta)\|_{\mathbf H}$, where $\Pi_\Lambda(\theta)$ is the closest element in $\Lambda$ to $\theta$ (i.e., the projection of $\theta$ onto $\Lambda$), and $\|\cdot\|_{\mathbf H}$ is some $L^2$-norm. While the fruitful analytic properties of the projection operator enable \citet{FangSeo2019Shape} to develop a powerful test for a class of shape restrictions, there are regrettably other prominent choices of $\phi$ that fall beyond the scope of their framework.

First, there are other shape enforcing operators that have received increasing attention recently, notably the rearrangement
\citep{ChernozhukovFernandezGalichon2009Improving,ChernozhukovFernandezGalichon2010Crossing,ChernozhukovFernandezValLuo2018Sorted,ChenChernozhukovFernandezKostyshakLuo2021Shape} and the greatest convex minorization (GCM) or the least concave majorization (LCM) \citep{CarolanTebbs2005LR,DelgadoEscanciano2012SM,BeareMoon2015DR,Seo2018SM,ChenChernozhukovFernandezKostyshakLuo2021Shape}. The use of these two operators have been, however, confined to the construction of point and interval estimates or testing problems where the parameter $\theta_0$ is $\sqrt n$-estimable. Empirical settings of shape restrictions (e.g., nonparametric regression and density estimation), on the other hand, are often concerned with $\theta_0$ that is only estimable at slower than the $\sqrt n$-rate. Second, one may be interested in projection defined by non-$L^2$-norms such as the sup norm. Analogous to the comparison of the Cram\'{e}r-von Mises ($L^2$-type) test vs.\ the Kolmogorov-Smirnov (sup-type) test, different norms lead to tests that are powerful against different classes of alternatives, and there are settings where it may be desirable to employ a particular test---see Chapter 14 in \citet{TSH2005}, \citet{AndrewsShi2013CMI}, \citet{Armstrong2018Choice}, and references therein for related discussions. Unfortunately, the projection operator in general may not be well-defined/behaved with respect to non-$L^2$-norms, rendering the framework of \citet{FangSeo2019Shape} not directly applicable.

As the first contribution of this paper, we show that the attractive statistical properties of \citet{FangSeo2019Shape}'s test are not unique to the $L^2$-projection operator. Indeed, as far as forming a suitable Wald functional $\phi$ is concerned, it suffices to require convexity, positive homogeneity, and Lipschitz continuity. The importance of convexity has been noted in \citet{FangSantos2018HDD}, while positive homogeneity and Lipschitz continuity are mild and automatically fulfilled for the class of Wald functionals we consider. To accommodate non-$\sqrt n$ estimable parameters $\theta_0$, we appeal to strong approximations as in \citet{FangSeo2019Shape}. We show that our test has asymptotic uniform size control and is uniformly consistent (under regularity conditions).

In turn, with the testing framework in hand, we examine some concrete choices of the Wald functional. First, for a shape enforcing operator $\Upsilon$, we consider
\begin{align}\label{Eqn: test functional, operator}
\phi(\theta)=\|\theta-  \Upsilon(\theta)\|~,
\end{align}
where $\|\cdot\|$ is a norm that may not be $L^2$. We show that this strategy works for GCM and LCM, but not for the rearrangement as the resulting $\phi$ is not convex. Second, since projection may not be well-defined, one may instead employ
\begin{align}\label{Eqn: test functional, distance}
\phi(\theta)=\inf_{\lambda\in\Lambda}\|\theta-\lambda\|~.
\end{align}
The aforementioned analytic properties are then met whenever $\Lambda$ is a nonempty closed convex cone. Devising a powerful test, however, further demands that the restriction under the null be incorporated into the construction of critical values, as well understood in the literature. This may be accomplished by using $\Upsilon(\hat\theta_n)$ for \eqref{Eqn: test functional, operator}, but using $\Pi_\Lambda(\hat\theta_n)$ for \eqref{Eqn: test functional, distance} as in \citet{FangSeo2019Shape} is problematic as $\Pi_\Lambda$ may not be well-defined/behaved if $\|\cdot\|$ is not a $L^2$-norm. To circumvent this challenge, we make use of $\Upsilon(\hat\theta_n)$ for \eqref{Eqn: test functional, distance} as well, provided $\Upsilon$ is a well-defined shape enforcing operator. In particular, the rearrangement is well suited to this end, despite that it is not suitable in forming \eqref{Eqn: test functional, operator}. Together, these results broaden the use of GCM/LCM and rearrangement to a wide range of nonparametric settings, and extend the $L^2$-projection test in  \citet{FangSeo2019Shape} to a distance-based test (defined by a general norm).

We conduct Monte Carlo simulations to evaluate the performance of our test. The simulation results show that our framework yields tests that are competitive (in terms of size and power) compared to the prominent sup tests of \citet{ChernozhukovLeeRosen2013Intersection} and \citet{Chetverikov2018Monotonicity}. To showcase the empirical relevance of our framework, we test the shape restrictions on the relationship between weakly working hours and the annual wage growth in the high-end labor market. Overall, our empirical findings support the theoretical predictions of \citet{Gicheva2013Working} for both men and women. The application also highlights the importance of conducting formal statistical tests when it comes to shape restrictions, rather than just relying on eyeball inspection.

The literature on shape restrictions dates back to \citet{Hildreth1954Concave}, \citet{AyerBrunkEwingReidSilverman1955Incomplete}, \citet{Brunk1955MLEmonotone}, \citet{Eeden1956MLEorder}, and \citet{Grenander1956II}. We contribute to the problem of testing shape restrictions. Our paper extends \citet{FangSeo2019Shape} who employ \eqref{Eqn: test functional, distance} defined by a $L^2$-norm. We also build upon some of the analytic results and computational algorithms in \citet{ChenChernozhukovFernandezKostyshakLuo2021Shape}. These authors study the use of shape enforcing operators for point and interval estimation, which is different from our focus on testing. We note that recent tests of concavity based on LCM are limited to {\it univariate} settings where $\theta_0$ is $\sqrt n$-estimable---see aforementioned references as well as \citet{,DelgadoEscanciano2013ConditionalSD,DelgadoEscanciano2016ConditionalMI}, \citet{BeareSchmidt2016Pricing}, \citet{BeareShi2019DRO}, and \citet{Fang2019KW}. Other recent work include \citet{ChernozhukovNeweySantos2019CCMM} and \citet{Zhu2020Shape} who study moment restriction models with partial identification, \citet{FreybergerReeves2017Shape} and \citet{ChiangKatoSasakiUra2021LP} who construct shape-constrained confidence bands, \citet{BreunigChen2020Adaptive} who develop adaptive and rate-optimal tests in nonparametric instrumental variable models, \citet{KomarovaHidalgo2020Shape} who devise a pivotal test based on the Khmaladze's
transformation for univariate nonparametric regression models, and \citet{KostyshakLuo2021PMP} who propose the partial monotonicity parameter as a generalization of regression monotonicity. For brevity, we refer the reader to \citet{FangSeo2019Shape} for additional references.


We now introduce some notation and concepts. Set $\mathbf R_+=\{a\in\mathbf R: a\ge 0\}$. For a vector $a\in\mathbf R^k$, we let $a^{(j)}$ be its $j$th entry and set $\|a\|_p$ to equal $\{\sum_{j=1}^{k}|a^{(j)}|^p\}^{1/p}$ if $p\in[1,\infty)$ and $\max_{j=1}^k|a^{(j)}|$ if $p=\infty$. We denote by $\mathbf M^{m\times k}$ the space of $m\times k$ matrices. For a function $f:\mathcal Z\to\mathbf R$ with $\mathcal Z\subset\mathbf R^{d_z}$, we set its $L^p$ norm $\|f\|_p$ to be $\{\int_{\mathcal Z}|f(z)|^p\,\mathrm dz\}^{1/p}$ if $p\in[1,\infty)$ and $\sup_{z\in\mathcal Z}|f(z)|$ if $p=\infty$. In turn, we define $L^p(\mathcal Z)\equiv\{f:\mathcal Z\to\mathbf R: \|f\|_p<\infty\}$ for $p\in[1,\infty)$ and  $\ell^\infty(\mathcal Z)\equiv\{f:\mathcal Z\to\mathbf R: \|f\|_\infty<\infty\}$. For a vector space $\mathbf B$, we recall that a map $\phi:\mathbf B\to\mathbf R$ is said to be positively homogeneous if $\phi(a\theta)=a\phi(\theta)$ for all $a\ge 0$ and $\theta\in\mathbf B$. For generic families of distributions $\mathbf P_n$, a sequence $\{a_n\}$ of positive scalars, and a sequence $\{\mathbb X_n\}$ of random elements in a normed space $\mathbf B$ with norm $\|\cdot\|_{\mathbf B}$, write $\mathbb X_n=o_p(a_n)$ uniformly in $P\in\mathbf P_n$ if $\lim_{n\to\infty}\sup_{P\in\mathbf P_n}P(\|\mathbb X_n\|_{\mathbf B}>a_n\epsilon)= 0$ for any $\epsilon>0$, and $\mathbb X_n=O_p(a_n)$ uniformly in $P\in\mathbf P_n$ if $\lim_{M\to\infty}\limsup_{n\to\infty}\sup_{P\in\mathbf P_n}P(\|\mathbb X_n\|_{\mathbf B}>Ma_n)= 0$.

The remainder of the paper is structured as follows. In Section \ref{Sec: framework}, we develop the testing framework, investigates a number of possible Wald functionals, and provide some implementation guidance. Section \ref{Sec: simulation} conducts Monte Carlo simulation studies, while Section \ref{Sec: empirical} tests some shape restrictions concerning labor supply. Section \ref{Sec: conclusion} concludes. All proofs are relegated to the appendix.

\section{The Testing Framework}\label{Sec: framework}

As well understood in the literature \citep{ImbensManski2004PID,Mikusheva2007Uniform,AndrewsGuggenberger2009Validity}, it is imperative to ensure that testing procedures in nonstandard settings (such as the present one) be uniformly valid. To this end, we shall make explicit the dependence of the parameter interest $\theta_0$ on the underlying distribution $P$ by instead writing $\theta_P$ whenever appropriate. Moreover, we denote by $\mathbf P_0$ the model under the null and by $\mathbf P_1$ the model under the alternative. That is, $\mathbf P_0=\{P\in\mathbf P: \phi(\theta_P)=0\}$ and $\mathbf P_1=\mathbf P\backslash\mathbf P_0$, where $\mathbf P$ is the posited family of distributions possibly generating the data. We allow $\mathbf P_0$, $\mathbf P_1$ and $\mathbf P$ to depend on the sample size $n$, but such dependence is suppressed for notational simplicity.

\subsection{Size and Power}\label{Sec: statistical properties}

In order to present a unifying treatment, we assume throughout that the parameter of interest $\theta_0$ lives in an abstract Banach space $\mathbf B$ (i.e., a complete normed space) with a known norm $\|\cdot\|_{\mathbf B}$. The following assumption formalizes our main restrictions on the Wald functional $\phi: \mathbf B\to\mathbf R_+$.

\begin{ass}\label{Ass: space and map}
For a Banach space $\mathbf B$ with norm $\|\cdot\|_{\mathbf B}$, a known map $\phi: \mathbf B\to\mathbf R_+$ is (i) positively homogeneous; (ii) convex; (iii) Lipschitz continuous.
\end{ass}

Assumptions \ref{Ass: space and map}(i)(ii) effectively demand that the null parameter space $\Lambda\equiv\{\theta\in\mathbf B: \phi(\theta)=0\}$ be a nonempty convex cone, which must be closed under Assumption \ref{Ass: space and map}(iii). As illustrated in \citet{FangSeo2019Shape}, this is satisfied by a number of common shape restrictions such as nonnegativity, monotonicity, convexity/concavity, Slutsky restriction, supermodularity and any intersections of these restrictions. Nonetheless, some other prominent restrictions such as quasi-convexity/concavity are excluded. We note that, given Assumption \ref{Ass: space and map}(i), Assumption \ref{Ass: space and map}(ii) is equivalent to $\phi$ being subadditive, i.e., $\phi(\theta_1+\theta_2)\le\phi(\theta_1)+\phi(\theta_2)$ for all $\theta_1,\theta_2\in\mathbf B$. Assumption \ref{Ass: space and map}(iii) is a convenient mild condition for our distributional and bootstrap approximations. As shall be discussed in Section \ref{Sec: Wald functional}, Assumption \ref{Ass: space and map} is satisfied for the distance function generated by a closed convex cone and the map $\phi$ generated by GCM/LCM, but is violated for the map $\phi$ generated by the rearrangement operator.

Assumption \ref{Ass: space and map} implies that the map $a\mapsto\phi(h+a\theta_0)$ is weakly decreasing on $[0,\infty)$ for any $\theta_0\in\Lambda$ and $h\in\mathbf B$, a property shared by the $L^2$-distance map in \citet[Lemma D.1]{FangSeo2019Shape}. This in turn yields the following key relation that is fundamental in the development of our test: for an estimator $\hat\theta_n$ of $\theta_0$ and $0\le \kappa_n\le r_n$,
\begin{align}\label{Eqn: upper bound}
r_n\phi(\hat\theta_n) = \phi(r_n\{\hat\theta_n-\theta_0\}+r_n\theta_0)\le \phi(r_n\{\hat\theta_n-\theta_0\}+\kappa_n\theta_0)~.
\end{align}
The display \eqref{Eqn: upper bound} reveals both challenges and opportunities in devising a test based on $r_n\phi(\hat\theta_n)$. On one hand, while the law of $r_n\{\hat\theta_n-\theta_P\}$ may be consistently bootstrapped, the generic impossibility of consistently estimating $r_n\theta_P$ poses challenges for estimating the distribution of $r_n\phi(\hat\theta_n)$, an issue prevalent in other nonstandard settings \citep{AndrewsSoares2010,ChernozhukovNeweySantos2019CCMM}. On the other hand, consistently estimating the law of the upper bound in \eqref{Eqn: upper bound} is possible if $\kappa_n$ is suitably small, in view of the identity $\kappa_n\hat\theta-\kappa_n\theta_P=\kappa_n/r_n\cdot r_n\{\hat\theta_n-\theta_P\}$. Taken together, the inequality in \eqref{Eqn: upper bound} thus suggests that a test with $r_n\phi(\hat\theta_n)$ as the test statistic and critical values from the upper bound may control size in large samples.

To formalize the above idea, we introduce our second main assumption.
\begin{ass}\label{Ass: strong approx}
i) $\hat\theta_n: \{X_i\}_{i=1}^n\to\mathbf B$ satisfies $\|r_n\{\hat\theta_n-\theta_P\}-\mathbb Z_{n,P}\|_{\mathbf B}=o_p(c_n)$ uniformly in $P\in\mathbf P$ for some $r_n\to\infty$, $\mathbb Z_{n,P}\in\mathbf B$ and $c_n>0$ with $c_n=O(1)$; (ii) $\hat{\mathbb G}_n\in\mathbf B$ is a bootstrap estimator satisfying: $\|\hat{\mathbb G}_{n}-\bar{\mathbb Z}_{n,P}\|_{\mathbf B}=o_p(c_n)$ uniformly in $P\in\mathbf P$, for $\bar{\mathbb Z}_{n,P}$ a copy of $\mathbb Z_{n,P}$ that is independent of $\{X_i\}_{i=1}^n$.
\end{ass}

Assumption \ref{Ass: strong approx}(i) requires a uniform (in $P\in\mathbf P$) distributional approximation  of $r_n\{\hat\theta_n-\theta_P\}$ by a {\it sequence} of coupling variables $\mathbb Z_{n,P}$, in lieu of an {\it asymptotic} distribution. While one may work with the latter if $r_n\{\hat\theta_n-\theta_P\}$ does converge in distribution, this becomes problematic in general nonparametric settings in which $r_n\{\hat\theta_n-\theta_P\}$ fails to converge as a process. Nonetheless, \citet{ChernozhukovLeeRosen2013Intersection} and subsequently, \citet{BelloniChernozhukovChetverikovKato2015New}, \citet{ChernozhukovNeweySantos2019CCMM}, \citet{ChenChristensen2018SupNormOptimal}, \citet{BelloniChernozhukovChetverikovFernandez2019QR}, \citet{CattaneoFarrellFeng2018Partition}, and \citet{LiLiao2020Uniform}  show that approximations as in Assumption \ref{Ass: strong approx}(i) can be established under regularity conditions, thereby greatly facilitating inference on global constraints (e.g., shape) of nonparametric functions. Assumption \ref{Ass: strong approx}(ii) simply says that the law of $\bar{\mathbb Z}_{n,P}$ (or equivalently the law of $\mathbb Z_{n,P}$) can be consistently bootstrapped by $\hat{\mathbb G}_{n}$, which may be verified by the same machineries developed in the aforementioned work. We stress that Assumption \ref{Ass: strong approx} places no restrictions on the particular schemes underlying the estimator $\hat\theta_n$ or the bootstrap $\hat{\mathbb G}_n$. Thus, one may resort to sieve or kernel estimation in constructing $\hat\theta_n$ and various bootstrap/simulation methods for $\hat{\mathbb G}_n$. Finally, Assumption \ref{Ass: strong approx} implicitly entails certain smoothness on $\theta_P$, as typically required for nonparametric estimation.

Given the estimator $\hat\theta_n$ and the bootstrap $\hat{\mathbb G}_n$, we may employ $\phi(\hat{\mathbb G}_n+\kappa_n \hat\theta_n)$ as a bootstrap estimator for the upper bound in \eqref{Eqn: upper bound}. As well understood in analogous nonstandard settings \citep{AndrewsSoares2010,RomanoShaikhWolf2014TwoStep}, however, this may cause loss of power in finite samples because $\hat\theta_n$ (in the term $\kappa_n \hat\theta_n$) does not reflect the restriction $\theta_P\in\Lambda$ under the null. This motivates us to use the restricted estimator $\Gamma(\hat\theta_n)$, provided a suitable shape enforcing operator $\Gamma:\mathbf B\to\Lambda$ is available, an issue we shall revisit in Section \ref{Sec: Wald functional}. In turn, for a given significance level $\alpha\in(0,1)$, we may then obtain the critical value $\hat c_{n,1-\alpha}$ as:
\begin{align}\label{Eqn: critical value}
\hat c_{n,1-\alpha}\equiv \inf\{c\in\mathbf R: P(\phi(\hat{\mathbb G}_n+\kappa_n\Gamma(\hat\theta_n))\le c|\{X_i\}_{i=1}^n)\ge 1-\alpha\}~.
\end{align}
By construction, $\hat c_{n,1-\alpha}$ is an estimator of the $1-\alpha$ quantile, denoted $c_{n,P}(1-\alpha)$, of $\phi(\mathbb Z_{n,P}+\kappa_n\theta_P)$ (a distributional approximation of the upper bound in \eqref{Eqn: upper bound}). To justify the construction of $\hat c_{n,1-\alpha}$, we further impose:

\begin{ass}\label{Ass: enforce null}
An operator $\Gamma: \mathbf B\to\Lambda$ (i) satisfies $\Gamma(h)=h$ for all $h\in\Lambda$ where $\Lambda\equiv\{\theta\in\mathbf B: \phi(\theta)=0\}$; (ii) is Lipschitz continuous.
\end{ass}

\begin{ass}\label{Ass: critical value}
(i) $\mathbb Z_{n,P}$ is tight and centered Gaussian in $\mathbf B$ for each $n\in\mathbf N$ and $P\in\mathbf P$; (ii) $\sup_{P\in\mathbf P} E[\|\mathbb Z_{n,P}\|_{\mathbf B}]\le\zeta_n$ for some $\zeta_n\ge 1$; (iii) $c_{n,P}(1-\alpha-\varpi)\ge c_{n,P}(0.5)+\varsigma_n$ for some constants $\varpi,\varsigma_n>0$, each $n$ and $P\in\mathbf P_0$; (iv) $c_n\zeta_n/\varsigma_n^2=O(1)$ as $n\to\infty$.
\end{ass}

Assumption \ref{Ass: enforce null} demands mild requirements on $\Gamma$, which are satisfied by, e.g., the rearrangement for monotonicity, GCM for convexity, and a suitable composition of these two for monotonicity jointly with convexity---see Section \ref{Sec: Wald functional}. Assumption \ref{Ass: critical value}(i) is a mild technical restriction, while Assumption \ref{Ass: critical value}(iii)(iv) imply that the cdfs of $\psi_{\kappa_n,P}(\mathbb Z_{n,P})$ are suitably continuous around $c_{n,P}(1-\alpha)$. Assumption \ref{Ass: critical value}(ii) differs from \citet{FangSeo2019Shape} who assume uniform boundedness of $E[\|\mathbb Z_{n,P}\|_{\mathbf B}]$, because $E[\|\mathbb Z_{n,P}\|_{\mathbf B}]$ may grow with $n$ in our setting (e.g., when $\mathbf B=\ell^\infty([0,1])$).

Assumptions \ref{Ass: space and map}, \ref{Ass: strong approx}, \ref{Ass: enforce null}, and \ref{Ass: critical value} together deliver our first main result.

\begin{thm}\label{Thm: size and power}
Let Assumptions \ref{Ass: space and map} and \ref{Ass: strong approx} hold.
\begin{enumerate}
  \item[(i)] (Size) If Assumptions \ref{Ass: enforce null} and \ref{Ass: critical value} hold and $0\le\kappa_n\zeta_n/r_n=o(c_n)$, then
\begin{align}\label{Thm: size, aux1}
\limsup_{n\to\infty}\sup_{P\in\mathbf P_0} P( r_n \phi(\hat\theta_n)> \hat c_{n,1-\alpha})\le \alpha~,
\end{align}
and, for $\bar{\mathbf P}_0\equiv \{P\in\mathbf P_0: \phi(h+\theta_P)=\phi(h),\forall\, h\in\mathbf B\}$,
\begin{align}\label{Thm: size, aux2}
\limsup_{n\to\infty}\sup_{P\in\bar{\mathbf P}_0} |P( r_n \phi(\hat\theta_n)> \hat c_{n,1-\alpha})-\alpha|=0~.
\end{align}
 \item[(ii)] (Power) If Assumptions \ref{Ass: enforce null}(i) and \ref{Ass: critical value}(ii) hold and $\kappa_n\ge 0$, then
\begin{align}\label{Thm: power, aux}
\liminf_{\Delta\to\infty}\liminf_{n\to\infty}\inf_{P\in\mathbf P_{1,n}^\Delta} P( r_n \phi(\hat\theta_n)> \hat c_{n,1-\alpha})=1~,
\end{align}
where $\mathbf P_{1,n}^\Delta\equiv\{P\in\mathbf P_1: \phi(\theta_P)\ge \Delta\zeta_n/r_n\}$ for $\Delta>0$.
\end{enumerate}

\end{thm}

Theorem \ref{Thm: size and power} formally establishes the asymptotic size control of our test, under a proper choice of $\kappa_n$. While one may ignore $\zeta_n$ from $\kappa_n\zeta_n/r_n=o(c_n)$ if $\mathbf B$ is a Hilbert space \citep{FangSeo2019Shape}, it should be taken into account when $\mathbf B$ is endowed with, e.g., a uniform norm. In the nonparametric regression settings of \citet{ChernozhukovLeeRosen2013Intersection}, one may take $\zeta_n=\sqrt{\log n}$.\footnote{Thus, the rate $r_n$ may not necessarily be the convergence rate of $\hat\theta_n$ as $\mathbb Z_{n,P}$ may diverge.} Theorem \ref{Thm: size and power} further show that our test uniformly attains the nominal rejection rates in the limit over a class of null distributions that may heuristically be thought of as the ``boundary''. We also note that the classes of alternatives against which our test is uniformly consistent depends on the Wald functional $\phi$. In general, there does not exist a choice that leads to the most powerful test, and a different $\phi$ ``distributes'' power over a different region of the parameter space.

A key element in implementing our test is the choice of the tuning parameter $\kappa_n$. Intuitively, while $\kappa_n$ should be small as dictated in Theorem \ref{Thm: size and power}, it should not be ``too small'' in the sense of causing the upper bound in \eqref{Eqn: upper bound} overly crude. Following \citet{FangSeo2019Shape}, we provide a data-driven choice of $\kappa_n$ as follows. For some small $\gamma_n\in(0,1)$, set $\hat\kappa_n\equiv r_nc_n/\hat\tau_{n,1-\gamma_n}$ where
\begin{align}\label{Eqn: tuning parameter, data driven}
\hat\tau_{n,1-\gamma_n}\equiv \inf\{c\in\mathbf R: P(\|\hat{\mathbb G}_n\|_{\mathbf B}\le c|\{X_i\}_{i=1}^n)\ge 1-\gamma_n\}~.
\end{align}
In order to validate the use of $\hat\kappa_n$, we need to introduce our final assumption where
\begin{align}
\bar\sigma_{n,P}^2\equiv \sup_{b^*\in\mathbf B^*: \|b^*\|_{\mathbf B^*}\le 1} E[\langle b^*,\mathbb Z_{n,P}\rangle^2]~,
\end{align}
for $\mathbf B^*$ the space of continuous linear functions $b^*:\mathbf B\to\mathbf R$ endowed with the norm $\|b^*\|_{\mathbf B^*}\equiv\sup_{b\in\mathbf B: \|b\|_{\mathbf B}\le 1}|\langle b^*,b\rangle|$, and $\langle b^*,b\rangle\equiv b^*(b)$.

\begin{ass}\label{Ass: tuning}
$\liminf_{n\to\infty}\inf_{P\in\mathbf P_0} \{\bar\sigma_{n,P}/\zeta_n\}>0$.
\end{ass}

Assumption \ref{Ass: tuning} is in line with Assumption \ref{Ass: critical value}(ii) which allows $\{\|\mathbb Z_{n,P}\|_{\mathbf B}\}$ to diverge. Intuitively, Assumption \ref{Ass: tuning} requires that there be enough variations in $\{\mathbb Z_{n,P}\}$ even if the sequence $\{E[\|\mathbb Z_{n,P}\|_{\mathbf B}]\}$ of expected norms diverges.

\begin{pro}\label{Pro: tuning parameter}
Let Assumptions \ref{Ass: strong approx}, \ref{Ass: critical value}(i)(ii) and \ref{Ass: tuning} hold, and set $\hat\kappa_n\equiv r_nc_n/\hat\tau_{n,1-\gamma_n}$ with $\gamma_n\in(0,1)$ and $\hat\tau_{n,1-\gamma_n}$ as in \eqref{Eqn: tuning parameter, data driven}. If $\gamma_n\to 0$, then $\hat\kappa_n\zeta_n/r_n = o_p(c_n)$ uniformly in $P\in\mathbf P_0$. If $(r_nc_n)^{-2}\zeta_n^2\log\gamma_n\to 0$, then $\hat\kappa_n/\zeta_n\convp\infty$ uniformly in $P\in\mathbf P_0$.
\end{pro}

The condition $\gamma_n\to 0$ ensures that our data driven $\hat\kappa_n$ satisfies the rate condition in Theorem \ref{Thm: size and power} in probability, while $(r_nc_n)^{-2}\zeta_n^2\log\gamma_n\to 0$ formalizes the precise sense in which $\gamma_n$ and $\hat\kappa_n$ should not be ``too small.'' In line with prior findings in the literature \citep{FangSantos2018HDD,ChernozhukovNeweySantos2019CCMM,FangSeo2019Shape}, our Monte Carlo simulations show that the testing results are quite insensitive to the choice of $\gamma_n$. We recommend $\gamma_n=0.01/\log n$ or $1/n$ for practical implementations.

\subsection{The Wald Functional}\label{Sec: Wald functional}

We next introduce a number of concrete Wald functionals, and investigate their suitability in applying the previous framework.

\subsubsection{Rearrangement}
Rearrangment is an operation that enforces a specific shape, namely monotonicity---throughout monotonicity means ``weakly increasing.'' For $\theta: \mathcal Z\to\mathbf R$ a potentially non-monotonic function with some bounded set $\mathcal Z\subset\mathbf R$, then the rearrangement operator $\Upsilon$ monotonizes $\theta$ as follows: for any $z\in\mathcal Z$,
\begin{align}\label{Eqn: rearrangement, univariate}
\Upsilon(\theta)(z)=\inf\{u\in\mathbf R: \int_{\mathcal Z}1\{\theta(y)\le u\}\,\mathrm dy\ge z\}~.
\end{align} 
If $\theta$ is multivariate, then one may monotonize $\theta$ by applying \eqref{Eqn: rearrangement, univariate} along each argument of $\theta$---see Section \ref{Sec: implementation} for more details. While rerrangement has proven particularly convenient in obtaining restricted point and interval estimates  \citep{Fougeres1997Density,DetteNeumeyerPilz2006Simple,ChernozhukovFernandezGalichon2009Improving,ChernozhukovFernandezGalichon2010Crossing,ChernozhukovFernandezValLuo2018Sorted,ChenChernozhukovFernandezKostyshakLuo2021Shape}, it does not generate a convex $\phi$ as in \eqref{Eqn: test functional, operator}, a property that is crucial for implementing our test. This can be easily seen when $\theta_0\in\mathbf B$ is finite dimensional. For example, if $\mathbf B=\mathbf R^4$ is endowed with the max norm and $\phi(\theta)=\|\theta-\Upsilon(\theta)\|_\infty$ with $\Upsilon$ the rearrangement operator, then, for $\theta_1=[40,54,42,69]^\transpose$ and $\theta_2=[21,88,3,68]^\transpose$, simple calculations yield
\begin{align}
\phi(\theta_1+\theta_2)= 92 > 12 + 67 = \phi(\theta_1) + \phi(\theta_2)~,
\end{align}
implying that $\phi$ is not convex. The lack of convexity remains true for alternative (e.g., $L^1$ or $L^2$) norms. Nonetheless, rearrangement satisfies Assumption \ref{Ass: enforce null} \citep{LiebLoss2001Analysis,ChernozhukovFernandezGalichon2009Improving,ChenChernozhukovFernandezKostyshakLuo2021Shape}, and may thus be utilized to construct constrained estimators for the purpose of power improvement.

\subsubsection{Greatest Convex Minorization}

GCM is also specific to a particular shape restriction, namely, convexity. The greatest convex minorant (also abbreviated GCM) of a function $\theta$ is the pointwise supremum of convex functions lying below $\theta$. Analogously, the least concave majorant (LCM) of a function $\theta$ is the pointwise infimum of concave functions lying above $\theta$. In what follows, we shall focus on GCM, as the analysis of LCM is similar. Following \citet{ChenChernozhukovFernandezKostyshakLuo2021Shape}, we define GCM directly through the Legendre-Fenchel transform. Concretely, let $\mathcal Z\subset\mathbf R^{d_z}$ be a nonempty convex set, and $\theta:\mathcal Z\to\mathbf R$. Then the conjugate $\mathcal Z^\star$ of $\mathcal Z$ is $\mathcal Z^\star\equiv\{y\in\mathbf R^{d_z}: \sup_{z\in\mathcal Z}\{\langle y,z\rangle-\theta(z)\}<\infty\}$ which is a nonempty convex set, and the convex conjugate of $\theta$ is a map $\theta^\star:\mathcal Z^\star\to\mathbf R$ defined by
\begin{align}\label{Eqn: conjugate}
\theta^\star(y)=\sup_{z\in\mathcal Z}\{\langle y,z\rangle-\theta(z)\}~,
\end{align} 
for all $y\in\mathcal Z^\star$. In turn, the biconjugate $\theta^{\star\star}\equiv (\theta^\star)^\star: \mathcal Z\to\mathbf R$ of $\theta$ is
\begin{align}
\theta^{\star\star}(z)=\sup_{y\in\mathcal Z^\star}\{\langle y,z\rangle-\theta^\star(y)\}~.
\end{align}
Thus, the shape enforcing operator $\Upsilon$ in this case assigns each $\theta$ with its biconjugate $\theta^{\star\star}$. While well understood in mathematics (see, e.g., \citet[p.494]{Zeidler1990III}), the connection of GCM to the Legendre-Fenchel transform appears to be largely unnoticed in econometrics and statistics until the recent work by \citet{ChenChernozhukovFernandezKostyshakLuo2021Shape}. This connection enables \citet{ChenChernozhukovFernandezKostyshakLuo2021Shape} to propose a linear programming algorithm for the computation of $\theta^{\star\star}$---see Section \ref{Sec: implementation} for more details.

Following the literature, notably \citet{BeareMoon2015DR}, we may construct the Wald functional based on the GCM operator $\Upsilon$ as: for all $\theta\in\ell^\infty(\mathcal Z)$ and $p\in[1,\infty]$,
\begin{align}\label{Eqn: test functional, GCM}
\phi(\theta)=\|\theta-\Upsilon(\theta)\|_p~.
\end{align}
Our next theorem establishes the analytic properties of $\phi$.

\begin{thm}\label{Thm: Greatest convex minorant}
Let $\mathcal Z\subset\mathbf R^{d_z}$ be a nonempty bounded and convex set, and $p\in[1,\infty]$. Then map $\phi$ in \eqref{Eqn: test functional, GCM} satisfies Assumption \ref{Ass: space and map}.
\end{thm}

Positive homogeneity and Lipschitz continuity are well understood \citep{BeareFang2016Grenander,ChenChernozhukovFernandezKostyshakLuo2021Shape}, though proving convexity of $\phi$ is nontrivial (to us). Theorem \ref{Thm: Greatest convex minorant} remains true for concavity if we set $\Upsilon(\theta)=-(-\theta)^{\star\star}$ in the construction \eqref{Eqn: test functional, GCM}. We note that the boundedness of $\mathcal Z$ may be dispensed with at the cost of introducing a suitable weighting function in the definition of the $L^p$ norm. Theorems \ref{Thm: size and power} and \ref{Thm: Greatest convex minorant} together extend the use of GCM/LCM to nonparametric settings where $\theta_0$ may be convex/concave with respect to multiple variables and/or may not be $\sqrt n$-estimable. As GCM/LCM may be obtained through linear programming, the test based on \eqref{Eqn: test functional, GCM} may be more desirable than one based on $L^2$-projection (which requires quadratic programming), if computation cost is a binding constraint.

\subsubsection{Distance and Projection}\label{Sec: analytic properties}

If $\Lambda$ is the class of all elements in $\mathbf B$ satisfying the shape restriction in question, then it is natural to form $\phi$ as the distance function defined in \eqref{Eqn: test functional, distance} with $\|\cdot\|$ being $\|\cdot\|_{\mathbf B}$. By construction, statistical properties of the resulting test heavily depends on the shape of $\Lambda$ and the norm $\|\cdot\|_{\mathbf B}$. If $\Lambda$ is a nonempty closed convex set and $\mathbf B$ is a Hilbert space (a complete inner product space), then every $\theta\in\mathbf B$ admits a unique element in $\Lambda$, denoted $\Pi_{\Lambda}(\theta)$ and called the projection of $\theta$ onto $\Lambda$, which is closest to $\theta$. Thus, the map $\phi$ in \eqref{Eqn: test functional, distance} reduces to a particular instance of \eqref{Eqn: test functional, operator}: for any $\theta\in\mathbf B$,
\begin{align}\label{Eqn: test functional, distance L2}
\phi(\theta)=\|\theta-  \Pi_\Lambda(\theta)\|_{\mathbf B}~.
\end{align}
This is pursued in \citet{FangSantos2018HDD}, and further in \citet{FangSeo2019Shape} who additionally exploit $\Lambda$ being a cone. Projection onto closed convex sets/cones in Hilbert spaces enjoy elegant analytic properties that these authors extensively utilize to establish the statistical properties of their tests.

In particular settings, however, practitioners may wish to work with a different norm, such as the uniform norm if he/she wants to guard against alternatives uniformly deviating from the null. Such an extension is nontrivial for two reasons. First, the projection operator $\Pi_\Lambda$ in non-Hilbert spaces is in general not well-defined/behaved in the sense that it may be empty-valued, set-valued, discontinuous, or continuous but not uniformly so---see \citet[pp.211-2]{BarbuPrecupanu2012Convexity}, \citet[p.45]{DontchevZolezzi1993WellPosed}, and \citet{Alber1996Projection}. Second, employing the constrained estimator $\Pi_\Lambda(\hat\theta_n)$ as in \citet{FangSeo2019Shape} (to improve power) is also problematic in view of the erratic behaviors of $\Pi_\Lambda$. The first challenge prompts us to take a step back and work directly with the distance function \eqref{Eqn: test functional, distance}, which remains well-defined. Importantly, $\phi$ enjoys attractive analytic properties, as summarized in the following well known lemma.


\begin{lem}\label{Lem: distance}
If $\Lambda$ is a nonempty closed convex cone in a Banach space $\mathbf B$, then the map $\theta\mapsto\phi(\theta)=\inf_{\lambda\in\Lambda}\|\theta-\lambda\|_{\mathbf B}$ satisfies Assumption \ref{Ass: space and map}.
\end{lem}

To circumvent the second challenge, we employ shape enforcing operators designed for specific restrictions, such as the rearrangement for monotonicity and GCM/LCM for convexity/concavity. For the joint restriction of monotonicity and convexity/concavity, if $\Upsilon_1$ denotes rearrangement and $\Upsilon_2$ GCM, then the composition $\Upsilon_2\circ\Upsilon_1$ satisfies Assumption \ref{Ass: enforce null} but not $\Upsilon_1\circ\Upsilon_2$---see Remark 17 in \citet{ChenChernozhukovFernandezKostyshakLuo2021Shape}. Together with Theorem \ref{Thm: size and power}, Lemma \ref{Lem: distance} thus extends \citet{FangSeo2019Shape} to a distance-based test under alternative norms. In particular, the sup-norm yields a competing test to existing sup-type tests such as \citet{ChernozhukovLeeRosen2013Intersection} and \citet{Chetverikov2018Monotonicity}.

\subsection{Implementation}\label{Sec: implementation}


For the convenience of practitioners, we now provide an implementation guide.

\noindent\underline{\sc Step 1:} Compute the test statistic $r_n\phi(\hat\theta_n)$.

There are two aspects involved:  obtain the estimator $\hat\theta_n$ and its rate $r_n$, and compute $r_n\phi(\hat\theta_n)$ for a given $\hat\theta_n$. The former may be obtained by standard procedures such as kernel or sieve estimation---see \citet{FangSeo2019Shape} for more details. Given $\hat\theta_n$ and $r_n$, the key remains to compute $\phi(\hat\theta_n)$ which we demonstrate through two examples. Let $\mathbf B=\ell^\infty([0,1])$ and set $\vartheta=[\hat\theta_n(z_0),\ldots,\hat\theta_n(z_{N-1})]^\transpose$ for a large enough $N$ and $z_j\equiv j/(N-1)$. Consider first the set $\Lambda$ of weakly increasing functions in $\ell^\infty([0,1])$, and $\phi$ being the distance function. Then we may approximate $\phi(\hat\theta_n)$ by solving
\begin{align}\label{Eqn: distance, computation}
\max_{h\in\mathbf R^N: D_Nh\ge 0} \|h-\vartheta\|_\infty~,
\end{align}
where $D_N\in\mathbf M^{(N-1)\times N}$ is the matrix such that $D_Nh=[h^{(2)}-h^{(1)},\ldots, h^{(N)}-h^{(N-1)}]^\transpose$. Note that \eqref{Eqn: distance, computation} is a linear programming problem \citep[p.293]{BoydVandenberghe2004Convex}. Consider now the set $\Lambda$ consisting of convex functions in $\ell^\infty([0,1])$, and the Wald functional $\theta\mapsto\phi(\theta)=\|\theta-\theta^{\star\star}\|_\infty$. Then we may approximate $\phi(\hat\theta_n)$ by $\|\vartheta-\vartheta^{\star\star}\|_\infty$ where $l$th entry of $\vartheta^{\star\star}$ is given by (see, e.g., \citet[Lemma 1]{Carolan2002LCM}):
\begin{align}\label{Eqn: GCM, computation}
\min_{j=0}^{l-1} \min_{k=l-1}^{N-1}\frac{(z_k-z_{l-1})\hat\theta_n(z_j)+ (z_{l-1}-z_j)\hat\theta_n(z_k)}{z_k-z_j}~,
\end{align}
where the objective in \eqref{Eqn: GCM, computation} becomes $\hat\theta_n(z_{l-1})$ if $j=k=l-1$.

Both examples may be extended to multivariate settings. Concretely, let $\mathbf B=\ell^\infty([0,1]^{d_z})$, $\{z_j\}_{j=1}^N$ be a class of grid points over $[0,1]^{d_z}$, and $\vartheta=[\hat\theta_n(z_1),\ldots,\hat\theta_n(z_N)]^\transpose$. Then one may approximate $\phi(\hat\theta_n)$ in the first example by solving \eqref{Eqn: distance, computation} but now subject to $Ah\ge 0$ for some matrix $A$---see Example B.1 in \citet{FangSeo2019Shape}. For the second example,  following \citet{ChenChernozhukovFernandezKostyshakLuo2021Shape}, we approximate $\phi(\hat\theta_n)$ by $\|\vartheta-\vartheta^{\star\star}\|_\infty$ where $l$th entry of $\vartheta^{\star\star}$ is given by solving the linear programming problem:
\begin{align} 
\max_{v\in\mathbf R,\xi\in\mathbf R^{d_z}} \quad & v \notag\\
\textrm{s.t.} \quad &  v+\xi^\transpose(z_j-z_l)\le \hat\theta_n(z_j),\,j=1,\ldots,N~.
\end{align}

\noindent\underline{\sc Step 2:} Construct the critical value $\hat c_{n,1-\alpha}$ with $\alpha\in(0,1)$.

In this step, we presume that the practitioner is capable of computing/approximating $\phi(\theta)$ for any given $\theta\in\mathbf B$ (as described above).

\begin{enumerate}[label=(\roman*)]
  \item Generate a collection of bootstrap estimates $\{\hat{\mathbb G}_{n,b}\}_{b=1}^B$ for $\hat{\mathbb G}_n$ (e.g., $B=200$ or $1000$)---see \citet{FangSeo2019Shape} for more details.
  \item Set $\hat\kappa_n=r_nc_n/\hat\tau_{n,1-\gamma_n}$ where $\hat\tau_{n,1-\gamma_n}$ is the $(1-\gamma_n)$-quantile of $\|\hat{\mathbb G}_{n,1}\|_{\mathbf B},\ldots, \|\hat{\mathbb G}_{n,B}\|_{\mathbf B}$. We recommend $\gamma_n=0.01/\log n$ or $1/n$ as in \citet{FangSeo2019Shape} and note that $c_n=1/\log n$ in nonparametric regression models. In the examples of {\sc Step} 1, the $\|\cdot\|_{\mathbf B}$ norms may be approximated based on the grid points.
  \item Compute $\Gamma(\hat\theta_n)$. For the convexity examples in {\sc Step 1}, this requires no additional computation as one may simply let $\Gamma(\hat\theta_n)$ be $\vartheta^{\star\star}$. For the monotonicity example with $\mathbf B=\ell^\infty([0,1])$, one may let $\Gamma$ be the rearrangement operator so $\Gamma(\hat\theta_n)$ is just the sorted version of $\hat\theta_n$. If instead $\mathbf B=\ell^\infty([0,1]^{d_z})$, then rearrangement may be implemented as follows \citep{ChernozhukovFernandezGalichon2009Improving}:
      \begin{align}\label{Eqn: rearrangement, multivariate}
      \Gamma(\hat\theta_n)=\frac{1}{|\Pi|}\sum_{\pi\in\Pi}\mathcal M_\pi\hat\theta_n~,
      \end{align}
      where $\Pi$ consists of all permutations of $1,\ldots,d_z$, $|\Pi|$ is the cardinality of $\Pi$, and, for each permutation $\pi\equiv(\pi_1,\ldots,\pi_{d_z})$, $\mathcal M_\pi\hat\theta_n\equiv\mathcal M_{\pi_1}\circ\cdots\circ\mathcal M_{\pi_{d_z}}\hat\theta_n$ with $\mathcal M_j\hat\theta_n$ the sorted version of $\hat\theta_n$ viewed as a function of its $j$th argument (holding others fixed). The averaging in \eqref{Eqn: rearrangement, multivariate} is to eliminate the ambiguity caused by the order in which rearrangement is implemented.

  \item Approximate $\hat c_{n,1-\alpha}$ by the $(1-\alpha)$-quantile of the $B$ numbers
\begin{align}\label{Eqn: implementation, aux3}
\phi(\hat{\mathbb G}_{n,1}+\hat\kappa_n\Gamma(\hat\theta_n)), \ldots, \phi(\hat{\mathbb G}_{n,B}+\hat\kappa_n\Gamma(\hat\theta_n))~.
\end{align}
\end{enumerate}

\noindent\underline{\sc Step 3:} Reject $\mathrm H_0$ if and only if $r_n\phi(\hat\theta_n)>\hat c_{n,1-\alpha}$.

\section{Simulation Studies}\label{Sec: simulation}

We next conduct Monte Carlo simulations to examine the finite sample performance of our test, based on both univariate and bivariate designs of nonparametric regression models. Throughout, the significance level is  5\%, the number of Monte Carlo replications is 3000, and the number of bootstrap samples for each replication is 200. The tuning parameter $\kappa_n$ will be selected as in Proposition \ref{Pro: tuning parameter}, which entails a choice of $\gamma_n$. Since prior studies have repeatedly shown that testing results are quite insensitive to $\gamma_n$, we choose three values for $\gamma_n$: $0.01$, $0.01/\log n$, and $1/n$.

\subsection{Testing Monotonicity}\label{Sec: MC, Mon}

In this section, we are concerned with monotonicity. For the univariate designs, the regression function $\theta_0:[-1,1]\to\mathbf R$ under the null is of the form:
\begin{align}\label{Eqn: MC1,aux1}
\theta_0(z)=\mathsf az-\mathsf b\varphi(\mathsf cz)~,
\end{align}
where $\varphi$ is the standard normal pdf, and $(\mathsf a,\mathsf b,\mathsf c)$ equals $(0,0,0)$, $(0.1,0.5,0.5)$ or $(0.5,2,1)$, labeled D1, D2 and D3 respectively. We then consider two sets of alternatives, in order to showcase the relative advantages of our test. The first set of alternatives consists of functions as in \eqref{Eqn: MC1,aux1} but with $\{(\mathsf a,\mathsf b,\mathsf c): \mathsf a =\mathsf c=-\Delta\delta,\mathsf b = 0.2\delta,\delta=1,\ldots,10\}$ for $\Delta=0.05$. The second set of alternatives are defined by: for $\mathsf b= 0.5\delta$ with $\delta=1,2,\ldots,10$,
\begin{align}\label{Eqn: MC1,aux2}
\theta_0(z)=-\mathsf b\varphi(|z|^{1.5})~.
\end{align}
The specification in \eqref{Eqn: MC1,aux2} is inspired by \citet{AndrewsShi2013CMI} and \citet[Section L.1]{ChernozhukovLeeRosen2013Intersection}, but modified so that the empirical power is close to one for $\delta$ close to $10$. Figure \ref{Fig: MC} depicts the curves of $\theta_P$ based on these designs. Importantly, $\theta_0$ in Figure \ref{Fig: MC}-(b) becomes sharp V-shaped as $\delta$ increases, while $\theta_0$ in Figure \ref{Fig: MC}-(c) has a visually ``flat'' bottom for each $\delta$. Finally, we draw i.i.d.\ samples $\{Z_i^*,u_i\}_{i=1}^n$ with $n\in\{500,750,1000\}$ from the standard normal distribution in $\mathbf R^2$, and set $Z_i=-1+2\Phi(Z_i^*)\in[-1,1]$ and $Y_i=\theta_0(Z_i)+u_i$, with $\Phi$ the standard normal cdf.

{
\newlength\figurewidth
\setlength\figurewidth{0.35\textwidth}
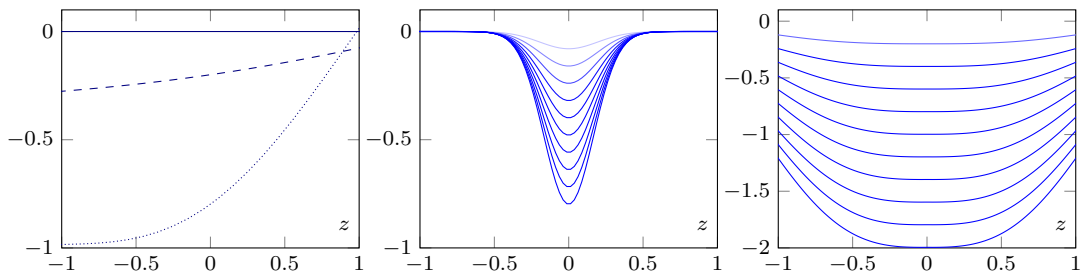
\begin{figure}[!h]
\centering\scriptsize
\begin{tikzpicture} 
\begin{groupplot}[group style={group name=myplots,group size=3 by 2,horizontal sep= 0.8cm,vertical sep=1.1cm},
    grid = minor,
    width = 0.375\textwidth,
    xmax=1,xmin=-1,
    xlabel=$z$,
    x label style={at={(axis description cs:0.95,0.04)},anchor=south},
    xtick={-1,-0.5,...,1},
    enlargelimits=false,
    tick label style={/pgf/number format/fixed},
    no markers,
    title style={at={(xticklabel cs:0.5)}, below=3ex, text width=\figurewidth},
]
\nextgroupplot[samples=100,ymax=0.1,ymin=-1, ytick={0,-0.5,-1},title={\subcaption{Designs in \eqref{Eqn: MC1,aux1} under $\mathrm H_0$.}}]
\addplot[NavyBlue]{0};
\addplot[NavyBlue,dashed]{0.1*x-0.5*npdf(0.5*x, 0, 1)};
\addplot[NavyBlue,densely dotted]{0.5*x-2*npdf(x, 0, 1)};
\nextgroupplot[samples=400,ymax=0.1,ymin=-1,title={\subcaption{Designs in \eqref{Eqn: MC1,aux1} under $\mathrm H_1$.}}]
\foreach \b in {0.2,0.4,...,2}{\pgfmathsetmacro{\k}{\b*120}
\edef\temp{\noexpand
\addplot+[solid,color=Blue!\k]{-\b*npdf((5+\b/2)*x, 0, 1)};
}\temp
}
\nextgroupplot[samples=400,ymax=0.1,ymin=-2,title={\subcaption{Designs in \eqref{Eqn: MC1,aux2} under $\mathrm H_1$.}}]
\foreach \b in {0.5,1,...,5}{\pgfmathsetmacro{\k}{\b*120}
\edef\temp{\noexpand
\addplot+[solid,color=Blue!\k]{-\b*npdf(abs(x)^1.5, 0, 1)};
}\temp
}
\end{groupplot}
\end{tikzpicture}
\caption{In the left figure, D1, D2 and D3 are solid, dashed and dotted respectively. In the middle and right figures, from top to bottom are curves corresponding to $\delta=1,2,\ldots,10$.} \label{Fig: MC}
\end{figure}
}

For the bivariate designs, the regression function $\theta_0:[0,1]^2\to\mathbf R$ is specified as in \citet{FangSeo2019Shape}: for some $(\mathsf a,\mathsf b,\mathsf c)\in\mathbf R^3$,
\begin{align}\label{Eqn: MC1,aux3}
\theta_0(z_1,z_2)=\mathsf a\big(\frac{1}{2}z_1^{\mathsf b}+\frac{1}{2}z_2^{\mathsf b}\big)^{1/\mathsf b}+\mathsf c\log(1+z_1+z_2)~,
\end{align}
where the first term on the right-hand side of \eqref{Eqn: MC1,aux3} equals $\mathsf a\sqrt{z_1z_2}$ when $\mathsf b=0$. We consider three choices of $(\mathsf a,\mathsf b,\mathsf c)$ under $\mathrm H_0$, namely $(0,0,0)$, $(0.2,1,0)$ and $(0.5,0,0.5)$ (labeled D1, D2, and D3, respectively), and, for the alternative, the collection $\{(\mathsf a,\mathsf b,\mathsf c): \mathsf a =\mathsf c=-\Delta\delta,\mathsf b = 0.2\delta,\delta=1,\ldots,10\}$ for $\Delta=0.05$. We draw i.i.d.\ samples $\{Z_{1i}^*,Z_{2i}^*,u_i\}_{i=1}^n$ with $n\in\{500,750,1000\}$ from the standard normal distribution in $\mathbf R^3$, and set $Y_i=\theta_0(Z_i)+u_i$ where $Z_i\equiv(Z_{1i},Z_{2i})$ with $Z_{ji}=\Phi(Z_{ji}^*)\in[0,1]$ for all $i$ and $j=1,2$.

The implementation of our test is based on series least squares estimation using B-splines. Specifically, we employ cubic B-splines with 3, 5, or 7 interior knots for the univariate designs, and quadratic as well as cubic B-splines each with one or zero knots for the bivariate designs. The knots in both cases are placed at the equispaced empirical quantiles of the regressors. We choose the test statistics based on the supremum distance as described in Section \ref{Sec: implementation}. In turn, $\hat{\mathbb G}_{n,b}$ is obtained by score bootstrap with i.i.d.\ weights from the standard normal distribution, and the coupling rate $c_n=1/\log n$. For ease of reference, we label our test with quadratic B-splines and $j$ knots as F-Q$j$; similarly, F-C$j$ is the implementation with cubic B-splines and $j$ knots.

For the sake of fair comparisons, we implement the sup-test of \citet{ChernozhukovLeeRosen2013Intersection} and the one-step test of \citet{Chetverikov2018Monotonicity} (labelled as C-OS) which is also of sup-type. For the former, we closely follow the steps articulated in Section 6.1 of \citet{ChernozhukovLeeRosen2013Intersection}, and label the resulting test as CLR-C$j$ if the estimation is based on cubic B-splines with $j$ interior knots---CLR-Q$j$ is similarly defined. The C-OS test is implemented as in \citet{FangSeo2019Shape}---see also \citet[p.749]{Chetverikov2018Monotonicity} and its working paper version for more details.

{
\setlength{\tabcolsep}{6.5pt}
\renewcommand{\arraystretch}{1.1}
\begin{table}[!ht]
\caption{Empirical Size of Monotonicity Tests for $\theta_0$ in \eqref{Eqn: MC1,aux1} at $\alpha=5\%$} \label{Tab: MonMC, size1} 
\centering\footnotesize
\begin{threeparttable}
\sisetup{table-number-alignment = center, table-format = 1.3} 
\begin{tabularx}{\linewidth}{@{} cc *{3}{S[round-mode = places,round-precision = 3]}  c *{3}{S[round-mode = places,round-precision = 3]}  c *{3}{S[round-mode = places,round-precision = 3]}@{}} 
\hline
\hline
 $n$ & $\gamma_n$ & {D1} & {D2} & {D3}  & & {D1} & {D2} & {D3} & & {D1} & {D2} & {D3}\\
\hline
\rule{0pt}{15pt}
&& \multicolumn{3}{c}{{F-C3: $k_n=7$}} && \multicolumn{3}{c}{{F-C5: $k_n=9$}} && \multicolumn{3}{c}{{F-C7: $k_n=11$}}\\
\multirow{3}{*}{$500$}   & $1/n$                  & 0.0587 & 0.0167 & 0.0097 & & 0.0643 & 0.0320 & 0.0157 & & 0.0647 & 0.0327 & 0.0143\\
                         & $0.01/\log n$          & 0.0587 & 0.0167 & 0.0097 & & 0.0643 & 0.0320 & 0.0157 & & 0.0647 & 0.0327 & 0.0143\\
                         & $0.01$                 & 0.0600 & 0.0173 & 0.0103 & & 0.0653 & 0.0327 & 0.0157 & & 0.0653 & 0.0330 & 0.0143\\
\rule{0pt}{12pt}
 \multirow{3}{*}{$750$}  & $1/n$                  & 0.0613 & 0.0203 & 0.0080 & & 0.0607 & 0.0213 & 0.0107 & & 0.0630 & 0.0273 & 0.0133\\
                         & $0.01/\log n$          & 0.0613 & 0.0203 & 0.0080 & & 0.0607 & 0.0213 & 0.0107 & & 0.0630 & 0.0273 & 0.0133\\
                         & $0.01$                 & 0.0633 & 0.0203 & 0.0083 & & 0.0610 & 0.0220 & 0.0110 & & 0.0637 & 0.0280 & 0.0133\\
\rule{0pt}{12pt}
 \multirow{3}{*}{$1000$} & $1/n$                  & 0.0647 & 0.0150 & 0.0057 & & 0.0690 & 0.0247 & 0.0113 & & 0.0587 & 0.0280 & 0.0127\\
                         & $0.01/\log n$          & 0.0647 & 0.0150 & 0.0057 & & 0.0690 & 0.0247 & 0.0113 & & 0.0587 & 0.0280 & 0.0127\\
                         & $0.01$                 & 0.0657 & 0.0153 & 0.0060 & & 0.0690 & 0.0250 & 0.0127 & & 0.0593 & 0.0287 & 0.0130\\
\rule{0pt}{15pt}
&& \multicolumn{3}{c}{CLR-C3: $k_n=7$} && \multicolumn{3}{c}{CLR-C5: $k_n=9$} && \multicolumn{3}{c}{CLR-C7: $k_n=11$}\\
$500$ &                      & 0.0603 & 0.0387 & 0.0173 & & 0.0650 & 0.0490 & 0.0233 & & 0.0703 & 0.0503 & 0.0260\\
$750$ &                      & 0.0623 & 0.0403 & 0.0170 & & 0.0630 & 0.0433 & 0.0170 & & 0.0677 & 0.0487 & 0.0210\\
$1000$&                      & 0.0590 & 0.0230 & 0.0097 & & 0.0630 & 0.0427 & 0.0173 & & 0.0607 & 0.0433 & 0.0200\\
\rule{0pt}{15pt}
&& \multicolumn{3}{c}{C-OS: $n=500$} && \multicolumn{3}{c}{C-OS: $n=750$} && \multicolumn{3}{c}{C-OS: $n=1000$}\\
 \multicolumn{2}{c}{ }                        & 0.0550 & 0.0430 & 0.0143 & & 0.0543 & 0.0407 & 0.0130 & & 0.0563 & 0.0377 & 0.0103\\
\hline
\hline
\end{tabularx}
\begin{tablenotes}[flushleft]
\item {\it Note:} The parameter $\gamma_n$ determines $\hat\kappa_n$ as in Proposition \ref{Pro: tuning parameter} with $c_n=1/\log n$ and $r_n=(n/k_n)^{1/2}$.
\end{tablenotes}
\end{threeparttable}
\end{table}
}

{
\setlength{\tabcolsep}{3pt}
\renewcommand{\arraystretch}{1.1}
\begin{table}[!ht]
\caption{Empirical Size of Monotonicity Tests for $\theta_0$ in \eqref{Eqn: MC1,aux3} at $\alpha=5\%$} \label{Tab: MonMC, size2}
\centering\footnotesize
\begin{threeparttable}
\sisetup{table-number-alignment = center, table-format = 1.3} 
\begin{tabularx}{\linewidth}{@{}cc *{3}{S[round-mode = places,round-precision = 3]} c *{3}{S[round-mode = places,round-precision = 3]} c *{3}{S[round-mode = places,round-precision = 3]} c *{3}{S[round-mode = places,round-precision = 3]}@{}} 
\hline
\hline
$n$ & $\gamma_n$ & {D1} & {D2} & {D3}  && {D1} & {D2} & {D3} && {D1} & {D2} & {D3} && {D1} & {D2} & {D3}\\
\hline
\rule{0pt}{15pt}
&& \multicolumn{3}{c}{F-Q0: $k_n=9$} && \multicolumn{3}{c}{F-Q1: $k_n=16$} && \multicolumn{3}{c}{F-C0: $k_n=16$} && \multicolumn{3}{c}{F-C1: $k_n=25$}\\
\multirow{3}{*}{$500$} & $1/n$         & 0.0590 & 0.0283 & 0.0020 & & 0.0720 & 0.0360 & 0.0050 & & 0.0747 & 0.0363 & 0.0050 & & 0.0627 & 0.0413 & 0.0087\\
                       & $0.01/\log n$ & 0.0590 & 0.0283 & 0.0020 & & 0.0720 & 0.0360 & 0.0050 & & 0.0747 & 0.0363 & 0.0050 & & 0.0627 & 0.0413 & 0.0087\\
                       & $0.01$        & 0.0593 & 0.0283 & 0.0020 & & 0.0723 & 0.0360 & 0.0050 & & 0.0750 & 0.0367 & 0.0050 & & 0.0630 & 0.0423 & 0.0087\\
\rule{0pt}{12pt}
 \multirow{3}{*}{$750$}& $1/n$         & 0.0603 & 0.0240 & 0.0003 & & 0.0633 & 0.0270 & 0.0023 & & 0.0670 & 0.0297 & 0.0007 & & 0.0640 & 0.0347 & 0.0040\\
                       & $0.01/\log n$ & 0.0603 & 0.0240 & 0.0003 & & 0.0633 & 0.0270 & 0.0023 & & 0.0670 & 0.0297 & 0.0007 & & 0.0640 & 0.0347 & 0.0040\\
                       & $0.01$        & 0.0613 & 0.0240 & 0.0007 & & 0.0637 & 0.0273 & 0.0023 & & 0.0677 & 0.0303 & 0.0010 & & 0.0647 & 0.0347 & 0.0040\\
\rule{0pt}{12pt}
\multirow{3}{*}{$1000$}& $1/n$         & 0.0563 & 0.0203 & 0.0003 & & 0.0550 & 0.0240 & 0.0003 & & 0.0567 & 0.0247 & 0.0003 & & 0.0600 & 0.0287 & 0.0037\\
                       & $0.01/\log n$ & 0.0563 & 0.0203 & 0.0003 & & 0.0550 & 0.0240 & 0.0003 & & 0.0567 & 0.0247 & 0.0003 & & 0.0600 & 0.0287 & 0.0037\\
                       & $0.01$        & 0.0570 & 0.0210 & 0.0003 & & 0.0550 & 0.0243 & 0.0003 & & 0.0570 & 0.0253 & 0.0003 & & 0.0600 & 0.0287 & 0.0037\\
\rule{0pt}{15pt}
&& \multicolumn{3}{c}{CLR-Q0: $k_n=9$} && \multicolumn{3}{c}{CLR-Q1: $k_n=16$} && \multicolumn{3}{c}{CLR-C0: $k_n=16$} && \multicolumn{3}{c}{CLR-C1: $k_n=25$}\\
$500$ &               & 0.0720 & 0.0377 & 0.0060 && 0.1027 & 0.0730 & 0.0180 && 0.1003 & 0.0723 & 0.0190 && 0.1337 & 0.1120 & 0.0460\\
$750$ &               & 0.0713 & 0.0333 & 0.0023 && 0.0897 & 0.0563 & 0.0100 && 0.0923 & 0.0610 & 0.0093 && 0.1023 & 0.0767 & 0.0310\\
$1000$&               & 0.0630 & 0.0270 & 0.0023 && 0.0833 & 0.0493 & 0.0080 && 0.0797 & 0.0500 & 0.0110 && 0.0930 & 0.0660 & 0.0177\\
\rule{0pt}{15pt}
&& \multicolumn{3}{c}{C-OS: $n=500$} && \multicolumn{3}{c}{C-OS: $n=750$} && \multicolumn{3}{c}{C-OS: $n=1000$} && \multicolumn{3}{c}{ }\\
\multicolumn{2}{c}{ }                & 0.0593 & 0.0460 & 0.0217 && 0.0560 & 0.0430 & 0.0157 && 0.0587 & 0.0407 & 0.0200 &&   &  &  \\
\hline
\hline
\end{tabularx}
\begin{tablenotes}[flushleft]
\item {\it Note:} The parameter $\gamma_n$ determines $\hat\kappa_n$ as in Proposition \ref{Pro: tuning parameter} with $c_n=1/\log n$ and $r_n=(n/k_n)^{1/2}$.
\end{tablenotes}
\end{threeparttable}
\end{table}
}

Tables \ref{Tab: MonMC, size1} and \ref{Tab: MonMC, size2} present the empirical sizes. In the univariate designs, all tests seem to control size reasonably well, though our tests and the CLR tests slightly over-reject. In the bivariate designs, while the rejections rates of our tests and the C-OS test are close to the nominal level (at least in large samples), the CLR tests are notably over-sized even in large samples. One possible explanation is that they are derivative-based tests and so the slow rate of convergence \citep{Stone1982Global} is exacerbated in the bivariate designs. Figure \ref{Fig: MC,Mon} in turn depicts the power curves, where we only present our test with $\gamma_n=0.01/\log n$ as other choices lead to very similar results (here and below). Overall, these curves demonstrate that our test is competitive to the tests of \citet{ChernozhukovLeeRosen2013Intersection} and \citet{Chetverikov2018Monotonicity} in terms of power as well. We note that, while the C-OS test enjoys some adaptivity and optimality properties in univariate settings, its bivariate version only captures part of the discordance between the outcome and the regressors so that its power in our bivariate designs is relatively low.

\pgfplotstableread{ 
delta alpha MonFiveKn3 MonFiveKn5 MonFiveKn7 MonSevenKn3 MonSevenKn5 MonSevenKn7 MonTenKn3 MonTenKn5 MonTenKn7
0     0.05    0.0587     0.0643     0.0647     0.0613      0.0607      0.0630      0.0647    0.0690    0.0587
1     0.05    0.0713     0.0727     0.0667     0.0847      0.0777      0.0737      0.0860    0.0810    0.0753
2     0.05    0.1177     0.1017     0.0923     0.1527      0.1350      0.1120      0.1747    0.1577    0.1380
3     0.05    0.1947     0.1667     0.1453     0.2823      0.2290      0.2010      0.3730    0.3143    0.2590
4     0.05    0.3227     0.2727     0.2293     0.4827      0.4023      0.3533      0.6190    0.5437    0.4580
5     0.05    0.4820     0.4290     0.3557     0.6917      0.6213      0.5437      0.8363    0.7690    0.6733
6     0.05    0.6650     0.5900     0.5033     0.8567      0.8033      0.7263      0.9397    0.9173    0.8600
7     0.05    0.8070     0.7460     0.6587     0.9513      0.9217      0.8697      0.9907    0.9780    0.9537
8     0.05    0.9080     0.8637     0.8027     0.9883      0.9783      0.9533      0.9987    0.9963    0.9863
9     0.05    0.9630     0.9420     0.8950     0.9970      0.9973      0.9860      1.0000    0.9997    0.9983
10    0.05    0.9890     0.9810     0.9560     0.9997      0.9990      0.9983      1.0000    1.0000    0.9997
}\UniMona

\pgfplotstableread{ 
delta alpha MonFiveKn3 MonFiveKn5 MonFiveKn7 MonSevenKn3 MonSevenKn5 MonSevenKn7 MonTenKn3 MonTenKn5 MonTenKn7
0     0.05    0.0587     0.0643     0.0647     0.0613      0.0607      0.0630      0.0647    0.0690    0.0587
1     0.05    0.0663     0.0713     0.0720     0.0787      0.0667      0.0700      0.0793    0.0777    0.0717
2     0.05    0.1033     0.0900     0.0917     0.1310      0.0967      0.0923      0.1357    0.1157    0.1080
3     0.05    0.1587     0.1307     0.1210     0.2170      0.1593      0.1453      0.2543    0.2000    0.1727
4     0.05    0.2433     0.1847     0.1690     0.3473      0.2570      0.2177      0.4357    0.3213    0.2727
5     0.05    0.3543     0.2703     0.2247     0.5260      0.3900      0.3213      0.6427    0.4940    0.4177
6     0.05    0.4883     0.3757     0.3147     0.6877      0.5433      0.4473      0.8013    0.6683    0.5677
7     0.05    0.6213     0.4950     0.4143     0.8317      0.6867      0.5883      0.9167    0.8130    0.7100
8     0.05    0.7470     0.6110     0.5217     0.9260      0.8177      0.7157      0.9767    0.9153    0.8260
9     0.05    0.8500     0.7323     0.6303     0.9747      0.9117      0.8270      0.9943    0.9680    0.9140
10    0.05    0.9237     0.8203     0.7260     0.9923      0.9623      0.9020      0.9990    0.9933    0.9703
}\UniMonb

\pgfplotstableread{ 
delta alpha MonFiveQKn0 MonFiveQKn1 MonSevenQKn0 MonSevenQKn1 MonTenQKn0 MonTenQKn1
0     0.05    0.0590      0.0720      0.0603       0.0633       0.0563     0.0550
1     0.05    0.0803      0.0967      0.0877       0.0987       0.0847     0.0873
2     0.05    0.1113      0.1207      0.1307       0.1360       0.1337     0.1340
3     0.05    0.1560      0.1557      0.1977       0.1883       0.2117     0.2100
4     0.05    0.2107      0.2030      0.2723       0.2520       0.3120     0.2913
5     0.05    0.2773      0.2573      0.3783       0.3370       0.4463     0.4190
6     0.05    0.3650      0.3280      0.4923       0.4463       0.5893     0.5457
7     0.05    0.4650      0.4047      0.6070       0.5597       0.7283     0.6883
8     0.05    0.5700      0.5003      0.7190       0.6760       0.8410     0.8067
9     0.05    0.6633      0.5993      0.8223       0.7787       0.9257     0.8920
10    0.05    0.7407      0.6817      0.8957       0.8537       0.9703     0.9460
}\BiMonQ

\pgfplotstableread{ 
delta alpha MonFiveCKn0 MonFiveCKn1 MonSevenCKn0 MonSevenCKn1 MonTenCKn0 MonTenCKn1
0     0.05    0.0747      0.0627      0.0670        0.0640      0.0567     0.0600
1     0.05    0.1000      0.0830      0.1020        0.0863      0.0903     0.0823
2     0.05    0.1283      0.0997      0.1437        0.1110      0.1370     0.1140
3     0.05    0.1637      0.1293      0.1953        0.1497      0.2167     0.1640
4     0.05    0.2203      0.1627      0.2670        0.2007      0.3083     0.2333
5     0.05    0.2733      0.2077      0.3583        0.2700      0.4303     0.3103
6     0.05    0.3440      0.2577      0.4740        0.3437      0.5670     0.4197
7     0.05    0.4273      0.3237      0.5853        0.4453      0.7067     0.5370
8     0.05    0.5317      0.3983      0.7000        0.5473      0.8230     0.6567
9     0.05    0.6270      0.4710      0.8020        0.6523      0.9037     0.7600
10    0.05    0.7093      0.5497      0.8723        0.7390      0.9527     0.8533
}\BiMonC

\pgfplotstableread{ 
delta alpha MonFiveKn3 MonFiveKn5 MonFiveKn7 MonSevenKn3 MonSevenKn5 MonSevenKn7 MonTenKn3 MonTenKn5 MonTenKn7
0     0.05   0.0603      0.0650     0.0703     0.0623      0.0630      0.0677      0.0590    0.0630    0.0607
1     0.05   0.0737      0.0737     0.0700     0.0817      0.0703      0.0717      0.0737    0.0817    0.0663
2     0.05   0.1167      0.1007     0.0823     0.1447      0.1113      0.0937      0.1687    0.1400    0.1033
3     0.05   0.1997      0.1530     0.1107     0.2863      0.1953      0.1350      0.3623    0.2520    0.1660
4     0.05   0.3307      0.2320     0.1560     0.4723      0.3317      0.2173      0.6157    0.4490    0.2753
5     0.05   0.4883      0.3507     0.2327     0.6740      0.5187      0.3357      0.8210    0.6710    0.4493
6     0.05   0.6617      0.5020     0.3247     0.8537      0.7080      0.4847      0.9370    0.8370    0.6377
7     0.05   0.8057      0.6510     0.4440     0.9477      0.8473      0.6480      0.9870    0.9397    0.7913
8     0.05   0.9053      0.7817     0.5787     0.9840      0.9343      0.7873      0.9977    0.9813    0.8993
9     0.05   0.9593      0.8850     0.6967     0.9960      0.9807      0.8910      0.9997    0.9973    0.9580
10    0.05   0.9850      0.9450     0.8070     1.0000      0.9953      0.9523      1.0000    1.0000    0.9907
}\UniMonCLRa

\pgfplotstableread{ 
delta alpha MonFiveKn3 MonFiveKn5 MonFiveKn7 MonSevenKn3 MonSevenKn5 MonSevenKn7 MonTenKn3 MonTenKn5 MonTenKn7
0     0.05    0.0603      0.0650    0.0703     0.0623      0.0630      0.0677      0.0590    0.0630    0.0607
1     0.05    0.0663      0.0710    0.0707     0.0690      0.0663      0.0677      0.0653    0.0677    0.0640
2     0.05    0.0790      0.0790    0.0723     0.0913      0.0770      0.0773      0.0933    0.0827    0.0700
3     0.05    0.1043      0.0853    0.0753     0.1223      0.0940      0.0867      0.1417    0.1007    0.0850
4     0.05    0.1397      0.1060    0.0840     0.1770      0.1220      0.0983      0.2237    0.1397    0.1050
5     0.05    0.1927      0.1313    0.0960     0.2633      0.1583      0.1217      0.3357    0.1797    0.1253
6     0.05    0.2523      0.1607    0.1157     0.3590      0.2113      0.1490      0.4623    0.2457    0.1700
7     0.05    0.3320      0.2033    0.1350     0.4810      0.2663      0.1837      0.6093    0.3370    0.2130
8     0.05    0.4230      0.2520    0.1633     0.6123      0.3550      0.2290      0.7400    0.4347    0.2747
9     0.05    0.5177      0.3077    0.1943     0.7337      0.4287      0.2740      0.8483    0.5487    0.3377
10    0.05    0.6287      0.3720    0.2233     0.8240      0.5267      0.3227      0.9287    0.6570    0.4057
}\UniMonCLRb

\pgfplotstableread{ 
delta alpha MonFiveQKn0 MonFiveQKn1 MonSevenQKn0 MonSevenQKn1 MonTenQKn0 MonTenQKn1
0     0.05    0.0720      0.1027       0.0713       0.0897      0.0630     0.0833
1     0.05    0.0933      0.1193       0.0947       0.1107      0.0937     0.1000
2     0.05    0.1277      0.1353       0.1330       0.1323      0.1397     0.1237
3     0.05    0.1717      0.1587       0.1800       0.1573      0.1983     0.1547
4     0.05    0.2143      0.1850       0.2550       0.1900      0.2870     0.1870
5     0.05    0.2733      0.2140       0.3437       0.2247      0.3923     0.2350
6     0.05    0.3533      0.2470       0.4457       0.2653      0.5123     0.2893
7     0.05    0.4287      0.2817       0.5530       0.3267      0.6513     0.3517
8     0.05    0.5110      0.3263       0.6563       0.3767      0.7727     0.4260
9     0.05    0.5873      0.3763       0.7587       0.4370      0.8660     0.5033
10    0.05    0.6767      0.4150       0.8420       0.5020      0.9283     0.5813
}\BiMonQCLR

\pgfplotstableread{ 
delta alpha MonFiveCKn0 MonFiveCKn1 MonSevenCKn0 MonSevenCKn1 MonTenCKn0 MonTenCKn1
0     0.05    0.1003      0.1337      0.0923        0.1023      0.0797     0.0930
1     0.05    0.1193      0.1440      0.1110        0.1143      0.0997     0.1053
2     0.05    0.1377      0.1530      0.1363        0.1337      0.1190     0.1147
3     0.05    0.1577      0.1717      0.1570        0.1443      0.1453     0.1340
4     0.05    0.1797      0.1840      0.1850        0.1617      0.1770     0.1543
5     0.05    0.2047      0.2013      0.2160        0.1773      0.2177     0.1697
6     0.05    0.2370      0.2277      0.2473        0.2057      0.2670     0.1967
7     0.05    0.2657      0.2403      0.2927        0.2273      0.3147     0.2243
8     0.05    0.3003      0.2567      0.3420        0.2487      0.3790     0.2627
9     0.05    0.3307      0.2850      0.3877        0.2813      0.4453     0.2990
10    0.05    0.3747      0.3043      0.4403        0.3100      0.5157     0.3287
}\BiMonCCLR

\pgfplotstableread{ 
delta alpha FivePI FiveOS FiveSD SevenPI SevenOS SevenSD TenPI   TenOS   TenSD
0     0.05  0.0550 0.0550 0.0550 0.0543  0.0543  0.0543  0.0563  0.0563  0.0563
1     0.05  0.0630 0.0630 0.0630 0.0630  0.0630  0.0630  0.0613  0.0613  0.0613
2     0.05  0.0800 0.0800 0.0800 0.0937  0.0937  0.0937  0.1150  0.1150  0.1150
3     0.05  0.1177 0.1177 0.1177 0.1743  0.1743  0.1743  0.2350  0.2350  0.2350
4     0.05  0.2050 0.2050 0.2050 0.3240  0.3240  0.3240  0.4357  0.4357  0.4357
5     0.05  0.3290 0.3290 0.3290 0.5253  0.5253  0.5253  0.6920  0.6920  0.6920
6     0.05  0.4830 0.4830 0.4830 0.7183  0.7183  0.7183  0.8627  0.8627  0.8627
7     0.05  0.6457 0.6457 0.6457 0.8737  0.8737  0.8737  0.9527  0.9527  0.9527
8     0.05  0.7807 0.7807 0.7807 0.9503  0.9503  0.9503  0.9873  0.9873  0.9873
9     0.05  0.8860 0.8860 0.8860 0.9873  0.9873  0.9873  0.9990  0.9990  0.9990
10    0.05  0.9480 0.9480 0.9480 0.9963  0.9963  0.9963  1.0000  1.0000  1.0000
}\UniMonCheta

\pgfplotstableread{ 
delta alpha FivePI FiveOS FiveSD SevenPI SevenOS SevenSD TenPI   TenOS   TenSD
0     0.05  0.0550 0.0550 0.0550 0.0543  0.0543  0.0543  0.0563  0.0563  0.0563
1     0.05  0.0523 0.0523 0.0523 0.0547  0.0547  0.0547  0.0643  0.0643  0.0643
2     0.05  0.0647 0.0647 0.0647 0.0697  0.0697  0.0697  0.0917  0.0917  0.0917
3     0.05  0.0877 0.0877 0.0877 0.1083  0.1083  0.1083  0.1540  0.1540  0.1540
4     0.05  0.1220 0.1220 0.1220 0.1930  0.1930  0.1930  0.2603  0.2603  0.2603
5     0.05  0.1900 0.1900 0.1900 0.3207  0.3207  0.3207  0.4310  0.4310  0.4310
6     0.05  0.2910 0.2910 0.2910 0.4757  0.4757  0.4757  0.6323  0.6323  0.6323
7     0.05  0.4053 0.4053 0.4053 0.6427  0.6427  0.6427  0.7970  0.7970  0.7970
8     0.05  0.5387 0.5387 0.5387 0.7940  0.7940  0.7940  0.9140  0.9140  0.9140
9     0.05  0.6750 0.6750 0.6750 0.8993  0.8993  0.8993  0.9703  0.9703  0.9703
10    0.05  0.7820 0.7820 0.7820 0.9623  0.9623  0.9623  0.9927  0.9927  0.9927
}\UniMonChetb

\pgfplotstableread{ 
delta alpha FivePI FiveOS FiveSD SevenPI SevenOS SevenSD TenPI   TenOS   TenSD
0     0.05  0.0593 0.0593 0.0593 0.0560  0.0560  0.0560  0.0587  0.0587  0.0587
1     0.05  0.0657 0.0657 0.0657 0.0660  0.0660  0.0660  0.0697  0.0697  0.0697
2     0.05  0.0703 0.0703 0.0703 0.0733  0.0733  0.0733  0.0833  0.0833  0.0833
3     0.05  0.0793 0.0793 0.0793 0.0877  0.0877  0.0877  0.0960  0.0960  0.0960
4     0.05  0.0853 0.0853 0.0853 0.1050  0.1050  0.1050  0.1187  0.1187  0.1187
5     0.05  0.0973 0.0973 0.0973 0.1270  0.1270  0.1270  0.1427  0.1427  0.1427
6     0.05  0.1127 0.1127 0.1127 0.1523  0.1523  0.1523  0.1823  0.1823  0.1823
7     0.05  0.1357 0.1357 0.1357 0.1820  0.1820  0.1820  0.2300  0.2300  0.2300
8     0.05  0.1543 0.1543 0.1543 0.2317  0.2317  0.2317  0.2913  0.2913  0.2913
9     0.05  0.1853 0.1853 0.1853 0.2900  0.2900  0.2900  0.3727  0.3727  0.3727
10    0.05  0.2147 0.2147 0.2147 0.3520  0.3520  0.3520  0.4643  0.4643  0.4643
}\BiMonChet

\begin{figure}[!h]
\centering\scriptsize
\begin{tikzpicture} 
\begin{groupplot}[group style={group name=myplots,group size=3 by 3,horizontal sep= 0.8cm,vertical sep=1.1cm},
    grid = minor,
    width = 0.375\textwidth,
    xmax=10,xmin=0,
    ymax=1,ymin=0,
    every axis title/.style={below,at={(0.2,0.8)}},
    xlabel=$\delta$,
    x label style={at={(axis description cs:0.95,0.04)},anchor=south},
    xtick={0,2,...,10},
    ytick={0.05,0.5,1},
    tick label style={/pgf/number format/fixed},
    legend style={font=\tiny,text=black,cells={align=center},row sep = 3pt,legend columns = -1, draw=none,fill=none},
    cycle list={
{smooth,tension=0.5,color=OrRd-G, mark=star,mark size=1.75pt,line width=0.5pt},
{smooth,tension=0.5,color=BuGn-M, mark=halfsquare*,every mark/.append style={rotate=90},mark size=1.75pt,line width=0.5pt},
{smooth,tension=0.5,color=BuGn-K, mark=halfsquare*,every mark/.append style={rotate=180},mark size=1.75pt,line width=0.5pt}, 
{smooth,tension=0.5,color=BuGn-I, mark=halfsquare*,every mark/.append style={rotate=270},mark size=1.75pt,line width=0.5pt}, 
{smooth,tension=0.5,color=BuGn-G, mark=halfsquare*,every mark/.append style={rotate=360},mark size=1.75pt,line width=0.5pt}, 
{smooth,tension=0.5,color=BuPu-M, mark=halfcircle*,every mark/.append style={rotate=270},mark size=1.75pt,line width=0.5pt}, 
{smooth,tension=0.5,color=BuPu-K, mark=halfcircle*,every mark/.append style={rotate=360},mark size=1.75pt,line width=0.5pt}, 
{smooth,tension=0.5,color=BuPu-J, mark=halfcircle*,every mark/.append style={rotate=90},mark size=1.75pt,line width=0.5pt},
{smooth,tension=0.5,color=BuPu-I, mark=halfcircle*,every mark/.append style={rotate=180},mark size=1.75pt,line width=0.5pt},
}
]
\nextgroupplot[legend style = {column sep = 3pt, legend to name = LegendMon1}]
\addplot[smooth,tension=0.5,color=NavyBlue, no markers,line width=0.25pt, densely dotted,forget plot] table[x = delta,y=alpha] from \UniMona;
\addplot table[x = delta,y=FiveOS] from \UniMonCheta;
\addplot table[x = delta,y=MonFiveKn3] from \UniMonCLRa;
\addplot table[x = delta,y=MonFiveKn5] from \UniMonCLRa;
\addplot table[x = delta,y=MonFiveKn7] from \UniMonCLRa;
\pgfplotsset{cycle list shift=1}
\addplot table[x = delta,y=MonFiveKn3] from \UniMona;
\addplot table[x = delta,y=MonFiveKn5] from \UniMona;
\addplot table[x = delta,y=MonFiveKn7] from \UniMona;
\node[anchor=north,align=center,font=\fontsize{5}{4}\selectfont] at (axis description cs: 0.25,  0.95) {Design \eqref{Eqn: MC1,aux1}\\$n=500$};
\addlegendentry{C-OS};
\addlegendentry{CLR-C3};
\addlegendentry{CLR-C5};
\addlegendentry{CLR-C7};
\addlegendentry{F-C3};
\addlegendentry{F-C5};
\addlegendentry{F-C7};
\nextgroupplot
\node[anchor=north,align=center,font=\fontsize{5}{4}\selectfont] at (axis description cs: 0.25,  0.95) {Design \eqref{Eqn: MC1,aux1}\\$n=750$};
\addplot[smooth,tension=0.5,color=NavyBlue, no markers,line width=0.25pt, densely dotted,forget plot] table[x = delta,y=alpha] from \UniMona;
\addplot table[x = delta,y=SevenOS] from \UniMonCheta;
\addplot table[x = delta,y=MonSevenKn3] from \UniMonCLRa;
\addplot table[x = delta,y=MonSevenKn5] from \UniMonCLRa;
\addplot table[x = delta,y=MonSevenKn7] from \UniMonCLRa;
\pgfplotsset{cycle list shift=1}
\addplot table[x = delta,y=MonSevenKn3] from \UniMona;
\addplot table[x = delta,y=MonSevenKn5] from \UniMona;
\addplot table[x = delta,y=MonSevenKn7] from \UniMona;
\nextgroupplot
\node[anchor=north,align=center,font=\fontsize{5}{4}\selectfont] at (axis description cs: 0.25,  0.95) {Design \eqref{Eqn: MC1,aux1}\\$n=1000$};
\addplot[smooth,tension=0.5,color=NavyBlue, no markers,line width=0.25pt, densely dotted,forget plot] table[x = delta,y=alpha] from \UniMona;
\addplot table[x = delta,y=TenOS] from \UniMonCheta;
\addplot table[x = delta,y=MonTenKn3] from \UniMonCLRa;
\addplot table[x = delta,y=MonTenKn5] from \UniMonCLRa;
\addplot table[x = delta,y=MonTenKn7] from \UniMonCLRa;
\pgfplotsset{cycle list shift=1}
\addplot table[x = delta,y=MonTenKn3] from \UniMona;
\addplot table[x = delta,y=MonTenKn5] from \UniMona;
\addplot table[x = delta,y=MonTenKn7] from \UniMona;
\nextgroupplot[legend style = {column sep = 3pt, legend to name = LegendMon2}]
\addplot[smooth,tension=0.5,color=NavyBlue, no markers,line width=0.25pt, densely dotted,forget plot] table[x = delta,y=alpha] from \UniMonb;
\addplot table[x = delta,y=FiveOS] from \UniMonChetb;
\addplot table[x = delta,y=MonFiveKn3] from \UniMonCLRb;
\addplot table[x = delta,y=MonFiveKn5] from \UniMonCLRb;
\addplot table[x = delta,y=MonFiveKn7] from \UniMonCLRb;
\pgfplotsset{cycle list shift=1}
\addplot table[x = delta,y=MonFiveKn3] from \UniMonb;
\addplot table[x = delta,y=MonFiveKn5] from \UniMonb;
\addplot table[x = delta,y=MonFiveKn7] from \UniMonb;
\node[anchor=north,align=center,font=\fontsize{5}{4}\selectfont] at (axis description cs: 0.25,  0.95) {Design \eqref{Eqn: MC1,aux2}\\$n=500$};
\addlegendentry{C-OS};
\addlegendentry{CLR-C3};
\addlegendentry{CLR-C5};
\addlegendentry{CLR-C7};
\addlegendentry{F-C3};
\addlegendentry{F-C5};
\addlegendentry{F-C7};
\nextgroupplot
\node[anchor=north,align=center,font=\fontsize{5}{4}\selectfont] at (axis description cs: 0.25,  0.95) {Design \eqref{Eqn: MC1,aux2}\\$n=750$};
\addplot[smooth,tension=0.5,color=NavyBlue, no markers,line width=0.25pt, densely dotted,forget plot] table[x = delta,y=alpha] from \UniMonb;
\addplot table[x = delta,y=SevenOS] from \UniMonChetb;
\addplot table[x = delta,y=MonSevenKn3] from \UniMonCLRb;
\addplot table[x = delta,y=MonSevenKn5] from \UniMonCLRb;
\addplot table[x = delta,y=MonSevenKn7] from \UniMonCLRb;
\pgfplotsset{cycle list shift=1}
\addplot table[x = delta,y=MonSevenKn3] from \UniMonb;
\addplot table[x = delta,y=MonSevenKn5] from \UniMonb;
\addplot table[x = delta,y=MonSevenKn7] from \UniMonb;
\nextgroupplot
\node[anchor=north,align=center,font=\fontsize{5}{4}\selectfont] at (axis description cs: 0.25,  0.95) {Design \eqref{Eqn: MC1,aux2}\\$n=1000$};
\addplot[smooth,tension=0.5,color=NavyBlue, no markers,line width=0.25pt, densely dotted,forget plot] table[x = delta,y=alpha] from \UniMonb;
\addplot table[x = delta,y=TenOS] from \UniMonChetb;
\addplot table[x = delta,y=MonTenKn3] from \UniMonCLRb;
\addplot table[x = delta,y=MonTenKn5] from \UniMonCLRb;
\addplot table[x = delta,y=MonTenKn7] from \UniMonCLRb;
\pgfplotsset{cycle list shift=1}
\addplot table[x = delta,y=MonTenKn3] from \UniMonb;
\addplot table[x = delta,y=MonTenKn5] from \UniMonb;
\addplot table[x = delta,y=MonTenKn7] from \UniMonb;
\nextgroupplot[legend style = {column sep = 0.5pt, legend to name = LegendMon3}]
\node[anchor=north,align=center,font=\fontsize{5}{4}\selectfont] at (axis description cs: 0.25,  0.95) {Design \eqref{Eqn: MC1,aux3}\\$n=500$};
\addplot[smooth,tension=0.5,color=NavyBlue, no markers,line width=0.25pt, densely dotted,forget plot] table[x = delta,y=alpha] from \BiMonQ;
\addplot table[x = delta,y=FiveOS] from \BiMonChet;
\addplot table[x = delta,y=MonFiveQKn0] from \BiMonQCLR;
\addplot table[x = delta,y=MonFiveQKn1] from \BiMonQCLR;
\addplot table[x = delta,y=MonFiveCKn0] from \BiMonCCLR;
\addplot table[x = delta,y=MonFiveCKn1] from \BiMonCCLR;
\addplot table[x = delta,y=MonFiveQKn0] from \BiMonQ;
\addplot table[x = delta,y=MonFiveQKn1] from \BiMonQ;
\addplot table[x = delta,y=MonFiveCKn0] from \BiMonC;
\addplot table[x = delta,y=MonFiveCKn1] from \BiMonC;
\addlegendentry{C-OS};
\addlegendentry{CLR-Q0};
\addlegendentry{CLR-Q1};
\addlegendentry{CLR-C0};
\addlegendentry{CLR-C1};
\addlegendentry{F-Q0};
\addlegendentry{F-Q1};
\addlegendentry{F-C0};
\addlegendentry{F-C1};
\nextgroupplot
\node[anchor=north,align=center,font=\fontsize{5}{4}\selectfont] at (axis description cs: 0.25,  0.95) {Design \eqref{Eqn: MC1,aux3}\\$n=750$};
\addplot[smooth,tension=0.5,color=NavyBlue, no markers,line width=0.25pt, densely dotted,forget plot] table[x = delta,y=alpha] from \BiMonQ;
\addplot table[x = delta,y=SevenOS] from \BiMonChet;
\addplot table[x = delta,y=MonSevenQKn0] from \BiMonQCLR;
\addplot table[x = delta,y=MonSevenQKn1] from \BiMonQCLR;
\addplot table[x = delta,y=MonSevenCKn0] from \BiMonCCLR;
\addplot table[x = delta,y=MonSevenCKn1] from \BiMonCCLR;
\addplot table[x = delta,y=MonSevenQKn0] from \BiMonQ;
\addplot table[x = delta,y=MonSevenQKn1] from \BiMonQ;
\addplot table[x = delta,y=MonSevenCKn0] from \BiMonC;
\addplot table[x = delta,y=MonSevenCKn1] from \BiMonC;
\nextgroupplot
\node[anchor=north,align=center,font=\fontsize{5}{4}\selectfont] at (axis description cs: 0.25,  0.95) {Design \eqref{Eqn: MC1,aux3}\\$n=1000$};
\addplot[smooth,tension=0.5,color=NavyBlue, no markers,line width=0.25pt, densely dotted,forget plot] table[x = delta,y=alpha] from \BiMonQ;
\addplot table[x = delta,y=TenOS] from \BiMonChet;
\addplot table[x = delta,y=MonTenQKn0] from \BiMonQCLR;
\addplot table[x = delta,y=MonTenQKn1] from \BiMonQCLR;
\addplot table[x = delta,y=MonTenCKn0] from \BiMonCCLR;
\addplot table[x = delta,y=MonTenCKn1] from \BiMonCCLR;
\addplot table[x = delta,y=MonTenQKn0] from \BiMonQ;
\addplot table[x = delta,y=MonTenQKn1] from \BiMonQ;
\addplot table[x = delta,y=MonTenCKn0] from \BiMonC;
\addplot table[x = delta,y=MonTenCKn1] from \BiMonC;
\end{groupplot}
\node at ($(myplots c2r1) + (0,-2.25cm)$) {\ref{LegendMon1}};
\node at ($(myplots c2r2) + (0,-2.25cm)$) {\ref{LegendMon2}};
\node[align=center] at ($(myplots c2r3) + (0,-2.25cm)$) {\ref{LegendMon3}};
\end{tikzpicture}
\caption{Testing monotonicity: Empirical power of our test, the CLR test, and the C-OS test for the designs \eqref{Eqn: MC1,aux1}, \eqref{Eqn: MC1,aux2}, and \eqref{Eqn: MC1,aux3}, where corresponding to $\delta=0$ are the empirical sizes under D1. In the first row, C-OS, CLR-C5, and F-C7 nearly overlap one another, and so do CLR-C3 and F-C3; in the third row, CLR-Q0 and F-Q1 nearly overlap each other.} \label{Fig: MC,Mon}
\end{figure}
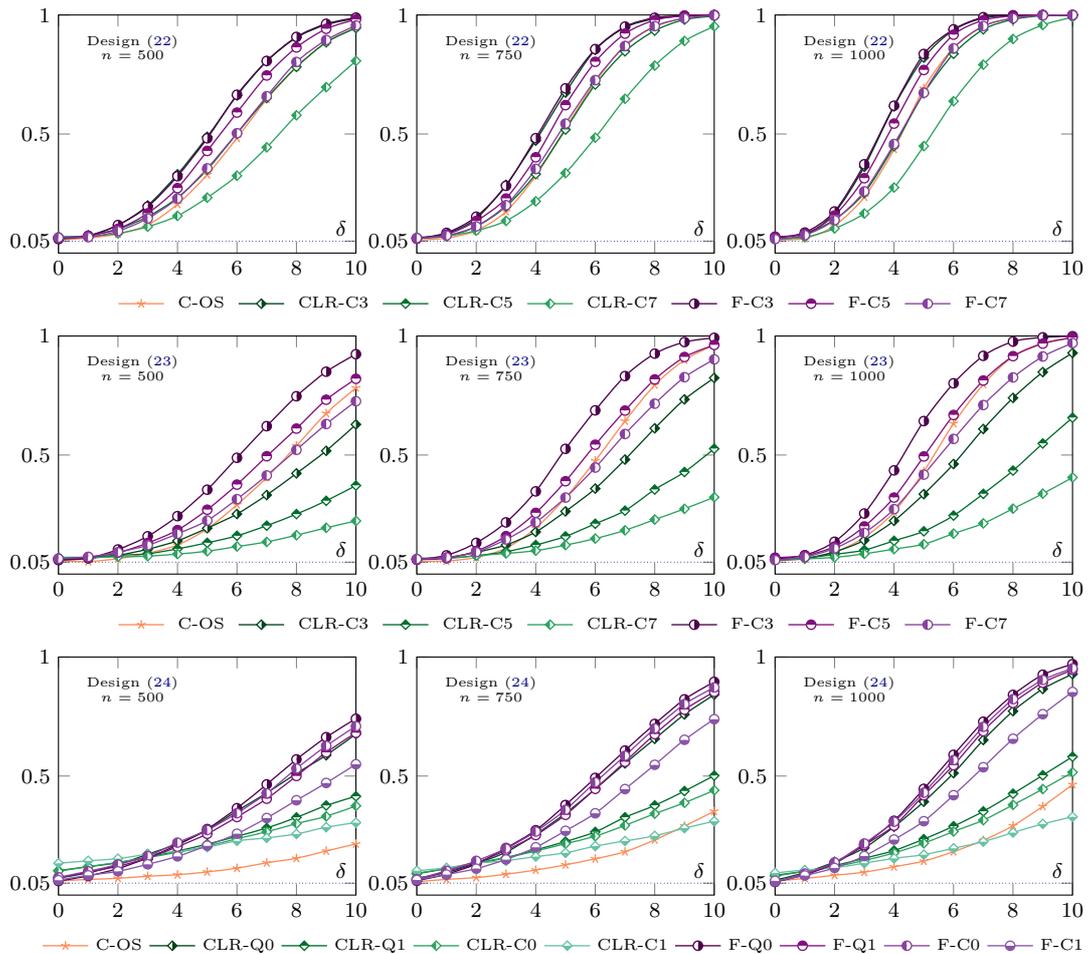

\subsection{Testing Convexity/Concavity}\label{Sec: simulations, con}

We adopt the same univariate designs as in Section \ref{Sec: MC, Mon}, with the only change being that the specification in \eqref{Eqn: MC1,aux2} is replaced by
\begin{align}\label{Eqn: MC2,aux2}
\theta_P(z)=\mathsf b\varphi(|z|^{1.5})~.
\end{align}
The null hypothesis in this case is that $\theta_0: [-1,1]\to\mathbf R$ is convex. In bivariate designs, we instead test the concavity of $\theta_0$ by employing designs that are slight variations of \eqref{Eqn: MC1,aux3} (so that the power curves get close to one as $\delta$ increases):
\begin{align}\label{Eqn: MC2,aux3}
\theta_0(z_1,z_2)=\mathsf a\big(\frac{1}{2}z_1^{\mathsf b}+\frac{1}{2}z_2^{\mathsf b}\big)^{1/\mathsf b}+\mathsf c\log(1+5(z_1+z_2))~,
\end{align}
where $(\mathsf a,\mathsf b,\mathsf c)$ is chosen to be the same as for \eqref{Eqn: MC1,aux3} except $\Delta=0.2$. We implement our test based on the GCM/LCM operators (i.e., based on \eqref{Eqn: test functional, GCM} with $p=\infty$), and compare it to the sup-test of \citet{ChernozhukovLeeRosen2013Intersection}. The estimation and bootstrap steps are the same as those in Section \ref{Sec: MC, Mon}.

{
\setlength{\tabcolsep}{6.5pt}
\renewcommand{\arraystretch}{1.1}
\begin{table}[!ht]
\caption{Empirical Size of Convexity Tests for $\theta_0$ in \eqref{Eqn: MC1,aux1} at $\alpha=5\%$} \label{Tab: ConMC, size1}
\centering\footnotesize
\begin{threeparttable}
\sisetup{table-number-alignment = center, table-format = 1.3} 
\begin{tabularx}{\linewidth}{@{} cc *{3}{S[round-mode = places,round-precision = 3]}  c *{3}{S[round-mode = places,round-precision = 3]}  c *{3}{S[round-mode = places,round-precision = 3]}@{}} 
\hline
\hline
 $n$ & $\gamma_n$ & {D1} & {D2} & {D3}  & & {D1} & {D2} & {D3} & & {D1} & {D2} & {D3}\\
\hline
\rule{0pt}{15pt}
&& \multicolumn{3}{c}{{F-C3: $k_n=7$}} && \multicolumn{3}{c}{{F-C5: $k_n=9$}} && \multicolumn{3}{c}{{F-C7: $k_n=11$}}\\
\multirow{3}{*}{$500$}   & $1/n$                  & 0.0557 & 0.0477 & 0.0117 & & 0.0523 & 0.0483 & 0.0187 & & 0.0547 & 0.0490 & 0.0217\\
                         & $0.01/\log n$          & 0.0557 & 0.0477 & 0.0117 & & 0.0523 & 0.0483 & 0.0187 & & 0.0547 & 0.0490 & 0.0217\\
                         & $0.01$                 & 0.0557 & 0.0480 & 0.0117 & & 0.0523 & 0.0483 & 0.0190 & & 0.0547 & 0.0490 & 0.0223\\
\rule{0pt}{12pt}
 \multirow{3}{*}{$750$}  & $1/n$                  & 0.0560 & 0.0483 & 0.0107 & & 0.0553 & 0.0490 & 0.0183 & & 0.0607 & 0.0557 & 0.0240\\
                         & $0.01/\log n$          & 0.0560 & 0.0483 & 0.0107 & & 0.0553 & 0.0490 & 0.0183 & & 0.0607 & 0.0557 & 0.0240\\
                         & $0.01$                 & 0.0563 & 0.0487 & 0.0107 & & 0.0567 & 0.0493 & 0.0183 & & 0.0607 & 0.0557 & 0.0240\\
\rule{0pt}{12pt}
 \multirow{3}{*}{$1000$} & $1/n$                  & 0.0620 & 0.0527 & 0.0070 & & 0.0627 & 0.0563 & 0.0147 & & 0.0613 & 0.0560 & 0.0223\\
                         & $0.01/\log n$          & 0.0620 & 0.0527 & 0.0070 & & 0.0627 & 0.0563 & 0.0147 & & 0.0613 & 0.0560 & 0.0223\\
                         & $0.01$                 & 0.0623 & 0.0530 & 0.0073 & & 0.0630 & 0.0563 & 0.0147 & & 0.0613 & 0.0567 & 0.0223\\
\rule{0pt}{15pt}
&& \multicolumn{3}{c}{CLR-C3: $k_n=7$} && \multicolumn{3}{c}{CLR-C5: $k_n=9$} && \multicolumn{3}{c}{CLR-C7: $k_n=11$}\\
$500$ &                       & 0.0543 & 0.0553 & 0.0330 && 0.0560 & 0.0553 & 0.0437 && 0.0643 & 0.0617 & 0.0583\\
$750$ &                       & 0.0610 & 0.0573 & 0.0337 && 0.0630 & 0.0643 & 0.0467 && 0.0650 & 0.0637 & 0.0537\\
$1000$&                       & 0.0577 & 0.0567 & 0.0320 && 0.0650 & 0.0640 & 0.0470 && 0.0613 & 0.0600 & 0.0520\\
\hline
\hline
\end{tabularx}
\begin{tablenotes}[flushleft]
\item {\it Note:} The parameter $\gamma_n$ determines $\hat\kappa_n$ as in Proposition \ref{Pro: tuning parameter} with $c_n=1/\log n$ and $r_n=(n/k_n)^{1/2}$.
\end{tablenotes}
\end{threeparttable}
\end{table}
}

Tables \ref{Tab: ConMC, size1} and \ref{Tab: ConMC, size2} summarize the empirical sizes. In the univariate designs, both our tests and the CLR tests have reasonable size control, while in the bivariate designs the CLR tests once again exhibit notable over-rejections across the sample sizes and the sieve spaces. Note that both D1 and D2 are at the ``boundaries'' of the parameter spaces, and thus the rejections rates are expected to be close to $\alpha=5\%$. Figure \ref{Fig: MC,Con} shows that our test remains competitive to the sup-test of \citet{ChernozhukovLeeRosen2013Intersection} as far as power is concerned. This is particularly the case for the designs based on \eqref{Eqn: MC2,aux2} in which $\theta_0$ has a flat region around $z=0$ for each $\delta$, in line with the discussions in \citet{ChernozhukovLeeRosen2013Intersection}. One possible explanation is that these are settings where the set estimation step in implementing the CLR tests is statistically challenging.

{ 
\setlength{\tabcolsep}{3pt}
\renewcommand{\arraystretch}{1.1}
\begin{table}[!ht]
\caption{Empirical Size of Concavity Tests for $\theta_0$ in \eqref{Eqn: MC2,aux3} at $\alpha=5\%$} \label{Tab: ConMC, size2}
\centering\footnotesize
\begin{threeparttable}
\sisetup{table-number-alignment = center, table-format = 1.3} 
\begin{tabularx}{\linewidth}{@{}cc *{3}{S[round-mode = places,round-precision = 3]} c *{3}{S[round-mode = places,round-precision = 3]} c *{3}{S[round-mode = places,round-precision = 3]} c *{3}{S[round-mode = places,round-precision = 3]}@{}} 
\hline
\hline
$n$ & $\gamma_n$ & {D1} & {D2} & {D3}  && {D1} & {D2} & {D3} && {D1} & {D2} & {D3} && {D1} & {D2} & {D3}\\
\hline
\rule{0pt}{15pt}
&& \multicolumn{3}{c}{F-Q0: $k_n=9$} && \multicolumn{3}{c}{F-Q1: $k_n=16$} && \multicolumn{3}{c}{F-C0: $k_n=16$} && \multicolumn{3}{c}{F-C1: $k_n=25$}\\
\multirow{3}{*}{$500$} & $1/n$         & 0.0573 & 0.0557 & 0.0207 & & 0.0643 & 0.0630 & 0.0287 & & 0.0650 & 0.0653 & 0.0293 & & 0.0643 & 0.0617 & 0.0340\\
                       & $0.01/\log n$ & 0.0573 & 0.0557 & 0.0207 & & 0.0643 & 0.0630 & 0.0287 & & 0.0650 & 0.0653 & 0.0293 & & 0.0643 & 0.0617 & 0.0340\\
                       & $0.01$        & 0.0583 & 0.0563 & 0.0207 & & 0.0643 & 0.0640 & 0.0287 & & 0.0650 & 0.0657 & 0.0293 & & 0.0647 & 0.0617 & 0.0343\\
\rule{0pt}{12pt}
 \multirow{3}{*}{$750$}& $1/n$         & 0.0603 & 0.0597 & 0.0147 & & 0.0683 & 0.0667 & 0.0273 & & 0.0730 & 0.0703 & 0.0247 & & 0.0700 & 0.0683 & 0.0370\\
                       & $0.01/\log n$ & 0.0603 & 0.0597 & 0.0147 & & 0.0683 & 0.0667 & 0.0273 & & 0.0730 & 0.0703 & 0.0247 & & 0.0700 & 0.0683 & 0.0370\\
                       & $0.01$        & 0.0607 & 0.0600 & 0.0147 & & 0.0687 & 0.0667 & 0.0273 & & 0.0733 & 0.0717 & 0.0247 & & 0.0710 & 0.0687 & 0.0373\\
\rule{0pt}{12pt}
\multirow{3}{*}{$1000$}& $1/n$         & 0.0563 & 0.0607 & 0.0097 & & 0.0627 & 0.0630 & 0.0180 & & 0.0637 & 0.0643 & 0.0153 & & 0.0603 & 0.0587 & 0.0247\\
                       & $0.01/\log n$ & 0.0563 & 0.0607 & 0.0097 & & 0.0627 & 0.0630 & 0.0180 & & 0.0637 & 0.0643 & 0.0153 & & 0.0603 & 0.0587 & 0.0247\\
                       & $0.01$        & 0.0563 & 0.0607 & 0.0097 & & 0.0630 & 0.0637 & 0.0180 & & 0.0643 & 0.0643 & 0.0153 & & 0.0603 & 0.0587 & 0.0247\\
\rule{0pt}{15pt}
&& \multicolumn{3}{c}{CLR-Q0: $k_n=9$} && \multicolumn{3}{c}{CLR-Q1: $k_n=16$} && \multicolumn{3}{c}{CLR-C0: $k_n=16$} && \multicolumn{3}{c}{CLR-C1: $k_n=25$}\\
$500$ &               & 0.0713 & 0.0733 & 0.0293 && 0.0873 & 0.0890 & 0.0683 && 0.1010 & 0.0967 & 0.0657 && 0.1360 & 0.1450 & 0.1150\\
$750$ &               & 0.0627 & 0.0623 & 0.0270 && 0.0800 & 0.0810 & 0.0553 && 0.0863 & 0.0853 & 0.0527 && 0.1030 & 0.1053 & 0.0750\\
$1000$&               & 0.0653 & 0.0590 & 0.0153 && 0.0800 & 0.0740 & 0.0437 && 0.0763 & 0.0733 & 0.0390 && 0.0897 & 0.0863 & 0.0607\\
\hline
\hline
\end{tabularx}
\begin{tablenotes}[flushleft]
\item {\it Note:} The parameter $\gamma_n$ determines $\hat\kappa_n$ as in Proposition \ref{Pro: tuning parameter} with $c_n=1/\log n$ and $r_n=(n/k_n)^{1/2}$.
\end{tablenotes}
\end{threeparttable}
\end{table}
}

\pgfplotstableread{ 
delta alpha ConFiveKn3 ConFiveKn5 ConFiveKn7 ConSevenKn3 ConSevenKn5 ConSevenKn7 ConTenKn3 ConTenKn5 ConTenKn7
0     0.05    0.0557     0.0523     0.0547     0.0560       0.0553     0.0607      0.0620    0.0627    0.0613
1     0.05    0.0867     0.0713     0.0683     0.1043       0.0853     0.0823      0.1127    0.1047    0.0920
2     0.05    0.1613     0.1273     0.1103     0.2277       0.1740     0.1420      0.2703    0.2253    0.1923
3     0.05    0.3027     0.2373     0.1903     0.4180       0.3317     0.2693      0.5497    0.4443    0.3587
4     0.05    0.4770     0.3883     0.3157     0.6530       0.5580     0.4727      0.7900    0.7007    0.6107
5     0.05    0.6713     0.5687     0.4810     0.8530       0.7683     0.6890      0.9423    0.8927    0.8220
6     0.05    0.8287     0.7430     0.6570     0.9550       0.9130     0.8473      0.9857    0.9713    0.9440
7     0.05    0.9247     0.8763     0.8083     0.9907       0.9793     0.9497      0.9983    0.9960    0.9893
8     0.05    0.9747     0.9527     0.9070     0.9980       0.9980     0.9910      0.9993    0.9990    0.9987
9     0.05    0.9930     0.9840     0.9673     1.0000       1.0000     0.9990      1.0000    0.9997    0.9997
10    0.05    0.9987     0.9960     0.9900     1.0000       1.0000     1.0000      1.0000    1.0000    1.0000
}\UniCona

\pgfplotstableread{ 
delta alpha ConFiveKn3 ConFiveKn5 ConFiveKn7 ConSevenKn3 ConSevenKn5 ConSevenKn7 ConTenKn3 ConTenKn5 ConTenKn7
0     0.05    0.0557     0.0523     0.0547     0.0560       0.0553     0.0607      0.0620    0.0627    0.0613
1     0.05    0.0947     0.0780     0.0750     0.1063       0.0887     0.0840      0.1163    0.1073    0.0917
2     0.05    0.1543     0.1207     0.1037     0.1890       0.1503     0.1303      0.2257    0.1957    0.1540
3     0.05    0.2623     0.1823     0.1517     0.3387       0.2433     0.2083      0.4180    0.3170    0.2593
4     0.05    0.3877     0.2720     0.2210     0.5277       0.3917     0.3163      0.6613    0.4987    0.4117
5     0.05    0.5330     0.3937     0.3120     0.7183       0.5583     0.4633      0.8480    0.6983    0.5997
6     0.05    0.6920     0.5273     0.4250     0.8740       0.7290     0.6130      0.9557    0.8663    0.7640
7     0.05    0.8320     0.6670     0.5430     0.9510       0.8683     0.7627      0.9910    0.9620    0.8897
8     0.05    0.9170     0.8087     0.6753     0.9863       0.9450     0.8787      0.9993    0.9900    0.9663
9     0.05    0.9690     0.9013     0.7953     0.9953       0.9800     0.9427      1.0000    0.9980    0.9893
10    0.05    0.9907     0.9527     0.8827     0.9997       0.9947     0.9793      1.0000    1.0000    0.9980
}\UniConb

\pgfplotstableread{ 
delta alpha ConFiveQKn0 ConFiveQKn1 ConSevenQKn0 ConSevenQKn1 ConTenQKn0 ConTenQKn1
0     0.05    0.0573      0.0643       0.0603       0.0683      0.0563     0.0627
1     0.05    0.0820      0.0827       0.0893       0.0937      0.0883     0.0923
2     0.05    0.1127      0.1043       0.1327       0.1200      0.1440     0.1353
3     0.05    0.1553      0.1357       0.1880       0.1673      0.2157     0.1937
4     0.05    0.2143      0.1827       0.2623       0.2327      0.3157     0.2743
5     0.05    0.2860      0.2377       0.3710       0.3123      0.4393     0.3820
6     0.05    0.3673      0.3007       0.4690       0.3953      0.5740     0.4967
7     0.05    0.4497      0.3823       0.5913       0.4877      0.7037     0.6163
8     0.05    0.5553      0.4617       0.7070       0.5973      0.8220     0.7340
9     0.05    0.6467      0.5483       0.8100       0.7073      0.9027     0.8337
10    0.05    0.7283      0.6247       0.8830       0.7953      0.9527     0.9063
}\BiConQ

\pgfplotstableread{ 
delta alpha ConFiveCKn0 ConFiveCKn1 ConSevenCKn0 ConSevenCKn1 ConTenCKn0 ConTenCKn1
0     0.05    0.0650      0.0643       0.0730       0.0700      0.0637     0.0603
1     0.05    0.0853      0.0767       0.0980       0.0917      0.0940     0.0767
2     0.05    0.1153      0.0913       0.1340       0.1077      0.1440     0.0967
3     0.05    0.1537      0.1073       0.1820       0.1347      0.2047     0.1313
4     0.05    0.2050      0.1327       0.2500       0.1627      0.2970     0.1720
5     0.05    0.2587      0.1523       0.3363       0.2050      0.4063     0.2337
6     0.05    0.3347      0.1897       0.4243       0.2530      0.5260     0.2963
7     0.05    0.4170      0.2290       0.5303       0.3127      0.6560     0.3683
8     0.05    0.5007      0.2753       0.6407       0.3810      0.7683     0.4640
9     0.05    0.5863      0.3130       0.7343       0.4557      0.8643     0.5513
10    0.05    0.6680      0.3847       0.8303       0.5303      0.9290     0.6537
}\BiConC

\pgfplotstableread{ 
delta alpha ConFiveKn3 ConFiveKn5 ConFiveKn7 ConSevenKn3 ConSevenKn5 ConSevenKn7 ConTenKn3 ConTenKn5 ConTenKn7
0     0.05    0.0543     0.0560     0.0643     0.0610      0.0630      0.0650      0.0577    0.0650    0.0613
1     0.05    0.0707     0.0637     0.0667     0.0823      0.0667      0.0670      0.0833    0.0780    0.0707
2     0.05    0.1077     0.0883     0.0737     0.1477      0.1040      0.0843      0.1693    0.1323    0.0933
3     0.05    0.1907     0.1297     0.0957     0.2727      0.1727      0.1090      0.3643    0.2237    0.1353
4     0.05    0.3150     0.1980     0.1317     0.4663      0.2837      0.1647      0.5900    0.3847    0.2080
5     0.05    0.4700     0.3107     0.1790     0.6590      0.4443      0.2493      0.7940    0.5803    0.3197
6     0.05    0.6430     0.4330     0.2573     0.8253      0.6267      0.3550      0.9233    0.7630    0.4777
7     0.05    0.7817     0.5870     0.3453     0.9373      0.7780      0.5043      0.9800    0.8873    0.6350
8     0.05    0.8893     0.7227     0.4563     0.9797      0.8923      0.6413      0.9947    0.9590    0.7820
9     0.05    0.9470     0.8343     0.5757     0.9940      0.9577      0.7830      0.9997    0.9873    0.8917
10    0.05    0.9763     0.9113     0.6837     0.9993      0.9863      0.8787      1.0000    0.9990    0.9537
}\UniConCLRa

\pgfplotstableread{ 
delta alpha ConFiveKn3 ConFiveKn5 ConFiveKn7 ConSevenKn3 ConSevenKn5 ConSevenKn7 ConTenKn3 ConTenKn5 ConTenKn7
0     0.05    0.0543     0.0560     0.0643     0.0610      0.0630      0.0650      0.0577    0.0650    0.0613
1     0.05    0.0663     0.0610     0.0663     0.0727      0.0700      0.0707      0.0720    0.0743    0.0660
2     0.05    0.0813     0.0703     0.0733     0.0913      0.0807      0.0723      0.1007    0.0817    0.0713
3     0.05    0.1020     0.0750     0.0730     0.1170      0.0883      0.0807      0.1230    0.0953    0.0743
4     0.05    0.1183     0.0853     0.0797     0.1490      0.0997      0.0827      0.1643    0.1123    0.0867
5     0.05    0.1373     0.0937     0.0837     0.1820      0.1123      0.0927      0.2043    0.1257    0.0937
6     0.05    0.1687     0.1063     0.0857     0.2213      0.1303      0.1007      0.2607    0.1440    0.1007
7     0.05    0.1977     0.1223     0.0943     0.2717      0.1493      0.1050      0.3257    0.1693    0.1063
8     0.05    0.2457     0.1297     0.0973     0.3240      0.1630      0.1120      0.3967    0.1863    0.1183
9     0.05    0.2853     0.1473     0.1067     0.3833      0.1847      0.1157      0.4800    0.2173    0.1277
10    0.05    0.3357     0.1543     0.1140     0.4593      0.2047      0.1273      0.5717    0.2497    0.1410
}\UniConCLRb

\pgfplotstableread{ 
delta alpha ConFiveQKn0 ConFiveQKn1 ConSevenQKn0 ConSevenQKn1 ConTenQKn0 ConTenQKn1
0     0.05    0.0713      0.0873       0.0627      0.0800       0.0653     0.0800
1     0.05    0.0977      0.0990       0.0897      0.0907       0.0863     0.0847
2     0.05    0.1200      0.1093       0.1257      0.1057       0.1223     0.1013
3     0.05    0.1590      0.1220       0.1690      0.1207       0.1697     0.1173
4     0.05    0.2003      0.1360       0.2273      0.1350       0.2357     0.1343
5     0.05    0.2387      0.1480       0.2877      0.1517       0.3120     0.1637
6     0.05    0.2933      0.1670       0.3623      0.1727       0.4187     0.1847
7     0.05    0.3523      0.1923       0.4420      0.1977       0.5320     0.2223
8     0.05    0.4217      0.2073       0.5293      0.2233       0.6350     0.2503
9     0.05    0.5067      0.2387       0.6213      0.2567       0.7330     0.2900
10    0.05    0.5730      0.2593       0.7073      0.2837       0.8157     0.3277
}\BiConQCLR

\pgfplotstableread{ 
delta alpha ConFiveCKn0 ConFiveCKn1 ConSevenCKn0 ConSevenCKn1 ConTenCKn0 ConTenCKn1
0     0.05    0.1010      0.1360      0.0863        0.1030      0.0763     0.0897
1     0.05    0.1143      0.1480      0.1047        0.1143      0.0927     0.0973
2     0.05    0.1367      0.1587      0.1193        0.1177      0.1137     0.1127
3     0.05    0.1593      0.1723      0.1463        0.1367      0.1470     0.1180
4     0.05    0.1847      0.1833      0.1833        0.1490      0.1823     0.1413
5     0.05    0.2157      0.1950      0.2190        0.1600      0.2333     0.1507
6     0.05    0.2463      0.2067      0.2650        0.1793      0.2957     0.1710
7     0.05    0.2893      0.2257      0.3190        0.1967      0.3683     0.1990
8     0.05    0.3307      0.2433      0.3803        0.2230      0.4537     0.2287
9     0.05    0.3787      0.2660      0.4477        0.2450      0.5373     0.2587
10    0.05    0.4347      0.2820      0.5250        0.2787      0.6230     0.2933
}\BiConCCLR


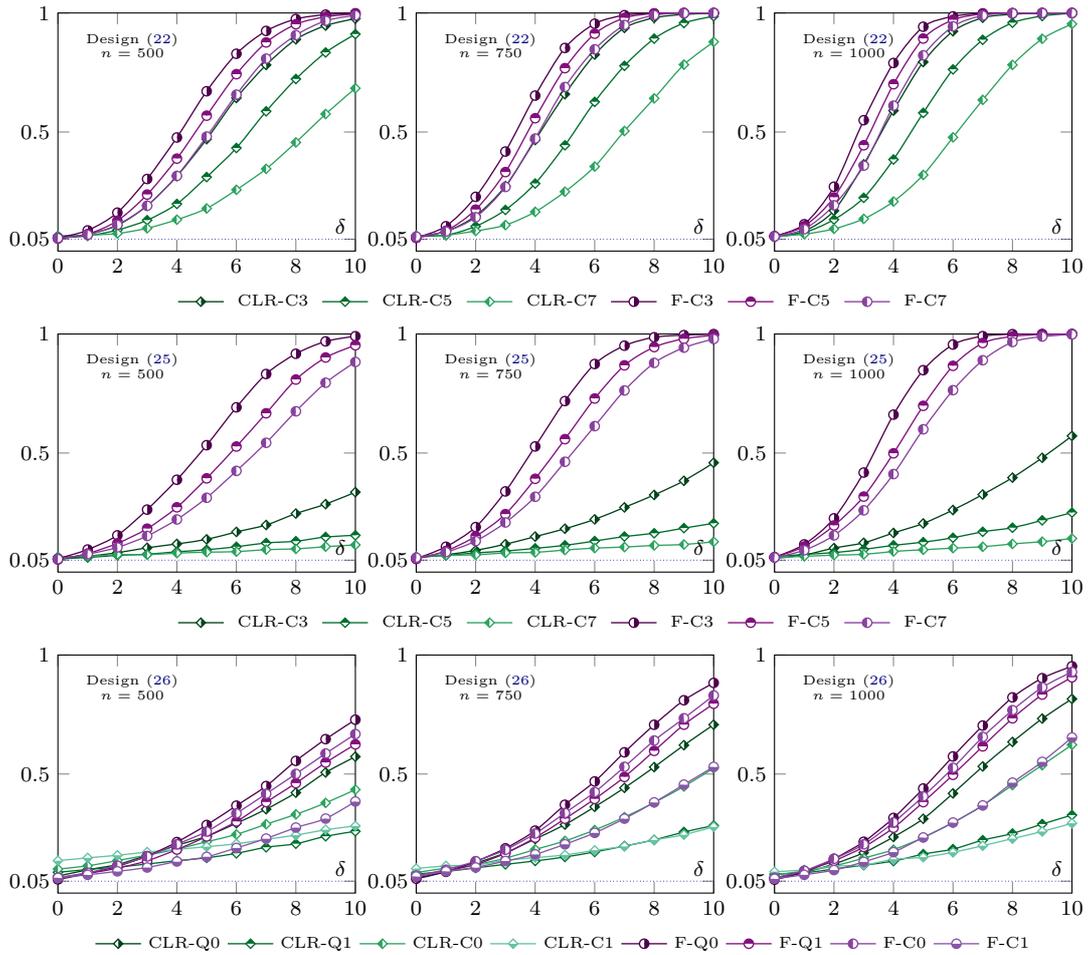
\begin{figure}[!h]
\centering\scriptsize
\begin{tikzpicture} 
\begin{groupplot}[group style={group name=myplots,group size=3 by 3,horizontal sep= 0.8cm,vertical sep=1.1cm},
    grid = minor,
    width = 0.375\textwidth,
    xmax=10,xmin=0,
    ymax=1,ymin=0,
    every axis title/.style={below,at={(0.2,0.8)}},
    xlabel=$\delta$,
    x label style={at={(axis description cs:0.95,0.04)},anchor=south},
    xtick={0,2,...,10},
    ytick={0.05,0.5,1},
    tick label style={/pgf/number format/fixed},
    legend style={font=\tiny,text=black,cells={align=center},row sep = 3pt,legend columns = -1, draw=none,fill=none},
    cycle list={
{smooth,tension=0.5,color=BuGn-M, mark=halfsquare*,every mark/.append style={rotate=90},mark size=1.75pt,line width=0.5pt},
{smooth,tension=0.5,color=BuGn-K, mark=halfsquare*,every mark/.append style={rotate=180},mark size=1.75pt,line width=0.5pt}, 
{smooth,tension=0.5,color=BuGn-I, mark=halfsquare*,every mark/.append style={rotate=270},mark size=1.75pt,line width=0.5pt}, 
{smooth,tension=0.5,color=BuGn-G, mark=halfsquare*,every mark/.append style={rotate=360},mark size=1.75pt,line width=0.5pt}, 
{smooth,tension=0.5,color=BuPu-M, mark=halfcircle*,every mark/.append style={rotate=270},mark size=1.75pt,line width=0.5pt}, 
{smooth,tension=0.5,color=BuPu-K, mark=halfcircle*,every mark/.append style={rotate=360},mark size=1.75pt,line width=0.5pt}, 
{smooth,tension=0.5,color=BuPu-J, mark=halfcircle*,every mark/.append style={rotate=90},mark size=1.75pt,line width=0.5pt},
{smooth,tension=0.5,color=BuPu-I, mark=halfcircle*,every mark/.append style={rotate=180},mark size=1.75pt,line width=0.5pt},
}
]
\nextgroupplot[legend style = {column sep = 3pt, legend to name = LegendCon1}]
\addplot[smooth,tension=0.5,color=NavyBlue, no markers,line width=0.25pt, densely dotted,forget plot] table[x = delta,y=alpha] from \UniCona;
\addplot table[x = delta,y=ConFiveKn3] from \UniConCLRa;
\addplot table[x = delta,y=ConFiveKn5] from \UniConCLRa;
\addplot table[x = delta,y=ConFiveKn7] from \UniConCLRa;
\pgfplotsset{cycle list shift=1}
\addplot table[x = delta,y=ConFiveKn3] from \UniCona;
\addplot table[x = delta,y=ConFiveKn5] from \UniCona;
\addplot table[x = delta,y=ConFiveKn7] from \UniCona;
\node[anchor=north,align=center,font=\fontsize{5}{4}\selectfont] at (axis description cs: 0.25,  0.95) {Design \eqref{Eqn: MC1,aux1}\\$n=500$};
\addlegendentry{CLR-C3};
\addlegendentry{CLR-C5};
\addlegendentry{CLR-C7};
\addlegendentry{F-C3};
\addlegendentry{F-C5};
\addlegendentry{F-C7};
\nextgroupplot
\node[anchor=north,align=center,font=\fontsize{5}{4}\selectfont] at (axis description cs: 0.25,  0.95) {Design \eqref{Eqn: MC1,aux1}\\$n=750$};
\addplot[smooth,tension=0.5,color=NavyBlue, no markers,line width=0.25pt, densely dotted,forget plot] table[x = delta,y=alpha] from \UniCona;
\addplot table[x = delta,y=ConSevenKn3] from \UniConCLRa;
\addplot table[x = delta,y=ConSevenKn5] from \UniConCLRa;
\addplot table[x = delta,y=ConSevenKn7] from \UniConCLRa;
\pgfplotsset{cycle list shift=1}
\addplot table[x = delta,y=ConSevenKn3] from \UniCona;
\addplot table[x = delta,y=ConSevenKn5] from \UniCona;
\addplot table[x = delta,y=ConSevenKn7] from \UniCona;
\nextgroupplot
\node[anchor=north,align=center,font=\fontsize{5}{4}\selectfont] at (axis description cs: 0.25,  0.95) {Design \eqref{Eqn: MC1,aux1}\\$n=1000$};
\addplot[smooth,tension=0.5,color=NavyBlue, no markers,line width=0.25pt, densely dotted,forget plot] table[x = delta,y=alpha] from \UniCona;
\addplot table[x = delta,y=ConTenKn3] from \UniConCLRa;
\addplot table[x = delta,y=ConTenKn5] from \UniConCLRa;
\addplot table[x = delta,y=ConTenKn7] from \UniConCLRa;
\pgfplotsset{cycle list shift=1}
\addplot table[x = delta,y=ConTenKn3] from \UniCona;
\addplot table[x = delta,y=ConTenKn5] from \UniCona;
\addplot table[x = delta,y=ConTenKn7] from \UniCona;
\nextgroupplot[legend style = {column sep = 3pt, legend to name = LegendCon2}]
\addplot[smooth,tension=0.5,color=NavyBlue, no markers,line width=0.25pt, densely dotted,forget plot] table[x = delta,y=alpha] from \UniConb;
\addplot table[x = delta,y=ConFiveKn3] from \UniConCLRb;
\addplot table[x = delta,y=ConFiveKn5] from \UniConCLRb;
\addplot table[x = delta,y=ConFiveKn7] from \UniConCLRb;
\pgfplotsset{cycle list shift=1}
\addplot table[x = delta,y=ConFiveKn3] from \UniConb;
\addplot table[x = delta,y=ConFiveKn5] from \UniConb;
\addplot table[x = delta,y=ConFiveKn7] from \UniConb;
\node[anchor=north,align=center,font=\fontsize{5}{4}\selectfont] at (axis description cs: 0.25,  0.95) {Design \eqref{Eqn: MC2,aux2}\\$n=500$};
\addlegendentry{CLR-C3};
\addlegendentry{CLR-C5};
\addlegendentry{CLR-C7};
\addlegendentry{F-C3};
\addlegendentry{F-C5};
\addlegendentry{F-C7};
\nextgroupplot
\node[anchor=north,align=center,font=\fontsize{5}{4}\selectfont] at (axis description cs: 0.25,  0.95) {Design \eqref{Eqn: MC2,aux2}\\$n=750$};
\addplot[smooth,tension=0.5,color=NavyBlue, no markers,line width=0.25pt, densely dotted,forget plot] table[x = delta,y=alpha] from \UniConb;
\addplot table[x = delta,y=ConSevenKn3] from \UniConCLRb;
\addplot table[x = delta,y=ConSevenKn5] from \UniConCLRb;
\addplot table[x = delta,y=ConSevenKn7] from \UniConCLRb;
\pgfplotsset{cycle list shift=1}
\addplot table[x = delta,y=ConSevenKn3] from \UniConb;
\addplot table[x = delta,y=ConSevenKn5] from \UniConb;
\addplot table[x = delta,y=ConSevenKn7] from \UniConb;
\nextgroupplot
\node[anchor=north,align=center,font=\fontsize{5}{4}\selectfont] at (axis description cs: 0.25,  0.95) {Design \eqref{Eqn: MC2,aux2}\\$n=1000$};
\addplot[smooth,tension=0.5,color=NavyBlue, no markers,line width=0.25pt, densely dotted,forget plot] table[x = delta,y=alpha] from \UniConb;
\addplot table[x = delta,y=ConTenKn3] from \UniConCLRb;
\addplot table[x = delta,y=ConTenKn5] from \UniConCLRb;
\addplot table[x = delta,y=ConTenKn7] from \UniConCLRb;
\pgfplotsset{cycle list shift=1}
\addplot table[x = delta,y=ConTenKn3] from \UniConb;
\addplot table[x = delta,y=ConTenKn5] from \UniConb;
\addplot table[x = delta,y=ConTenKn7] from \UniConb;
\nextgroupplot[legend style = {column sep = 0.5pt, legend to name = LegendCon3}]
\node[anchor=north,align=center,font=\fontsize{5}{4}\selectfont] at (axis description cs: 0.25,  0.95) {Design \eqref{Eqn: MC2,aux3}\\$n=500$};
\addplot[smooth,tension=0.5,color=NavyBlue, no markers,line width=0.25pt, densely dotted,forget plot] table[x = delta,y=alpha] from \BiConQ;
\addplot table[x = delta,y=ConFiveQKn0] from \BiConQCLR;
\addplot table[x = delta,y=ConFiveQKn1] from \BiConQCLR;
\addplot table[x = delta,y=ConFiveCKn0] from \BiConCCLR;
\addplot table[x = delta,y=ConFiveCKn1] from \BiConCCLR;
\addplot table[x = delta,y=ConFiveQKn0] from \BiConQ;
\addplot table[x = delta,y=ConFiveQKn1] from \BiConQ;
\addplot table[x = delta,y=ConFiveCKn0] from \BiConC;
\addplot table[x = delta,y=ConFiveCKn1] from \BiConC;
\addlegendentry{CLR-Q0};
\addlegendentry{CLR-Q1};
\addlegendentry{CLR-C0};
\addlegendentry{CLR-C1};
\addlegendentry{F-Q0};
\addlegendentry{F-Q1};
\addlegendentry{F-C0};
\addlegendentry{F-C1};
\nextgroupplot
\node[anchor=north,align=center,font=\fontsize{5}{4}\selectfont] at (axis description cs: 0.25,  0.95) {Design \eqref{Eqn: MC2,aux3}\\$n=750$};
\addplot[smooth,tension=0.5,color=NavyBlue, no markers,line width=0.25pt, densely dotted,forget plot] table[x = delta,y=alpha] from \BiConQ;
\addplot table[x = delta,y=ConSevenQKn0] from \BiConQCLR;
\addplot table[x = delta,y=ConSevenQKn1] from \BiConQCLR;
\addplot table[x = delta,y=ConSevenCKn0] from \BiConCCLR;
\addplot table[x = delta,y=ConSevenCKn1] from \BiConCCLR;
\addplot table[x = delta,y=ConSevenQKn0] from \BiConQ;
\addplot table[x = delta,y=ConSevenQKn1] from \BiConQ;
\addplot table[x = delta,y=ConSevenCKn0] from \BiConC;
\addplot table[x = delta,y=ConSevenCKn1] from \BiConC;
\nextgroupplot
\node[anchor=north,align=center,font=\fontsize{5}{4}\selectfont] at (axis description cs: 0.25,  0.95) {Design \eqref{Eqn: MC2,aux3}\\$n=1000$};
\addplot[smooth,tension=0.5,color=NavyBlue, no markers,line width=0.25pt, densely dotted,forget plot] table[x = delta,y=alpha] from \BiConQ;
\addplot table[x = delta,y=ConTenQKn0] from \BiConQCLR;
\addplot table[x = delta,y=ConTenQKn1] from \BiConQCLR;
\addplot table[x = delta,y=ConTenCKn0] from \BiConCCLR;
\addplot table[x = delta,y=ConTenCKn1] from \BiConCCLR;
\addplot table[x = delta,y=ConTenQKn0] from \BiConQ;
\addplot table[x = delta,y=ConTenQKn1] from \BiConQ;
\addplot table[x = delta,y=ConTenCKn0] from \BiConC;
\addplot table[x = delta,y=ConTenCKn1] from \BiConC;
\end{groupplot}
\node at ($(myplots c2r1) + (0,-2.25cm)$) {\ref{LegendCon1}};
\node at ($(myplots c2r2) + (0,-2.25cm)$) {\ref{LegendCon2}};
\node[align=center] at ($(myplots c2r3) + (0,-2.25cm)$) {\ref{LegendCon3}};
\end{tikzpicture}
\caption{Testing convexity/concavity: Empirical power of our test and the CLR test for the designs \eqref{Eqn: MC1,aux1}, \eqref{Eqn: MC2,aux2}, and \eqref{Eqn: MC2,aux3}, where corresponding to $\delta=0$ are the empirical sizes under D1. In the first row, CLR-C3, and F-C7 nearly overlap each other.} \label{Fig: MC,Con}
\end{figure}

\subsection{Testing Joint Restrictions}

Finally, we test the joint restrictions of monotonicity and convexity/concavity, based on the same designs as those in Section \ref{Sec: simulations, con}. We implement our test based on the supremum distance, coupled with the $\Gamma$ operator (see Assumption \ref{Ass: enforce null}) obtained by taking the composition $\Gamma=\Upsilon_2\circ\Upsilon_1$, where $\Upsilon_1$ is the rearrangement operator and $\Upsilon_2$ is the GCM/LCM operator. That is, we apply rearrangement first and then the GCM/LCM operation. The remaining steps of our tests as well as the CLR tests are the same as before, beyond the need of incorporating the joint restrictions.

{
\setlength{\tabcolsep}{6.5pt}
\renewcommand{\arraystretch}{1.1}
\begin{table}[!ht]
\caption{Empirical Size of Monotonicity-Convexity Tests for $\theta_0$ in \eqref{Eqn: MC1,aux1} at $\alpha=5\%$} \label{Tab: MonConMC, size1}
\centering\footnotesize
\begin{threeparttable}
\sisetup{table-number-alignment = center, table-format = 1.3} 
\begin{tabularx}{\linewidth}{@{} cc *{3}{S[round-mode = places,round-precision = 3]}  c *{3}{S[round-mode = places,round-precision = 3]}  c *{3}{S[round-mode = places,round-precision = 3]}@{}} 
\hline
\hline
 $n$ & $\gamma_n$ & {D1} & {D2} & {D3}  & & {D1} & {D2} & {D3} & & {D1} & {D2} & {D3}\\
\hline
\rule{0pt}{15pt}
&& \multicolumn{3}{c}{{F-C3: $k_n=7$}} && \multicolumn{3}{c}{{F-C5: $k_n=9$}} && \multicolumn{3}{c}{{F-C7: $k_n=11$}}\\
\multirow{3}{*}{$500$}   & $1/n$                  & 0.0553 & 0.0453 & 0.0120 & & 0.0517 & 0.0463 & 0.0157 & & 0.0533 & 0.0443 & 0.0237\\
                         & $0.01/\log n$          & 0.0553 & 0.0453 & 0.0120 & & 0.0517 & 0.0463 & 0.0157 & & 0.0533 & 0.0443 & 0.0237\\
                         & $0.01$                 & 0.0553 & 0.0453 & 0.0120 & & 0.0517 & 0.0463 & 0.0157 & & 0.0533 & 0.0443 & 0.0237\\
\rule{0pt}{12pt}
 \multirow{3}{*}{$750$}  & $1/n$                  & 0.0547 & 0.0463 & 0.0100 & & 0.0547 & 0.0477 & 0.0180 & & 0.0583 & 0.0530 & 0.0220\\
                         & $0.01/\log n$          & 0.0547 & 0.0463 & 0.0100 & & 0.0547 & 0.0477 & 0.0180 & & 0.0583 & 0.0530 & 0.0220\\
                         & $0.01$                 & 0.0550 & 0.0463 & 0.0100 & & 0.0547 & 0.0477 & 0.0180 & & 0.0583 & 0.0533 & 0.0220\\
\rule{0pt}{12pt}
 \multirow{3}{*}{$1000$} & $1/n$                  & 0.0603 & 0.0507 & 0.0063 & & 0.0600 & 0.0483 & 0.0137 & & 0.0600 & 0.0520 & 0.0217\\
                         & $0.01/\log n$          & 0.0603 & 0.0507 & 0.0063 & & 0.0600 & 0.0483 & 0.0137 & & 0.0600 & 0.0520 & 0.0217\\
                         & $0.01$                 & 0.0607 & 0.0507 & 0.0070 & & 0.0600 & 0.0487 & 0.0137 & & 0.0600 & 0.0527 & 0.0217\\
\rule{0pt}{15pt}
&& \multicolumn{3}{c}{CLR-C3: $k_n=7$} && \multicolumn{3}{c}{CLR-C5: $k_n=9$} && \multicolumn{3}{c}{CLR-C7: $k_n=11$}\\
$500$ &                       & 0.0590 & 0.0467 & 0.0300 && 0.0640 & 0.0527 & 0.0387 && 0.0723 & 0.0573 & 0.0470\\
$750$ &                       & 0.0653 & 0.0503 & 0.0320 && 0.0670 & 0.0563 & 0.0387 && 0.0667 & 0.0557 & 0.0433\\
$1000$&                       & 0.0583 & 0.0433 & 0.0247 && 0.0663 & 0.0533 & 0.0373 && 0.0663 & 0.0550 & 0.0427\\
\hline
\hline
\end{tabularx}
\begin{tablenotes}[flushleft]
\item {\it Note:} The parameter $\gamma_n$ determines $\hat\kappa_n$ as in Proposition \ref{Pro: tuning parameter} with $c_n=1/\log n$ and $r_n=(n/k_n)^{1/2}$.
\end{tablenotes}
\end{threeparttable}
\end{table}
}

Tables \ref{Tab: MonConMC, size1} and \ref{Tab: MonConMC, size2} summarize the empirical sizes, while Figure \ref{Fig: MC,Con} presents the power curves. Consistent with our previous findings, our tests control size reasonably well in both univariate and bivariate designs, across sample sizes and sieve spaces, while the CLR tests tend to over-reject in bivariate designs (but otherwise perform well in terms of size). Moreover, our tests enjoy competitive power compared to the CLR tests, especially when $\theta_0$ has a flat region.

To conclude our simulations, we stress that our intention is not to show that our test is uniformly better than existing tests. As well understood in the literature, uniformly most powerful tests typically do not exist in nonparametric (and also many parametric) settings, and thus no single test would dominate all others in terms of both size and power. Indeed, our numerical results show that, depending on the sample size, the functional form of $\theta_0$, and the sieve space, the CLR tests and the C-OS test may perform better than our tests (in terms of size or power). Instead, we hope to convey the message that our tests may serve as competitive alternatives, and may perform better in particular settings. In addition, compared to tests such as \citet{Chetverikov2018Monotonicity} that are designed for specific shapes, our framework readily accommodates additional/joint restrictions. Our numerical exercises also confirm the applicability and usefulness of shape enforcing operators in either forming the Wald functinal or enforcing the null restriction for the purpose of power improvement.


{
\setlength{\tabcolsep}{3pt}
\renewcommand{\arraystretch}{1.1}
\begin{table}[!ht]
\caption{Empirical Size of Monotonicity-Concavity Tests for $\theta_0$ in \eqref{Eqn: MC2,aux3} at $\alpha=5\%$} \label{Tab: MonConMC, size2}
\centering\footnotesize
\begin{threeparttable}
\sisetup{table-number-alignment = center, table-format = 1.3} 
\begin{tabularx}{\linewidth}{@{}cc *{3}{S[round-mode = places,round-precision = 3]} c *{3}{S[round-mode = places,round-precision = 3]} c *{3}{S[round-mode = places,round-precision = 3]} c *{3}{S[round-mode = places,round-precision = 3]}@{}} 
\hline
\hline
$n$ & $\gamma_n$ & {D1} & {D2} & {D3}  && {D1} & {D2} & {D3} && {D1} & {D2} & {D3} && {D1} & {D2} & {D3}\\
\hline
\rule{0pt}{15pt}
&& \multicolumn{3}{c}{F-Q0: $k_n=9$} && \multicolumn{3}{c}{F-Q1: $k_n=16$} && \multicolumn{3}{c}{F-C0: $k_n=16$} && \multicolumn{3}{c}{F-C1: $k_n=25$}\\
\multirow{3}{*}{$500$} & $1/n$         & 0.0597 & 0.0313 & 0.0007 & & 0.0693 & 0.0367 & 0.0050 & & 0.0760 & 0.0373 & 0.0047 & & 0.0673 & 0.0447 & 0.0087\\
                       & $0.01/\log n$ & 0.0597 & 0.0313 & 0.0007 & & 0.0693 & 0.0367 & 0.0050 & & 0.0760 & 0.0373 & 0.0047 & & 0.0673 & 0.0447 & 0.0087\\
                       & $0.01$        & 0.0597 & 0.0313 & 0.0007 & & 0.0693 & 0.0367 & 0.0053 & & 0.0767 & 0.0380 & 0.0767 & & 0.0677 & 0.0453 & 0.0087\\
\rule{0pt}{12pt}
 \multirow{3}{*}{$750$}& $1/n$         & 0.0597 & 0.0253 & 0.0013 & & 0.0660 & 0.0323 & 0.0033 & & 0.0700 & 0.0333 & 0.0027 & & 0.0677 & 0.0393 & 0.0060\\
                       & $0.01/\log n$ & 0.0597 & 0.0253 & 0.0013 & & 0.0660 & 0.0323 & 0.0033 & & 0.0700 & 0.0333 & 0.0027 & & 0.0677 & 0.0393 & 0.0060\\
                       & $0.01$        & 0.0607 & 0.0253 & 0.0013 & & 0.0663 & 0.0323 & 0.0037 & & 0.0707 & 0.0333 & 0.0027 & & 0.0680 & 0.0397 & 0.0060\\
\rule{0pt}{12pt}
\multirow{3}{*}{$1000$}& $1/n$         & 0.0590 & 0.0203 & 0.0003 & & 0.0547 & 0.0253 & 0.0010 & & 0.0570 & 0.0253 & 0.0010 & & 0.0587 & 0.0303 & 0.0057\\
                       & $0.01/\log n$ & 0.0590 & 0.0203 & 0.0003 & & 0.0547 & 0.0253 & 0.0010 & & 0.0570 & 0.0253 & 0.0010 & & 0.0587 & 0.0303 & 0.0057\\
                       & $0.01$        & 0.0593 & 0.0207 & 0.0003 & & 0.0553 & 0.0260 & 0.0010 & & 0.0577 & 0.0257 & 0.0010 & & 0.0587 & 0.0303 & 0.0057\\
\rule{0pt}{15pt}
&& \multicolumn{3}{c}{CLR-Q0: $k_n=9$} && \multicolumn{3}{c}{CLR-Q1: $k_n=16$} && \multicolumn{3}{c}{CLR-C0: $k_n=16$} && \multicolumn{3}{c}{CLR-C1: $k_n=25$}\\
$500$ &                & 0.0703 & 0.0527 & 0.0173 && 0.1023 & 0.0900 & 0.0547 && 0.1113 & 0.0947 & 0.0533 && 0.1523 & 0.1520 & 0.1037\\
$750$ &                & 0.0697 & 0.0480 & 0.0153 && 0.0907 & 0.0763 & 0.0373 && 0.0913 & 0.0787 & 0.0417 && 0.1127 & 0.1030 & 0.0663\\
$1000$&                & 0.0653 & 0.0437 & 0.0087 && 0.0830 & 0.0667 & 0.0323 && 0.0793 & 0.0677 & 0.0330 && 0.0973 & 0.0873 & 0.0520\\
\hline
\hline
\end{tabularx}
\begin{tablenotes}[flushleft]
\item {\it Note:} The parameter $\gamma_n$ determines $\hat\kappa_n$ as in Proposition \ref{Pro: tuning parameter} with $c_n=1/\log n$ and $r_n=(n/k_n)^{1/2}$.
\end{tablenotes}
\end{threeparttable}
\end{table}
}

\pgfplotstableread{ 
delta alpha MConFiveKn3 MConFiveKn5 MConFiveKn7 MConSevenKn3 MConSevenKn5 MConSevenKn7 MConTenKn3 MConTenKn5 MConTenKn7
0     0.05     0.0553      0.0517      0.0533      0.0547       0.0547       0.0583       0.0603     0.0600     0.0600
1     0.05     0.0840      0.0700      0.0667      0.1007       0.0820       0.0810       0.1103     0.1030     0.0907
2     0.05     0.1590      0.1247      0.1070      0.2233       0.1727       0.1393       0.2663     0.2237     0.1897
3     0.05     0.2980      0.2333      0.1870      0.4150       0.3263       0.2660       0.5480     0.4420     0.3563
4     0.05     0.4720      0.3873      0.3117      0.6497       0.5557       0.4693       0.7880     0.6993     0.6060
5     0.05     0.6677      0.5657      0.4767      0.8517       0.7660       0.6873       0.9420     0.8900     0.8200
6     0.05     0.8263      0.7420      0.6520      0.9547       0.9120       0.8470       0.9857     0.9710     0.9430
7     0.05     0.9243      0.8740      0.8083      0.9907       0.9787       0.9493       0.9983     0.9960     0.9893
8     0.05     0.9747      0.9510      0.9063      0.9980       0.9980       0.9910       0.9993     0.9990     0.9990
9     0.05     0.9933      0.9833      0.9680      1.0000       1.0000       0.9987       1.0000     0.9997     0.9997
10    0.05     0.9987      0.9960      0.9893      1.0000       1.0000       1.0000       1.0000     1.0000     1.0000
}\UniMCona

\pgfplotstableread{ 
delta alpha MConFiveKn3 MConFiveKn5 MConFiveKn7 MConSevenKn3 MConSevenKn5 MConSevenKn7 MConTenKn3 MConTenKn5 MConTenKn7
0     0.05     0.0553      0.0517      0.0533      0.0547       0.0547       0.0583       0.0603     0.0600     0.0600
1     0.05     0.0927      0.0753      0.0727      0.1053       0.0873       0.0840       0.1153     0.1063     0.0913
2     0.05     0.1527      0.1203      0.1023      0.1880       0.1483       0.1297       0.2257     0.1920     0.1527
3     0.05     0.2610      0.1810      0.1500      0.3380       0.2423       0.2073       0.4170     0.3167     0.2563
4     0.05     0.3880      0.2700      0.2193      0.5263       0.3923       0.3143       0.6620     0.4993     0.4093
5     0.05     0.5327      0.3930      0.3120      0.7177       0.5577       0.4627       0.8480     0.6987     0.5990
6     0.05     0.6920      0.5277      0.4243      0.8737       0.7283       0.6113       0.9557     0.8660     0.7637
7     0.05     0.8323      0.6667      0.5420      0.9507       0.8683       0.7607       0.9910     0.9620     0.8893
8     0.05     0.9170      0.8077      0.6737      0.9863       0.9450       0.8797       0.9993     0.9900     0.9663
9     0.05     0.9693      0.9010      0.7953      0.9953       0.9800       0.9433       1.0000     0.9980     0.9893
10    0.05     0.9907      0.9520      0.8823      0.9997       0.9947       0.9790       1.0000     1.0000     0.9980
}\UniMConb

\pgfplotstableread{ 
delta alpha MConFiveQKn0 MConFiveQKn1 MConSevenQKn0 MConSevenQKn1 MConTenQKn0 MConTenQKn1
0     0.05     0.0597       0.0693       0.0597        0.0660        0.0590      0.0547
1     0.05     0.2857       0.2637       0.3840        0.3407        0.4533      0.4307
2     0.05     0.7570       0.6973       0.9040        0.8647        0.9727      0.9493
3     0.05     0.9823       0.9647       0.9993        0.9970        1.0000      1.0000
4     0.05     0.9997       0.9983       1.0000        1.0000        1.0000      1.0000
5     0.05     1.0000       1.0000       1.0000        1.0000        1.0000      1.0000
6     0.05     1.0000       1.0000       1.0000        1.0000        1.0000      1.0000
7     0.05     1.0000       1.0000       1.0000        1.0000        1.0000      1.0000
8     0.05     1.0000       1.0000       1.0000        1.0000        1.0000      1.0000
9     0.05     1.0000       1.0000       1.0000        1.0000        1.0000      1.0000
10    0.05     1.0000       1.0000       1.0000        1.0000        1.0000      1.0000
}\BiMConQ

\pgfplotstableread{ 
delta alpha MConFiveCKn0 MConFiveCKn1 MConSevenCKn0 MConSevenCKn1 MConTenCKn0 MConTenCKn1
0     0.05     0.0760       0.0673       0.0700        0.0677        0.0570     0.0587
1     0.05     0.2807       0.2060       0.3610        0.2683        0.4447     0.3210
2     0.05     0.7183       0.5583       0.8830        0.7593        0.9613     0.8640
3     0.05     0.9720       0.9063       0.9983        0.9883        1.0000     0.9993
4     0.05     0.9987       0.9940       1.0000        1.0000        1.0000     1.0000
5     0.05     1.0000       1.0000       1.0000        1.0000        1.0000     1.0000
6     0.05     1.0000       1.0000       1.0000        1.0000        1.0000     1.0000
7     0.05     1.0000       1.0000       1.0000        1.0000        1.0000     1.0000
8     0.05     1.0000       1.0000       1.0000        1.0000        1.0000     1.0000
9     0.05     1.0000       1.0000       1.0000        1.0000        1.0000     1.0000
10    0.05     1.0000       1.0000       1.0000        1.0000        1.0000     1.0000
}\BiMConC

\pgfplotstableread{ 
delta alpha MConFiveKn3 MConFiveKn5 MConFiveKn7 MConSevenKn3 MConSevenKn5 MConSevenKn7 MConTenKn3 MConTenKn5 MConTenKn7
0     0.05     0.0590      0.0640      0.0723      0.0653       0.0670       0.0667      0.0583     0.0663     0.0663
1     0.05     0.0760      0.0703      0.0713      0.0850       0.0763       0.0717      0.0787     0.0803     0.0763
2     0.05     0.1210      0.0997      0.0837      0.1543       0.1127       0.0967      0.1717     0.1413     0.1003
3     0.05     0.2107      0.1493      0.1163      0.2957       0.2077       0.1297      0.3733     0.2597     0.1647
4     0.05     0.3380      0.2240      0.1620      0.4897       0.3350       0.2203      0.6273     0.4577     0.2683
5     0.05     0.4963      0.3547      0.2303      0.6910       0.5223       0.3343      0.8290     0.6757     0.4380
6     0.05     0.6733      0.5067      0.3260      0.8623       0.7187       0.4907      0.9413     0.8420     0.6380
7     0.05     0.8107      0.6557      0.4447      0.9537       0.8497       0.6490      0.9893     0.9423     0.8017
8     0.05     0.9090      0.7970      0.5800      0.9860       0.9387       0.7970      0.9977     0.9847     0.9053
9     0.05     0.9623      0.8943      0.6987      0.9977       0.9817       0.8967      0.9997     0.9970     0.9610
10    0.05     0.9847      0.9533      0.8147      0.9993       0.9950       0.9560      1.0000     1.0000     0.9897
}\UniMConCLRa

\pgfplotstableread{ 
delta alpha MConFiveKn3 MConFiveKn5 MConFiveKn7 MConSevenKn3 MConSevenKn5 MConSevenKn7 MConTenKn3 MConTenKn5 MConTenKn7
0     0.05     0.0590      0.0640      0.0723      0.0653       0.0670       0.0667      0.0583     0.0663     0.0663
1     0.05     0.0633      0.0667      0.0673      0.0753       0.0750       0.0680      0.0690     0.0743     0.0687
2     0.05     0.0773      0.0740      0.0697      0.0980       0.0840       0.0770      0.0977     0.0863     0.0770
3     0.05     0.1003      0.0830      0.0793      0.1343       0.1037       0.0870      0.1547     0.1107     0.0863
4     0.05     0.1407      0.0983      0.0847      0.1880       0.1273       0.1047      0.2323     0.1497     0.1073
5     0.05     0.1897      0.1197      0.1003      0.2723       0.1653       0.1267      0.3373     0.1930     0.1287
6     0.05     0.2557      0.1520      0.1103      0.3677       0.2170       0.1470      0.4727     0.2523     0.1637
7     0.05     0.3407      0.1860      0.1277      0.4813       0.2647       0.1857      0.6213     0.3307     0.2010
8     0.05     0.4283      0.2290      0.1530      0.6150       0.3387       0.2207      0.7630     0.4347     0.2593
9     0.05     0.5233      0.2887      0.1803      0.7387       0.4230       0.2577      0.8610     0.5410     0.3160
10    0.05     0.6247      0.3527      0.2137      0.8347       0.5153       0.3030      0.9397     0.6440     0.3867
}\UniMConCLRb

\pgfplotstableread{ 
delta alpha MConFiveQKn0 MConFiveQKn1 MConSevenQKn0 MConSevenQKn1 MConTenQKn0 MConTenQKn1
0     0.05     0.0703       0.1023       0.0697        0.0907        0.0653      0.0830
1     0.05     0.2210       0.1733       0.2613        0.1680        0.3053      0.1760
2     0.05     0.5817       0.3103       0.7700        0.3613        0.8750      0.4287
3     0.05     0.9277       0.5380       0.9887        0.6757        0.9983      0.7973
4     0.05     0.9987       0.7923       1.0000        0.9190        1.0000      0.9820
5     0.05     1.0000       0.9447       1.0000        0.9957        1.0000      1.0000
6     0.05     1.0000       0.9927       1.0000        1.0000        1.0000      1.0000
7     0.05     1.0000       0.9997       1.0000        1.0000        1.0000      1.0000
8     0.05     1.0000       1.0000       1.0000        1.0000        1.0000      1.0000
9     0.05     1.0000       1.0000       1.0000        1.0000        1.0000      1.0000
10    0.05     1.0000       1.0000       1.0000        1.0000        1.0000      1.0000
}\BiMConQCLR

\pgfplotstableread{ 
delta alpha MConFiveCKn0 MConFiveCKn1 MConSevenCKn0 MConSevenCKn1 MConTenCKn0 MConTenCKn1
0     0.05     0.1113       0.1523       0.0913        0.1127        0.0793      0.0973
1     0.05     0.1647       0.1787       0.1620        0.1413        0.1517      0.1347
2     0.05     0.2720       0.2407       0.3157        0.2153        0.3620      0.2250
3     0.05     0.4603       0.3270       0.5763        0.3350        0.7047      0.3810
4     0.05     0.6907       0.4473       0.8490        0.5153        0.9513      0.6107
5     0.05     0.8930       0.5927       0.9747        0.7080        0.9980      0.8273
6     0.05     0.9790       0.7433       0.9990        0.8737        1.0000      0.9630
7     0.05     0.9983       0.8663       1.0000        0.9617        1.0000      0.9967
8     0.05     1.0000       0.9517       1.0000        0.9957        1.0000      1.0000
9     0.05     1.0000       0.9877       1.0000        1.0000        1.0000      1.0000
10    0.05     1.0000       0.9960       1.0000        1.0000        1.0000      1.0000
}\BiMConCCLR

\begin{figure}[!h]
\centering\scriptsize
\begin{tikzpicture} 
\begin{groupplot}[group style={group name=myplots,group size=3 by 3,horizontal sep= 0.8cm,vertical sep=1.1cm},
    grid = minor,
    width = 0.375\textwidth,
    xmax=10,xmin=0,
    ymax=1,ymin=0,
    every axis title/.style={below,at={(0.2,0.8)}},
    xlabel=$\delta$,
    x label style={at={(axis description cs:0.95,0.04)},anchor=south},
    xtick={0,2,...,10},
    ytick={0.05,0.5,1},
    tick label style={/pgf/number format/fixed},
    legend style={font=\tiny,text=black,cells={align=center},row sep = 3pt,legend columns = -1, draw=none,fill=none},
    cycle list={
{smooth,tension=0.5,color=BuGn-M, mark=halfsquare*,every mark/.append style={rotate=90},mark size=1.75pt,line width=0.5pt},
{smooth,tension=0.5,color=BuGn-K, mark=halfsquare*,every mark/.append style={rotate=180},mark size=1.75pt,line width=0.5pt}, 
{smooth,tension=0.5,color=BuGn-I, mark=halfsquare*,every mark/.append style={rotate=270},mark size=1.75pt,line width=0.5pt}, 
{smooth,tension=0.5,color=BuGn-G, mark=halfsquare*,every mark/.append style={rotate=360},mark size=1.75pt,line width=0.5pt}, 
{smooth,tension=0.5,color=BuPu-M, mark=halfcircle*,every mark/.append style={rotate=270},mark size=1.75pt,line width=0.5pt}, 
{smooth,tension=0.5,color=BuPu-K, mark=halfcircle*,every mark/.append style={rotate=360},mark size=1.75pt,line width=0.5pt}, 
{smooth,tension=0.5,color=BuPu-J, mark=halfcircle*,every mark/.append style={rotate=90},mark size=1.75pt,line width=0.5pt},
{smooth,tension=0.5,color=BuPu-I, mark=halfcircle*,every mark/.append style={rotate=180},mark size=1.75pt,line width=0.5pt},
}
]
\nextgroupplot[legend style = {column sep = 3pt, legend to name = LegendMCon1}]
\addplot[smooth,tension=0.5,color=NavyBlue, no markers,line width=0.25pt, densely dotted,forget plot] table[x = delta,y=alpha] from \UniMCona;
\addplot table[x = delta,y=MConFiveKn3] from \UniMConCLRa;
\addplot table[x = delta,y=MConFiveKn5] from \UniMConCLRa;
\addplot table[x = delta,y=MConFiveKn7] from \UniMConCLRa;
\pgfplotsset{cycle list shift=1}
\addplot table[x = delta,y=MConFiveKn3] from \UniMCona;
\addplot table[x = delta,y=MConFiveKn5] from \UniMCona;
\addplot table[x = delta,y=MConFiveKn7] from \UniMCona;
\node[anchor=north,align=center,font=\fontsize{5}{4}\selectfont] at (axis description cs: 0.25,  0.95) {Design \eqref{Eqn: MC1,aux1}\\$n=500$};
\addlegendentry{CLR-C3};
\addlegendentry{CLR-C5};
\addlegendentry{CLR-C7};
\addlegendentry{F-C3};
\addlegendentry{F-C5};
\addlegendentry{F-C7};
\nextgroupplot
\node[anchor=north,align=center,font=\fontsize{5}{4}\selectfont] at (axis description cs: 0.25,  0.95) {Design \eqref{Eqn: MC1,aux1}\\$n=750$};
\addplot[smooth,tension=0.5,color=NavyBlue, no markers,line width=0.25pt, densely dotted,forget plot] table[x = delta,y=alpha] from \UniMCona;
\addplot table[x = delta,y=MConSevenKn3] from \UniMConCLRa;
\addplot table[x = delta,y=MConSevenKn5] from \UniMConCLRa;
\addplot table[x = delta,y=MConSevenKn7] from \UniMConCLRa;
\pgfplotsset{cycle list shift=1}
\addplot table[x = delta,y=MConSevenKn3] from \UniMCona;
\addplot table[x = delta,y=MConSevenKn5] from \UniMCona;
\addplot table[x = delta,y=MConSevenKn7] from \UniMCona;
\nextgroupplot
\node[anchor=north,align=center,font=\fontsize{5}{4}\selectfont] at (axis description cs: 0.25,  0.95) {Design \eqref{Eqn: MC1,aux1}\\$n=1000$};
\addplot[smooth,tension=0.5,color=NavyBlue, no markers,line width=0.25pt, densely dotted,forget plot] table[x = delta,y=alpha] from \UniMCona;
\addplot table[x = delta,y=MConTenKn3] from \UniMConCLRa;
\addplot table[x = delta,y=MConTenKn5] from \UniMConCLRa;
\addplot table[x = delta,y=MConTenKn7] from \UniMConCLRa;
\pgfplotsset{cycle list shift=1}
\addplot table[x = delta,y=MConTenKn3] from \UniMCona;
\addplot table[x = delta,y=MConTenKn5] from \UniMCona;
\addplot table[x = delta,y=MConTenKn7] from \UniMCona;
\nextgroupplot[legend style = {column sep = 3pt, legend to name = LegendMCon2}]
\addplot[smooth,tension=0.5,color=NavyBlue, no markers,line width=0.25pt, densely dotted,forget plot] table[x = delta,y=alpha] from \UniMConb;
\addplot table[x = delta,y=MConFiveKn3] from \UniMConCLRb;
\addplot table[x = delta,y=MConFiveKn5] from \UniMConCLRb;
\addplot table[x = delta,y=MConFiveKn7] from \UniMConCLRb;
\pgfplotsset{cycle list shift=1}
\addplot table[x = delta,y=MConFiveKn3] from \UniMConb;
\addplot table[x = delta,y=MConFiveKn5] from \UniMConb;
\addplot table[x = delta,y=MConFiveKn7] from \UniMConb;
\node[anchor=north,align=center,font=\fontsize{5}{4}\selectfont] at (axis description cs: 0.25,  0.95) {Design \eqref{Eqn: MC2,aux2}\\$n=500$};
\addlegendentry{CLR-C3};
\addlegendentry{CLR-C5};
\addlegendentry{CLR-C7};
\addlegendentry{F-C3};
\addlegendentry{F-C5};
\addlegendentry{F-C7};
\nextgroupplot
\node[anchor=north,align=center,font=\fontsize{5}{4}\selectfont] at (axis description cs: 0.25,  0.95) {Design \eqref{Eqn: MC2,aux2}\\$n=750$};
\addplot[smooth,tension=0.5,color=NavyBlue, no markers,line width=0.25pt, densely dotted,forget plot] table[x = delta,y=alpha] from \UniMConb;
\addplot table[x = delta,y=MConSevenKn3] from \UniMConCLRb;
\addplot table[x = delta,y=MConSevenKn5] from \UniMConCLRb;
\addplot table[x = delta,y=MConSevenKn7] from \UniMConCLRb;
\pgfplotsset{cycle list shift=1}
\addplot table[x = delta,y=MConSevenKn3] from \UniMConb;
\addplot table[x = delta,y=MConSevenKn5] from \UniMConb;
\addplot table[x = delta,y=MConSevenKn7] from \UniMConb;
\nextgroupplot
\node[anchor=north,align=center,font=\fontsize{5}{4}\selectfont] at (axis description cs: 0.25,  0.95) {Design \eqref{Eqn: MC2,aux2}\\$n=1000$};
\addplot[smooth,tension=0.5,color=NavyBlue, no markers,line width=0.25pt, densely dotted,forget plot] table[x = delta,y=alpha] from \UniMConb;
\addplot table[x = delta,y=MConTenKn3] from \UniMConCLRb;
\addplot table[x = delta,y=MConTenKn5] from \UniMConCLRb;
\addplot table[x = delta,y=MConTenKn7] from \UniMConCLRb;
\pgfplotsset{cycle list shift=1}
\addplot table[x = delta,y=MConTenKn3] from \UniMConb;
\addplot table[x = delta,y=MConTenKn5] from \UniMConb;
\addplot table[x = delta,y=MConTenKn7] from \UniMConb;
\nextgroupplot[legend style = {column sep = 0.5pt, legend to name = LegendMCon3}]
\node[anchor=north,align=center,font=\fontsize{5}{4}\selectfont] at (axis description cs: 0.8, 0.4) {Design \eqref{Eqn: MC2,aux3}\\$n=500$};
\addplot[smooth,tension=0.5,color=NavyBlue, no markers,line width=0.25pt, densely dotted,forget plot] table[x = delta,y=alpha] from \BiMConQ;
\addplot table[x = delta,y=MConFiveQKn0] from \BiMConQCLR;
\addplot table[x = delta,y=MConFiveQKn1] from \BiMConQCLR;
\addplot table[x = delta,y=MConFiveCKn0] from \BiMConCCLR;
\addplot table[x = delta,y=MConFiveCKn1] from \BiMConCCLR;
\addplot table[x = delta,y=MConFiveQKn0] from \BiMConQ;
\addplot table[x = delta,y=MConFiveQKn1] from \BiMConQ;
\addplot table[x = delta,y=MConFiveCKn0] from \BiMConC;
\addplot table[x = delta,y=MConFiveCKn1] from \BiMConC;
\addlegendentry{CLR-Q0};
\addlegendentry{CLR-Q1};
\addlegendentry{CLR-C0};
\addlegendentry{CLR-C1};
\addlegendentry{F-Q0};
\addlegendentry{F-Q1};
\addlegendentry{F-C0};
\addlegendentry{F-C1};
\nextgroupplot
\node[anchor=north,align=center,font=\fontsize{5}{4}\selectfont] at (axis description cs: 0.8, 0.4) {Design \eqref{Eqn: MC2,aux3}\\$n=750$};
\addplot[smooth,tension=0.5,color=NavyBlue, no markers,line width=0.25pt, densely dotted,forget plot] table[x = delta,y=alpha] from \BiMConQ;
\addplot table[x = delta,y=MConSevenQKn0] from \BiMConQCLR;
\addplot table[x = delta,y=MConSevenQKn1] from \BiMConQCLR;
\addplot table[x = delta,y=MConSevenCKn0] from \BiMConCCLR;
\addplot table[x = delta,y=MConSevenCKn1] from \BiMConCCLR;
\addplot table[x = delta,y=MConSevenQKn0] from \BiMConQ;
\addplot table[x = delta,y=MConSevenQKn1] from \BiMConQ;
\addplot table[x = delta,y=MConSevenCKn0] from \BiMConC;
\addplot table[x = delta,y=MConSevenCKn1] from \BiMConC;
\nextgroupplot
\node[anchor=north,align=center,font=\fontsize{5}{4}\selectfont] at (axis description cs: 0.8, 0.4) {Design \eqref{Eqn: MC2,aux3}\\$n=1000$};
\addplot[smooth,tension=0.5,color=NavyBlue, no markers,line width=0.25pt, densely dotted,forget plot] table[x = delta,y=alpha] from \BiMConQ;
\addplot table[x = delta,y=MConTenQKn0] from \BiMConQCLR;
\addplot table[x = delta,y=MConTenQKn1] from \BiMConQCLR;
\addplot table[x = delta,y=MConTenCKn0] from \BiMConCCLR;
\addplot table[x = delta,y=MConTenCKn1] from \BiMConCCLR;
\addplot table[x = delta,y=MConTenQKn0] from \BiMConQ;
\addplot table[x = delta,y=MConTenQKn1] from \BiMConQ;
\addplot table[x = delta,y=MConTenCKn0] from \BiMConC;
\addplot table[x = delta,y=MConTenCKn1] from \BiMConC;
\end{groupplot}
\node at ($(myplots c2r1) + (0,-2.25cm)$) {\ref{LegendMCon1}};
\node at ($(myplots c2r2) + (0,-2.25cm)$) {\ref{LegendMCon2}};
\node[align=center] at ($(myplots c2r3) + (0,-2.25cm)$) {\ref{LegendMCon3}};
\end{tikzpicture}
\caption{Testing monotonicity jointly with convexity/concavity: Empirical power of our test and the CLR test for the designs \eqref{Eqn: MC1,aux1}, \ref{Eqn: MC2,aux2}, and \eqref{Eqn: MC2,aux3}, where corresponding to $\delta=0$ are the empirical sizes under D1.} \label{Fig: MC,MonCon}
\end{figure}
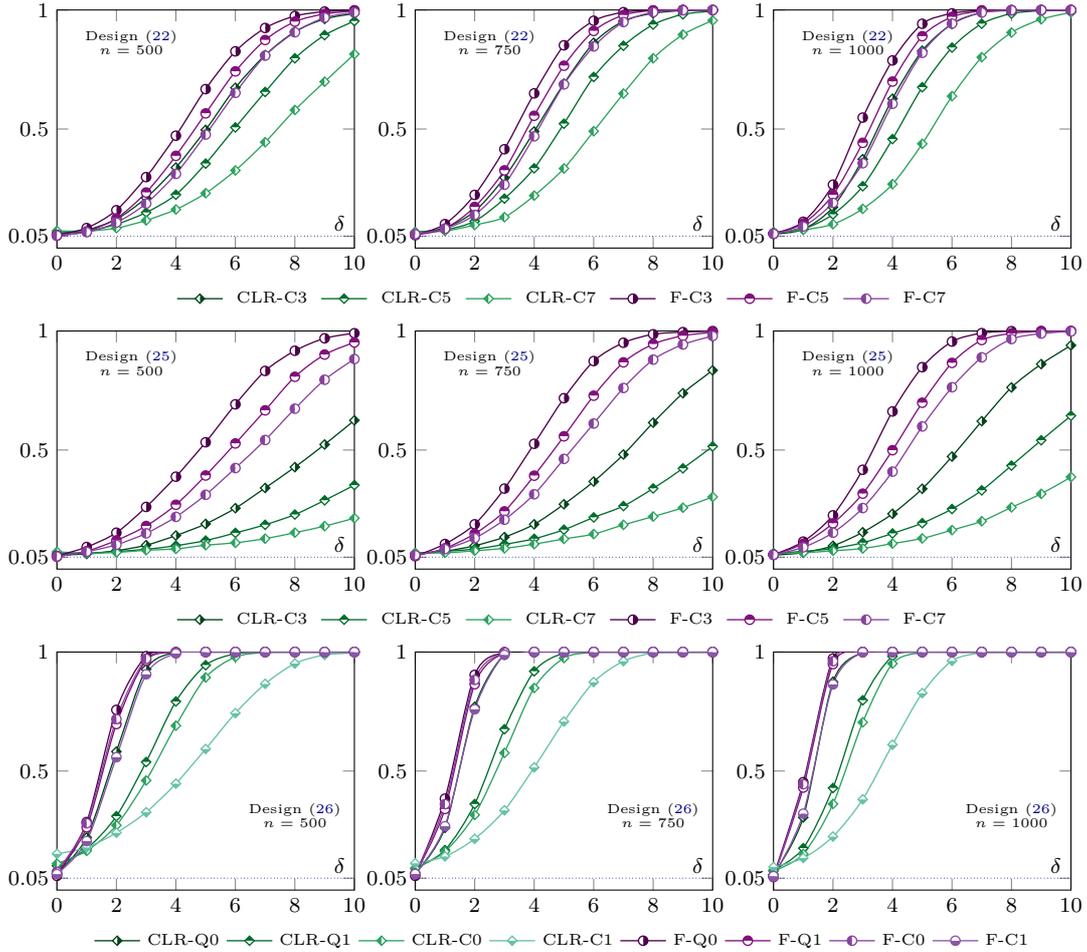

\section{Empirical Application}\label{Sec: empirical}

A central aspect of labor supply is the decision about working hours and its implications for earnings \citep{EhrenbergSmith2016Labor}. By incorporating the worker's labor supply decision and inherent ability into the dynamic complete information framework of \citet{GibbonsWaldman1999Wage}, \citet{Gicheva2013Working} builds up a model that predicts a positive and convex intertemporal relationship between working hours and wage growth. In turn, \citet{Gicheva2013Working} validates these shape predictions by analyzing the data set from a panel survey of registrants for the Graduate Management Admission Test (GMAT). Specifically, \citet{Gicheva2013Working} employs a partially linear model to estimate the relationship between working hours and wage growth, and then verifies the theoretical predictions through eyeball inspection. Below we complement her results by carrying out formal statistical tests of these restrictions based on the same data.

The GMAT sample involves individuals who registered to take the GMAT between June 1990 and March 1991, and were living in the United States at the time of registration. The survey was conducted in four waves of interviews: shortly after the registration, 15 months after registration, 3.5-4 years after registration, and 7 years registration. The sampled workers are concentrated in the high-end labor market: they are college educated, tend to work long hours, and have high earnings. The final cleaned sample consists of $1,911$ respondents of the survey; see Section III in \citet{Gicheva2013Working} for more details. Following \citet{Gicheva2013Working}, we base our analysis on the model:
\begin{align}
Y = \theta_0(Z) + W^\transpose\gamma + u~,
\end{align}
where $Y$ is the annual wage growth rate between the second and fourth interviews, $Z$ indicates weekly working hours, $W$ denotes a vector of demographic control variables (e.g., experience, education, gender, age, race, and family characteristics), and $E[u|Z,W]=0$. We then respectively test monotonicity, convexity, and monotonicity jointly with convexity of $\theta_0$, based on the full sample as in \citet{Gicheva2013Working} as well as each gender group ---there are $808$ women and $1,103$ men in the sample.

We employ the sup-distance test for monotonicity and the joint restriction, and the sup test based on GCM for convexity. For all three tests, the number of bootstrap samples is $5,000$, while the grid points for $Z$ is $3,...,90$ (based on the realized weakly working hours). The estimation and bootstrap steps are otherwise the same as the univariate designs in Section \ref{Sec: simulation}. Figure \ref{Fig: Empirical} depicts the curves of $\hat\theta_n$ obtained based on the full sample, women and men respectively, with the sieve dimension $k_n=9$ (i.e., cubic B-splines with 5 interior knots). While the aforementioned shape restrictions are ``overall'' present in the data, there are nonetheless local violations. It is therefore crucial to conduct hypothesis test in order to formally quantify the sampling uncertainty.


\pgfplotstableread{
Grid HatTheta
3     0.2764
4     0.2803
5     0.2838
6     0.2869
7     0.2895
8     0.2918
9     0.2937
10    0.2952
11    0.2965
12    0.2974
13    0.2980
14    0.2984
15    0.2985
16    0.2984
17    0.2980
18    0.2975
19    0.2969
20    0.2960
21    0.2951
22    0.2940
23    0.2929
24    0.2917
25    0.2905
26    0.2892
27    0.2880
28    0.2867
29    0.2855
30    0.2844
31    0.2833
32    0.2823
33    0.2815
34    0.2808
35    0.2802
36    0.2798
37    0.2797
38    0.2797
39    0.2803
40    0.2832
41    0.2862
42    0.2868
43    0.2857
44    0.2840
45    0.2825
46    0.2821
47    0.2826
48    0.2837
49    0.2853
50    0.2871
51    0.2887
52    0.2903
53    0.2918
54    0.2932
55    0.2946
56    0.2959
57    0.2971
58    0.2984
59    0.2996
60    0.3008
61    0.3021
62    0.3034
63    0.3047
64    0.3060
65    0.3074
66    0.3089
67    0.3105
68    0.3122
69    0.3140
70    0.3159
71    0.3179
72    0.3201
73    0.3225
74    0.3250
75    0.3277
76    0.3306
77    0.3337
78    0.3371
79    0.3406
80    0.3445
81    0.3485
82    0.3529
83    0.3575
84    0.3624
85    0.3676
86    0.3732
87    0.3791
88    0.3853
89    0.3919
90    0.3988
}\Full

\pgfplotstableread{
Grid HatTheta
6     0.1661
7     0.1740
8     0.1810
9     0.1871
10    0.1923
11    0.1967
12    0.2004
13    0.2033
14    0.2055
15    0.2071
16    0.2081
17    0.2086
18    0.2086
19    0.2081
20    0.2072
21    0.2059
22    0.2043
23    0.2024
24    0.2003
25    0.1981
26    0.1956
27    0.1931
28    0.1905
29    0.1880
30    0.1854
31    0.1829
32    0.1806
33    0.1785
34    0.1765
35    0.1748
36    0.1736
37    0.1737
38    0.1759
39    0.1814
40    0.1910
41    0.1980
42    0.1972
43    0.1924
44    0.1871
45    0.1843
46    0.1839
47    0.1853
48    0.1879
49    0.1908
50    0.1935
51    0.1953
52    0.1963
53    0.1965
54    0.1960
55    0.1949
56    0.1933
57    0.1913
58    0.1889
59    0.1862
60    0.1834
61    0.1805
62    0.1776
63    0.1748
64    0.1722
65    0.1698
66    0.1679
67    0.1663
68    0.1653
69    0.1649
70    0.1652
71    0.1664
72    0.1684
73    0.1713
74    0.1754
75    0.1805
76    0.1869
77    0.1947
78    0.2038
79    0.2145
80    0.2267
81    0.2406
82    0.2562
83    0.2737
84    0.2932
85    0.3146
86    0.3382
87    0.3640
88    0.3921
89    0.4225
90    0.4554
}\Woman

\pgfplotstableread{
Grid HatTheta
3     0.3692
4     0.3647
5     0.3607
6     0.3572
7     0.3541
8     0.3514
9     0.3491
10    0.3471
11    0.3455
12    0.3442
13    0.3431
14    0.3424
15    0.3418
16    0.3414
17    0.3413
18    0.3412
19    0.3413
20    0.3415
21    0.3418
22    0.3421
23    0.3425
24    0.3429
25    0.3432
26    0.3435
27    0.3437
28    0.3438
29    0.3438
30    0.3437
31    0.3433
32    0.3428
33    0.3420
34    0.3410
35    0.3397
36    0.3382
37    0.3362
38    0.3340
39    0.3313
40    0.3283
41    0.3269
42    0.3281
43    0.3307
44    0.3331
45    0.3341
46    0.3326
47    0.3303
48    0.3290
49    0.3301
50    0.3328
51    0.3357
52    0.3385
53    0.3413
54    0.3441
55    0.3468
56    0.3494
57    0.3520
58    0.3545
59    0.3569
60    0.3593
61    0.3616
62    0.3639
63    0.3661
64    0.3681
65    0.3702
66    0.3721
67    0.3739
68    0.3757
69    0.3774
70    0.3789
71    0.3804
72    0.3818
73    0.3831
74    0.3843
75    0.3854
76    0.3864
77    0.3872
78    0.3880
79    0.3886
80    0.3892
81    0.3896
82    0.3899
83    0.3901
84    0.3901
85    0.3900
86    0.3898
87    0.3894
88    0.3890
89    0.3883
90    0.3876
}\Man

{
\newlength\newfigurewidth
\setlength\newfigurewidth{0.35\textwidth}
\begin{figure}[!h]
\centering\scriptsize
\begin{tikzpicture} 
\begin{groupplot}[group style={group name=myplots,group size=3 by 1,horizontal sep= 0.8cm,vertical sep=0.75cm},
    grid = minor,
    enlargelimits=false,
    width = 0.375\textwidth,
    xmin=0,xmax=93,
    ymin=0.1,ymax=0.475,
    xlabel=$z$,
    x label style={at={(axis description cs:0.95,0.04)},anchor=south},
    xtick={0,15,...,90},
    every axis title/.style={below,at={(0.2,0.8)}},
    title style={at={(xticklabel cs:0.5)}, below=1ex, text width=\newfigurewidth},
]
\nextgroupplot[title={\subcaption{Full Sample}}]
\addplot[smooth,color=RoyalBlue1] table[x = Grid,y=HatTheta] from \Full;
\nextgroupplot[title={\subcaption{Woman}}]
\addplot[smooth,color=RoyalBlue1] table[x = Grid,y=HatTheta] from \Woman;
\nextgroupplot[title={\subcaption{Man}}]
\addplot[smooth,color=RoyalBlue1] table[x = Grid,y=HatTheta] from \Man;
\end{groupplot}
\end{tikzpicture}
\caption{The curves of $\hat\theta_n$ based on the full sample, women, and men, respectively.} \label{Fig: Empirical}
\end{figure}
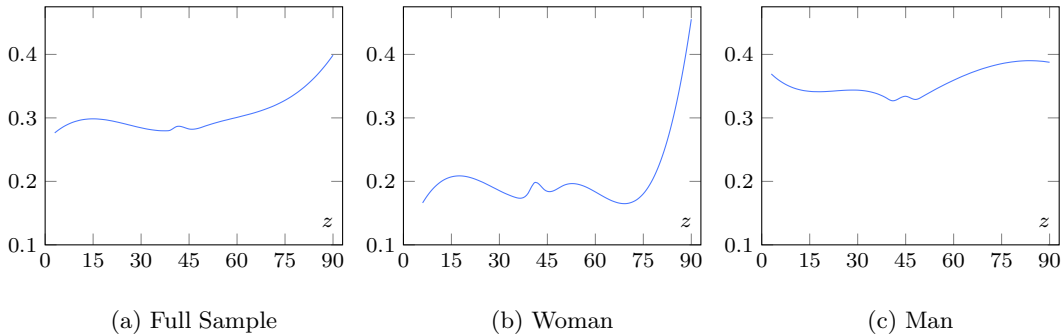
}

Table \ref{Tab: application} reports the $p$-values of our tests based on various choices of $k_n$ (the sieve dimension) and $\gamma_n$ (the tuning parameter that determines $\hat\kappa_n$). Consistent with \citet{Gicheva2013Working}, the shape restrictions are significant in the full sample and in men, at all three conventional significance levels. The evidence is, however, less strong among women. Inspecting Figure \ref{Fig: Empirical}-(b), we see that there are two segments of the working hours, namely $[18,36]$ and $[53,69]$, over which the wage growth rate declines. Following \citet{Gicheva2013Working}, we also re-compute the $p$-values by restricting to observations with working hours $Z\in[35,65]$. This eliminates 231 observations in total, 128 women, and 103 men. The presence of the shape restrictions remains strong, but more so for women. Overall, our findings are in line with the theoretical predictions in \citet{Gicheva2013Working}.

{
\setlength{\tabcolsep}{3.5pt}
\renewcommand{\arraystretch}{1.1}
\begin{table}[!ht]
\caption{Testing Shape Restrictions in Labor Supply: $p$-Values} \label{Tab: application}
\centering\footnotesize
\sisetup{table-number-alignment = center, table-format = 1.3} 
\begin{tabularx}{\linewidth}{@{} cc *{3}{S[round-mode = places,round-precision = 3]}  c *{3}{S[round-mode = places,round-precision = 3]}  c *{3}{S[round-mode = places,round-precision = 3]}@{}} 
\hline
\hline
\multirow{2}{*}{Sample} &  \multirow{2}{*}{Shape} & \multicolumn{3}{c}{$\gamma_n=0.01$} & & \multicolumn{3}{c}{$\gamma_n=0.01/\log n$} & & \multicolumn{3}{c}{$\gamma_n=1/n$}\\
\cline{3-5} \cline{7-9} \cline{11-13}
& & {$k_n=7$} & {$k_n=9$} & {$k_n=11$}  & & {$k_n=7$} & {$k_n=9$} & {$k_n=11$} & & {$k_n=7$} & {$k_n=9$} & {$k_n=11$}\\
\hline
\rule{0pt}{15pt}
& &   \multicolumn{11}{c}{Untrimmed Samples}  \\
\multirow{3}{*}{Full} & Mon          & 0.7170 & 0.7244 & 0.6820 & & 0.7224 & 0.7308 & 0.6886 & & 0.7256 & 0.7324 & 0.6910\\
                      & Con          & 0.5876 & 0.5668 & 0.8168 & & 0.5906 & 0.5726 & 0.8210 & & 0.5922 & 0.5742 & 0.8224\\
                      & Mon-Con      & 0.5908 & 0.5702 & 0.8266 & & 0.5938 & 0.5752 & 0.8286 & & 0.5958 & 0.5778 & 0.8298\\
\rule{0pt}{12pt}
\multirow{3}{*}{Woman}& Mon          & 0.4024 & 0.1340 & 0.1122 & & 0.4186 & 0.1424 & 0.1198 & & 0.4232 & 0.1430 & 0.1206\\
                      & Con          & 0.1074 & 0.0708 & 0.2366 & & 0.1114 & 0.0756 & 0.2526 & & 0.1128 & 0.0760 & 0.2532\\
                      & Mon-Con      & 0.1098 & 0.0752 & 0.2652 & & 0.1154 & 0.0802 & 0.2750 & & 0.1168 & 0.0814 & 0.2762\\
\rule{0pt}{12pt}
\multirow{3}{*}{Man}  & Mon          & 0.5952 & 0.3258 & 0.3594 & & 0.6024 & 0.3326 & 0.3640 & & 0.6040 & 0.3332 & 0.3646\\
                      & Con          & 0.3286 & 0.7138 & 0.9024 & & 0.3310 & 0.7196 & 0.9050 & & 0.3326 & 0.7212 & 0.9054\\
                      & Mon-Con      & 0.3364 & 0.7406 & 0.9146 & & 0.3384 & 0.7416 & 0.9146 & & 0.3392 & 0.7418 & 0.9148\\
\rule{0pt}{15pt}
                      & &   \multicolumn{11}{c}{Trimmed Samples}  \\
\multirow{3}{*}{Full} & Mon          & 0.9974 & 0.5486 & 0.6486 & & 0.9974 & 0.5584 & 0.6558 & & 0.9974 & 0.5596 & 0.6572\\
                      & Con          & 0.8946 & 0.2674 & 0.6124 & & 0.8962 & 0.2692 & 0.6154 & & 0.8970 & 0.2694 & 0.6160\\
                      & Mon-Con      & 0.8950 & 0.2738 & 0.6244 & & 0.8964 & 0.2752 & 0.6260 & & 0.8972 & 0.2754 & 0.6260\\
\rule{0pt}{12pt}
\multirow{3}{*}{Woman}& Mon          & 0.3248 & 0.5694 & 0.6596 & & 0.3294 & 0.5738 & 0.6672 & & 0.3294 & 0.5742 & 0.6672\\
                      & Con          & 0.1526 & 0.3742 & 0.4048 & & 0.1528 & 0.3742 & 0.4052 & & 0.1528 & 0.3742 & 0.4052\\
                      & Mon-Con      & 0.1528 & 0.3742 & 0.4060 & & 0.1528 & 0.3742 & 0.4062 & & 0.1528 & 0.3742 & 0.4062\\
\rule{0pt}{12pt}
\multirow{3}{*}{Man}  & Mon          & 0.6432 & 0.5348 & 0.7220 & & 0.6528 & 0.5430 & 0.7294 & & 0.6528 & 0.5430 & 0.7298\\
                      & Con          & 0.9740 & 0.9004 & 0.9850 & & 0.9754 & 0.9040 & 0.9854 & & 0.9754 & 0.9042 & 0.9854\\
                      & Mon-Con      & 0.9764 & 0.9098 & 0.9864 & & 0.9776 & 0.9120 & 0.9866 & & 0.9776 & 0.9120 & 0.9866\\
\hline
\hline
\end{tabularx}
\end{table}
}

\section{Conclusion}\label{Sec: conclusion}

This paper develops a unifying framework for testing shape restrictions, and investigates the suitability of some Wald functionals in applying the framework.
In particular, while the influential rearrangement operator has proven particularly useful in obtaining restricted point and interval estimates, it is inapplicable to our framework due to a lack of convexity. In contrast, the greatest convex minorization (resp.\ the least concave majorization) is shown to enjoy attractive analytic properties, and thus may be employed to test convexity (resp.\ concavity) in nonparametric settings that have not been explored previously. Finally, despite that the projection operator may not be well-defined/behaved in general Banach spaces, we show that one may nonetheless devise a powerful distance-based test by applying our framework.

\clearpage\newpage
\begin{appendices}
\titleformat{\section}{\Large\center}{{\sc Appendix} \thesection}{1em}{}
\setcounter{section}{0}
\renewcommand \thesection{\Alph{section}}
\numberwithin{equation}{section}
\numberwithin{ass}{section}
\numberwithin{figure}{section}
\numberwithin{table}{section}

\section{Proofs}

For ease of reference, we collect some notation in the table below.

{\renewcommand{\arraystretch}{1}
\begin{table}[h]
\begin{center}
\begin{tabularx}{\textwidth}{cX}
\hline\hline
$ a \lesssim b$                 & For some constant $M$ that is universal in the proof, $a\leq Mb$.\\
$\bar{\mathbf R}$               & The extended real number system, i.e., $\bar{\mathbf R}=\mathbf R\cup\{-\infty,\infty\}$.\\
$\|\theta\|_p$                  & For a function $\theta:\mathcal Z\to\mathbf R$ and $p\in[1,\infty)$, $\|\theta\|_p=\{\int_{\mathcal Z}|\theta(z)|^p\mathrm dz\}^{1/p}$.\\
$\|\theta\|_\infty$             & For a function $\theta:\mathcal Z\to\mathbf R$, $\|\theta\|_\infty=\sup_{z\in\mathcal Z}|\theta(z)|$.\\
$L^p(\mathcal Z)$               & For $p\in[1,\infty)$, $L^p(\mathcal Z) =\{\theta:\mathcal Z\to\mathbf R: \|\theta\|_p<\infty\}$.\\
$\ell^\infty(\mathcal Z)$       & The space of bounded functions, i.e., $\ell^\infty(\mathcal Z)=\{\theta:\mathcal Z\to\mathbf R: \|\theta\|_\infty<\infty\}$.\\
$\theta^\star$                  & The convex conjugate of $\theta:\mathcal Z\to\mathbf R$ (defined by \eqref{Eqn: conjugate} in the main text).\\
$\theta^{\star\star}$           & The biconjugate of $\theta:\mathcal Z\to\mathbf R$, i.e., $\theta^{\star\star}=(\theta^\star)^\star$.\\
$\mathbf B^*$                   & The space of linear continuous functions $f:\mathbf B\to\mathbf R$ for a normed space $\mathbf B$.\\
$\langle b^*,b\rangle$          & For $b\in\mathbf B$ (a normed space) and $b^*\in\mathbf B^*$ (the dual), $\langle b^*,b\rangle=b^*(b)$. \\
$\|\cdot\|_{\mathbf B^*}$       & For the dual space $\mathbf B^*$ of $\mathbf B$  and $b^*\in\mathbf B^*$, $\|b^*\|_{\mathbf B^*}=\sup_{b\in\mathbf B:\|b\|_{\mathbf B}\le 1}|\langle b^*,b\rangle|$.\\
\hline\hline
\end{tabularx}
\end{center}
\end{table}}

The following lemma plays a fundamental role in the proofs below.

\begin{lem}\label{Lem: phi representation}
If a lower-semicontinuous map $\phi: \mathbf B\to\mathbf R$ on a Banach space $\mathbf B$ satisfies Assumption \ref{Ass: space and map}(i)(ii), then it follows that, for all $\theta\in\mathbf B$,
\begin{align}
\phi(\theta)=\sup_{b^*\in \Lambda_\phi^\circ} \langle b^*, \theta\rangle~,
\end{align}
where $\Lambda_\phi^\circ\equiv\{b^*\in\mathbf B^*: \langle b^*,\theta\rangle\le \phi(\theta)\text{ for all }\theta\in\mathbf B\}$.
\end{lem}
\noindent{\sc Proof:} Let $\phi^*: \mathbf B^*\to\bar{\mathbf R}$ be the convex conjugate of $\phi$, i.e., for any $b^*\in\mathbf B^*$,
\begin{align}\label{Eqn: phi representation, aux1}
\phi^*(b^*)=\sup_{\theta\in\mathbf B}\{\langle b^*,\theta\rangle -\phi(\theta)\}~.
\end{align}
By Assumption \ref{Ass: space and map}(i), we must have $\phi(0) = 0$ and so, for any $b^*\in\mathbf B^*$,
\begin{align}\label{Eqn: phi representation, aux2}
\phi^*(b^*)=\sup_{\theta\in\mathbf B}\{\langle b^*,\theta\rangle-\phi(\theta)\}\ge  \langle b^*,0\rangle-\phi(0)=0~.
\end{align}
Moreover, Assumption \ref{Ass: space and map}(i) also implies that, for any $t>0$,
\begin{align}\label{Eqn: phi representation, aux3}
t\phi^*(b^*) = \sup_{\theta\in\mathbf B}\{\langle b^*,t\theta\rangle-\phi(t\theta)\} = \sup_{\vartheta\in\mathbf B}\{\langle b^*,\vartheta\rangle-\phi(\vartheta)\} = \phi^*(b^*)~.
\end{align}
We thus deduce from results \eqref{Eqn: phi representation, aux2} and \eqref{Eqn: phi representation, aux3} that $\phi^*(b^*)\in\{0,\infty\}$ for any $b^*\in\mathbf B^*$. By Assumptions \ref{Ass: space and map}(ii) and $\phi$ being lower-semicontinuous, we may invoke the Fenchel-Moreau theorem (see, e.g., Theorem 51.A in \citet{Zeidler1990III}) to conclude
\begin{align}\label{Eqn: phi representation, aux4}
\phi(\theta)= \sup_{b^*\in\mathbf B^*}\{\langle b^*,\theta\rangle-\phi^*(b^*)\} = \sup_{b^*\in\mathbf B^*: \phi^*(b^*)=0}\langle b^*,\theta\rangle ~,
\end{align}
for any $\theta\in\mathbf B$, where the second equality also exploits $\langle b^*,\theta\rangle\in\mathbf R$ for all $b^*\in\mathbf B^*$. By result \eqref{Eqn: phi representation, aux1} and $\phi(0)=0$, the restriction $\phi^*(b^*)=0$ is equivalent to, for all $\theta\in\mathbf B$,
\begin{align}\label{Eqn: phi representation, aux5}
\langle b^*,\theta\rangle\le \phi(\theta)~,
\end{align}
where the equality holds if $\theta=0$. The lemma now follows from \eqref{Eqn: phi representation, aux4} and \eqref{Eqn: phi representation, aux5}. \qed

\noindent{\sc Proof of Theorem \ref{Thm: size and power}:} We proceed by establishing the monotonicity of the map $a\mapsto\phi(h+a\theta_0)$ on $[0,\infty)$ for any $h\in\mathbf B$ and $\theta_0\in\Lambda$. To this end, fix $h\in\mathbf B$ and $\theta_0\in\Lambda$. By Lemma \ref{Lem: phi representation}, we obtain that, for $\Lambda_\phi^\circ\equiv\{b^*\in\mathbf B^*: \langle b^*,\theta\rangle\le \phi(\theta)\text{ for all }\theta\in\mathbf B\}$,
\begin{align}\label{Eqn: monotonicity, aux}
\phi(h+a\theta_0) = \sup_{b^*\in\Lambda_\phi^\circ} \langle b^*,h+a\theta_0\rangle =  \sup_{b^*\in\Lambda_\phi^\circ} \{\langle b^*,h\rangle + a\langle b^*,\theta_0\rangle\}~.
\end{align}
Since $\phi(\theta_0)=0$ by the definition of $\Lambda$, we have $\langle b^*,\theta_0\rangle\le 0$ for all $b^*\in\Lambda_\phi^\circ$. The conclusion of the lemma then immediately follows from the representation \eqref{Eqn: monotonicity, aux}.

Next, for notational simplicity, let $\mathbb G_{n,P}\equiv r_n\{\hat\theta_n-\theta_P\}$. By Assumptions \ref{Ass: space and map}(i)(iii) and \ref{Ass: strong approx}(i), we obtain that, uniformly in $P\in\mathbf P$,
\begin{align}\label{Eqn: size, aux1}
r_n\phi(\hat\theta_n)=\phi(\mathbb G_{n,P}+r_n\theta_P)=\phi(\mathbb Z_{n,P}+r_n\theta_P)+o_p(c_n)~.
\end{align}
By Assumptions \ref{Ass: space and map}(iii) and \ref{Ass: enforce null} and the triangle inequality, we have
\begin{multline}\label{Eqn: size, aux2}
|\phi(\hat{\mathbb G}_n+\kappa_n\Gamma(\hat\theta_n))- \phi(\hat{\mathbb G}_n+\kappa_n\theta_P)|
\lesssim \kappa_n\|\Gamma(\hat\theta_n)-\Gamma(\theta_P)\|_{\mathbf B} \\
\lesssim  \| \kappa_n \{\hat\theta_n-\theta_P\}\|_{\mathbf B}\le \frac{\kappa_n}{r_n}\{\|\mathbb G_{n,P}-\mathbb Z_{n,P}\|_{\mathbf B} + \|\mathbb Z_{n,P}\|_{\mathbf B}\} = o_p(c_n)~,
\end{multline}
uniformly in $P\in\mathbf P_0$, where the final step is due to Assumptions \ref{Ass: strong approx}(i) and \ref{Ass: critical value}(ii), Markov's inequality, $\zeta_n\ge 1$ and $\kappa_n\zeta_n/r_n=o(c_n)$ by hypothesis. By Assumptions \ref{Ass: space and map}(iii) and \ref{Ass: strong approx}(ii), we also have: uniformly in $P\in\mathbf P$,
\begin{align}\label{Eqn: size, aux3}
| \phi(\hat{\mathbb G}_n+\kappa_n\theta_P)-\phi(\bar{\mathbb Z}_{n,P}+\kappa_n\theta_P)|\le \|\hat{\mathbb G}_n- \bar{\mathbb Z}_{n,P}\|_{\mathbf B}=o_p(c_n)~.
\end{align}
It follows from \eqref{Eqn: size, aux2}, \eqref{Eqn: size, aux3} and the triangle inequality that, uniformly in $P\in\mathbf P_0$,
\begin{align}\label{Eqn: size, aux4}
\phi(\hat{\mathbb G}_n+\kappa_n\Gamma(\hat\theta_n)) = \phi(\bar{\mathbb Z}_{n,P}+\kappa_n\theta_P) +o_p(c_n)~.
\end{align}

Given results \eqref{Eqn: size, aux1} and \eqref{Eqn: size, aux4}, we may select a sequence $\{\epsilon_n\}$ of scalars such that $\epsilon_n=o(c_n)$ such that, as $n\to\infty$,
\begin{gather}
\sup_{P\in\mathbf P_0} P(|r_n\phi(\hat\theta_n)-\phi(\mathbb Z_{n,P}+r_n\theta_P)|>\epsilon_n)=o(1)~, \label{Eqn: size, aux5}\\
\sup_{P\in\mathbf P_0} P(|\phi(\hat{\mathbb G}_n+\kappa_n\Gamma(\hat\theta_n)) - \phi(\bar{\mathbb Z}_{n,P}+\kappa_n\theta_P)|>\epsilon_n)=o(1)~.\label{Eqn: size, aux6}
\end{gather}
By Markov's inequality (see, e.g., Lemma 6.10 in \citet{Kosorok2008}), Lemma 1.2.6 in \citet{Vaart1996} and \eqref{Eqn: size, aux6}, we have: for each $\eta>0$,
\begin{multline}\label{Eqn: size, aux7}
\sup_{P\in\mathbf P_0} P(P(|\phi(\hat{\mathbb G}_n+\kappa_n\Gamma(\hat\theta_n)) - \phi(\bar{\mathbb Z}_{n,P}+\kappa_n\theta_P)|>\epsilon_n|\{X_i\}_{i=1}^n)>\eta)\\
\le \sup_{P\in\mathbf P_0}\frac{1}{\eta} P(|\phi(\hat{\mathbb G}_n+\kappa_n\Gamma(\hat\theta_n)) - \phi(\bar{\mathbb Z}_{n,P}+\kappa_n\theta_P)|>\epsilon_n)=o(1)~.
\end{multline}
Thus, we may select positive scalars $\eta_n\downarrow 0$ such that: uniformly in $P\in\mathbf P_0$,
\begin{align}\label{Eqn: size, aux8}
P(|\phi(\hat{\mathbb G}_n+\kappa_n\Gamma(\hat\theta_n)) - \phi(\bar{\mathbb Z}_{n,P}+\kappa_n\theta_P)|>\epsilon_n|\{X_i\}_{i=1}^n)=o_p(\eta_n)~.
\end{align}
Since $\bar{\mathbb Z}_{n,P}$ is independent of $\{X_i\}_{i=1}^n$ by Assumption \ref{Ass: strong approx}(ii), the conditional cdf of $\phi(\bar{\mathbb Z}_{n,P}+\kappa_n\theta_P)$ given $\{X_i\}_{i=1}^n$ is precisely its unconditional analog. Thus, we may conclude by Lemma 11 in \citet{ChernozhukovLeeRosen2013Intersection} and result \eqref{Eqn: size, aux8} that
\begin{align}\label{Eqn: size, aux9}
\liminf_{n\to\infty}\inf_{P\in\mathbf P_0} P(\hat c_{n,1-\alpha}+\epsilon_n\ge c_{n,P}(1-\alpha-\eta_n))=1~.
\end{align}

By results \eqref{Eqn: size, aux5} and \eqref{Eqn: size, aux9}, we have:
\begin{align}\label{Eqn: size, aux15}
\limsup_{n\to\infty}\sup_{P\in\mathbf P_0}& P(r_n\phi(\hat\theta_n)>\hat c_{n,1-\alpha})\notag\\
&\le \limsup_{n\to\infty} \sup_{P\in\mathbf P_0}P(r_n\phi(\hat\theta_n)>\hat c_{n,1-\alpha},|r_n\phi(\hat\theta_n)-\phi(\mathbb Z_{n,P}+r_n\theta_P)|\le \epsilon_n)\notag\\
&\le \limsup_{n\to\infty}\sup_{P\in\mathbf P_0}P(\phi(\mathbb Z_{n,P}+r_n\theta_P)>c_{n,P}(1-\alpha-\eta_n)-2\epsilon_n)\notag\\
&\le \limsup_{n\to\infty}\sup_{P\in\mathbf P_0}P(\phi(\mathbb Z_{n,P}+\kappa_n\theta_P)>c_{n,P}(1-\alpha-\eta_n)-2\epsilon_n)~,
\end{align}
where the final step follows by the monotonicity established in the beginning and the fact $0\le\kappa_n\le r_n$ for all large $n$ (due to $\kappa_n/r_n=o(c_n/\zeta_n)$, $\zeta_n\ge 1$ and $c_n=O(1)$). In turn, $\epsilon_n=o(c_n)$ (by construction), Assumption \ref{Ass: critical value}(iii), Proposition \ref{Pro: anti concentration}, result \eqref{Eqn: size, aux15} and $\eta_n=o(1)$ imply that
\begin{multline}\label{Eqn: size, aux18}
\limsup_{n\to\infty}\sup_{P\in\mathbf P_0} P(r_n\phi(\hat\theta_n)>\hat c_{n,1-\alpha})\le \limsup_{n\to\infty}\sup_{P\in\mathbf P_0}P(\phi(\mathbb Z_{n,P}+\kappa_n\theta_P)>c_{n,P}(1-\alpha-\eta_n))\\
\le\limsup_{n\to\infty}\sup_{P\in\mathbf P_0}\{\alpha+\eta_n\}=\alpha~,
\end{multline}
as desired for the first claim of part (i). For the second claim, we note that
\begin{align}
\phi(\mathbb Z_{n,P}+r_n\theta_P) = \phi(\mathbb Z_{n,P}+\kappa_n\theta_P) =\phi(\mathbb Z_{n,P})
\end{align}
for all $P\in\bar{\mathbf P}_0$, by Assumption \ref{Ass: space and map}(i) and the definition of $\bar{\mathbf P}_0$. The remaining arguments proceed as in \citet{FangSeo2019Shape}, and we thus omit the details for brevity.

For part (ii), by Assumption \ref{Ass: enforce null}(i) and the established monotonicity, we have
\begin{multline}\label{Eqn: power, aux1}
\phi(\hat{\mathbb G}_n+\kappa_n\Gamma(\hat\theta_n)) \le  \phi(\hat{\mathbb G}_n)  = \phi(\hat{\mathbb G}_n) -\phi(0)\\
\lesssim \|\hat{\mathbb G}_n\|_{\mathbf B}\le \|\hat{\mathbb G}_n-\bar{\mathbb Z}_{n,P}\|_{\mathbf B}+\|\bar{\mathbb Z}_{n,P}\|_{\mathbf B}~,
\end{multline}
where the equality is by $\phi(0)=0$ (due to Assumption \ref{Ass: space and map}(i)), the second inequality is due to Assumption \ref{Ass: space and map}(iii), and the final step is by the triangle inequality. By Assumptions \ref{Ass: strong approx} and \ref{Ass: critical value}(ii) and Markov's inequality, we in turn have from \eqref{Eqn: power, aux1} that
\begin{align}\label{Eqn: power, aux2}
\phi(\hat{\mathbb G}_n+\kappa_n\Gamma(\hat\theta_n)) =o_p(c_n)+O_p(\zeta_n)=O_p(\zeta_n)~,
\end{align}
uniformly in $P\in\mathbf P$. By the definition of $\hat c_{n,1-\alpha}$, Markov's inequality and Lemma 1.2.6 in \citet{Vaart1996}, we note that, for any $M>0$,
\begin{multline}\label{Eqn: power, aux3}
P(\hat c_{n,1-\alpha}>M\zeta_n)\le P(P(\phi(\hat{\mathbb G}_n+\kappa_n\Gamma(\hat\theta_n))>M\zeta_n|\{X_i\}_{i=1}^n)>\alpha)\\
\le\frac{1}{\alpha} P(\phi(\hat{\mathbb G}_n+\kappa_n\Gamma(\hat\theta_n))>M\zeta_n) ~.
\end{multline}
Hence, results \eqref{Eqn: power, aux2} and \eqref{Eqn: power, aux3} yield $\hat c_{n,1-\alpha}=O_p(\zeta_n)$ uniformly in $P\in\mathbf P$.

Next, Assumption \ref{Ass: space and map}(iii) and the triangle inequality imply: uniformly in $P\in\mathbf P$,
\begin{multline}\label{Eqn: power, aux4}
|r_n\phi(\hat\theta_n)-r_n\phi(\theta_P)|\lesssim \|r_n\{\hat\theta_n-\theta_P\}\|_{\mathbf B}\\
\le \|r_n\{\hat\theta_n-\theta_P\}-\mathbb Z_{n,P}\|_{\mathbf B}+ \|\mathbb Z_{n,P}\|_{\mathbf B}\le o_p(c_n)+O_p(\zeta_n)=O_p(\zeta_n)~,
\end{multline}
where the third inequality follows by Assumptions \ref{Ass: strong approx}(i) and \ref{Ass: critical value}(ii), and the last step is due to $c_n=O(1)$ and $\zeta_n\ge 1$. Result \eqref{Eqn: power, aux4} and the definition of $\mathbf P_{1,n}^\Delta$ thus imply
\begin{align}\label{Eqn: power, aux5}
r_n\phi(\hat\theta_n)=r_n\phi(\theta_P)+r_n\phi(\hat\theta_n)-r_n\phi(\theta_P)\ge \Delta\zeta_n+O_p(\zeta_n)~,
\end{align}
uniformly in $P\in\mathbf P_{1,n}^\Delta$. The theorem thus follows from combining result \eqref{Eqn: power, aux5} and the order $\hat c_{n,1-\alpha}=O_p(\zeta_n)$ (uniformly in $P\in\mathbf P$) that we have established .\qed

\noindent{\sc Proof of Proposition \ref{Pro: tuning parameter}:} The proof is similar to the proof of Proposition 3.1 in \citet{FangSeo2019Shape} by working with $\|\hat{\mathbb G}_n/\zeta_n\|_{\mathbf B}$. We thus omit the details for brevity. \qed

\noindent{\sc Proof of Theorem \ref{Thm: Greatest convex minorant}:} By Proposition 13.23(i)(ii) in \citet{BauschkeCombettes2017Convex}, we have that, for any $a>0$ and $\theta\in\mathbf B\equiv\ell^\infty(\mathcal Z)$,
\begin{align}\label{Eqn: Greatest convex minorant, aux1}
\phi(a\theta) = \|a\theta-(a\theta)^{\star\star}\|_p = \|a\theta-(a\theta^\star(\cdot/a))^\star\|_p = \|a\theta-a\theta^{\star\star}\|_p=a\phi(\theta)~.
\end{align}
For $a=0$, note that $\phi(0)=\phi(2\cdot 0)=2\phi(0)$ by \eqref{Eqn: Greatest convex minorant, aux1} and hence
\begin{align}\label{Eqn: Greatest convex minorant, aux2}
\phi(0\cdot\theta) = \phi(0) = 0 = 0\phi(\theta)~.
\end{align}
Together, results \eqref{Eqn: Greatest convex minorant, aux1} and \eqref{Eqn: Greatest convex minorant, aux2} imply positive homogeneity of $\phi$.

Next, since any $\theta\in\mathbf B$ is real-valued and $\mathcal Z$ is nonempty, $\theta^\star$ and hence $\theta^{\star\star}$ do not attain $-\infty$. Fix any $a\in(0,1)$ and $\theta_1,\theta_2\in\mathbf B$. By Proposition 51.6(3) in \citet{Zeidler1990III} (the proposition also holds for maps defined on a subset of the entire space by inspecting the proof there), $a\theta_1^{\star\star}+(1-a)\theta_2^{\star\star}$ is convex and lower semicontinuous such that
\begin{align}\label{Eqn: Greatest convex minorant, aux3}
a\theta_1+(1-a)\theta_2\ge a\theta_1^{\star\star}+(1-a)\theta_2^{\star\star}~.
\end{align}
By Proposition 51.6(5) in \citet{Zeidler1990III}, it follows from \eqref{Eqn: Greatest convex minorant, aux3} that
\begin{align}\label{Eqn: Greatest convex minorant, aux4}
(a\theta_1+(1-a)\theta_2)^{\star\star}\ge a\theta_1^{\star\star}+(1-a)\theta_2^{\star\star}~.
\end{align}
For notational simplicity, set $\mathbf D = L^p(\mathcal Z)$ if $p\in[1,\infty)$ and $\mathbf D = \ell^\infty(\mathcal Z)$ if $p=\infty$. Clearly, $\mathbf B\subset\mathbf D$ since $\mathcal Z$ is bounded. Moreover, $\theta\ge 0$ with $\theta\in\mathbf B$ means $\theta(z)\ge 0$ for all $z\in\mathcal Z$. By Lemma A.1 in \citet{Fang2019KW} and the fact $\theta-\theta^{\star\star}\ge 0$ for any $\theta\in\mathbf B$ (see, e.g., Proposition 51.6(3) in \citet{Zeidler1990III}), we obtain that, for all $\theta\in\mathbf B$,
\begin{align}\label{Eqn: Greatest convex minorant, aux5}
\phi(\theta) = \sup_{f\in\mathbf D_1^*: f\ge 0} f(\theta-\theta^{\star\star}) = \sup_{f\in\mathbf D_1^*: f\ge 0} \{f(\theta)-f(\theta^{\star\star})\}~,
\end{align}
where $f\ge 0$ means $f(\vartheta)\ge 0$ whenever $\vartheta\ge 0$, and $\mathbf D_1^*\equiv\{h\in\mathbf D^*: \|h\|_{\mathbf D^*}= 1\}$ is the unit sphere in the topological dual $\mathbf D^*$ of $\mathbf D$. Result \eqref{Eqn: Greatest convex minorant, aux5} then implies
\begin{align}\label{Eqn: Greatest convex minorant, aux6}
\phi(a\theta_1+(1-a)\theta_2)& = \sup_{f\in\mathbf D_1^*: f\ge 0} \{f(a\theta_1+(1-a)\theta_2)-f(\{a\theta_1+(1-a)\theta_2\}^{\star\star})\}\notag\\
& \le \sup_{f\in\mathbf D_1^*: f\ge 0} \{f(a\theta_1+(1-a)\theta_2)-f(a\theta_1^{\star\star}+(1-a)\theta_2^{\star\star})\}\notag\\
&\le a\sup_{f\in\mathbf D_1^*: f\ge 0} \{f(\theta_1)-f(\theta_1^{\star\star})\} + (1-a)\sup_{f\in\mathbf D_1^*: f\ge 0} \{f(\theta_2)-f(\theta_2^{\star\star})\}\notag\\
& = a\phi(\theta_1)+(1-a)\phi(\theta_2)~,
\end{align}
where the first inequality follows by $f\ge 0$ and \eqref{Eqn: Greatest convex minorant, aux4}, the second inequality by the linearity of $f\in\mathbf D_1^*$ and the subadditivity of the supremum operator, and the last step by \eqref{Eqn: Greatest convex minorant, aux5}. It follows from result \eqref{Eqn: Greatest convex minorant, aux6} that $\phi$ is convex.

Finally, for any $\theta_1,\theta_2\in\mathbf B$, we have by the definition of conjugate that
\begin{align}\label{Eqn: Greatest convex minorant, aux8}
\|\theta_1^\star-\theta_2^\star\|_\infty=\sup_{y\in\mathcal Z^\star}|\theta_1^\star(y)-\theta_2^\star(y)|\le \sup_{z\in\mathcal Z}|\theta_1(z)-\theta_2(z)|=\|\theta_1-\theta_2\|_\infty~.
\end{align}
Result \eqref{Eqn: Greatest convex minorant, aux8} in turn implies that
\begin{align}\label{Eqn: Greatest convex minorant, aux9}
\|\theta_1^{\star\star}-\theta_2^{\star\star}\|_\infty\le\|\theta_1^\star-\theta_2^\star\|_\infty\le \|\theta_1-\theta_2\|_\infty~.
\end{align}
Lipschitz continuity of $\phi$ then readily follows from \eqref{Eqn: Greatest convex minorant, aux9}. \qed

\begin{pro}\label{Pro: anti concentration}
Let Assumptions \ref{Ass: space and map} and \ref{Ass: critical value}(i)(ii)(iii)(iv) hold. If $\{\epsilon_n\}$ is a sequence of scalars satisfying $\epsilon_n=o(c_n)$, then
\begin{align}\label{Eqn: anti concentration, aux}
\limsup_{n\to\infty}\sup_{P\in\mathbf P_0}\sup_{x\in [c_{n,P}(0.5)+\varsigma_n,\infty)} P(|\phi(\mathbb Z_{n,P}+\kappa_n\theta_P)-x|\le \epsilon_n)=0~.
\end{align}
\end{pro}
\noindent{\sc Proof:} By Lemma \ref{Lem: phi representation}, we may rewrite: for $e_{b^*}(n,P)\equiv \kappa_n \langle b^*,\theta_P\rangle$,
\begin{align}\label{Eqn: anti concentration, aux1}
\phi(\mathbb Z_{n,P}+\kappa_n\theta_P)=\sup_{b^*\in\Lambda_\phi^\circ}\{\langle b^*,\mathbb Z_{n,P}\rangle +e_{b^*}(n,P)\}~,
\end{align}
where $\Lambda_\phi^\circ\equiv\{b^*\in\mathbf B^*: \langle b^*,\theta\rangle\le \phi(\theta)\text{ for all }\theta\in\mathbf B\}$. Since $\phi(0)=0$ by Assumption \ref{Ass: space and map}(i), we have by Assumption \ref{Ass: space and map}(iii) that, for all $\theta\in\mathbf B$,
\begin{align}\label{Eqn: anti concentration, aux3}
\phi(\theta)=\phi(\theta)-\phi(0) \le C\|\theta\|_{\mathbf B}~,
\end{align}
where $C>0$ is an absolute constant throughout that may change at each appearance. Since $\phi(\mathbb Z_{n,P}+\kappa_n\theta_P)$ is nonnegative, the supremum in \eqref{Eqn: anti concentration, aux1} may be restricted to the set of $b^*\in\Lambda_\phi^\circ$ with $\langle b^*,\mathbb Z_{n,P}\rangle +e_{b^*}(n,P)\ge 0$ (which is nonempty since $0\in \Lambda_\phi^\circ$). For any such $b^*$, by the definition of $\Lambda_\phi^\circ$ and \eqref{Eqn: anti concentration, aux3}, we must have $e_{b^*}(n,P)\ge -CM\zeta_n$ whenever $\|\mathbb Z_{n,P}\|_{\mathbf B}\le M\zeta_n$. Define $\Lambda_{\phi,M}^\circ(n,P)\equiv\{b^*\in\Lambda_\phi^\circ: e_{b^*}(n,P)\ge -CM\zeta_n\}$. It follows that, whenever $\| \mathbb Z_{n,P}\|_{\mathbf B}\le M\zeta_n$,
\begin{align}\label{Eqn: anti concentration, aux4}
\phi(\mathbb Z_{n,P}+\kappa_n\theta_P) = \sup_{b^*\in \Lambda_{\phi,M}^\circ(n,P)}\{\langle b^*,\mathbb Z_{n,P}\rangle+e_{b^*}(n,P)\}~.
\end{align}
Fix a sequence $\{\epsilon_n\}$ of scalars such that $\epsilon_n=o(c_n)$. Without loss of generality, we may assume that $\{\epsilon_n\}$ are nonnegative. Then, for any $x\in\mathbf R$,
\begin{align}\label{Eqn: anti concentration, aux5}
P(|\phi(\mathbb Z_{n,P}&+\kappa_n\theta_P)-x|\le \epsilon_n)\notag\\
&\le P(|\sup_{b^*\in \Lambda_{\phi,M}^\circ(n,P)}\{\langle b^*,\mathbb Z_{n,P}\rangle+e_{b^*}(n,P)\}-x|\le \epsilon_n )+P(\|\mathbb Z_{n,P}\|_{\mathbf B}>M\zeta_n)\notag\\
&\le P(|\sup_{b^*\in \Lambda_{\phi,M}^\circ(n,P)}\{\langle b^*,\mathbb Z_{n,P}\rangle+e_{b^*}(n,P)\}-x|\le \epsilon_n )+\frac{1}{M}~,
\end{align}
where the second inequality is due to Markov's inequality and Assumption \ref{Ass: critical value}(ii).

Next, we proceed by following the steps in the proof of Proposition D.1 in \citet{FangSeo2019Shape}, but keep the arguments concise whenever appropriate for brevity. Let $F_{n,P,M}$ be the cdf of $T_{n,P}\equiv \sup_{b^*\in \Lambda_{\phi,M}^\circ(n,P)}\{\langle b^*,\mathbb Z_{n,P}\rangle+e_{b^*}(n,P)\}$, and $c_{n,P,M}(\tau)$ be its $\tau$-quantile with $m_{n,P,M}\equiv c_{n,P,M}(0.5)$. Fix $n$ and $P\in\mathbf P_0$. We claim that
\begin{align}\label{Eqn: anti concentration, aux6}
\bar\sigma_{n,P,M}^2\equiv\sup_{b^*\in \Lambda_{\phi,M}^\circ(n,P)} E[\langle b^*,\mathbb Z_{n,P}\rangle^2]>0~,
\end{align}
for all large $M>0$. To see this, we note that $\bar\sigma_{n,P}^2\equiv\sup_{t\in\Lambda_\phi^\circ} E[\langle b^*,\mathbb Z_{n,P}\rangle^2]>0$; otherwise $\phi(\mathbb Z_{n,P}+\kappa_n\theta_P)=0$ almost surely so that all quantiles of $\phi(\mathbb Z_{n,P}+\kappa_n\theta_P)$ collapse to zero, contradicting Assumption \ref{Ass: critical value}(iii) (see \citet{FangSeo2019Shape} for more formal arguments). Fix $\eta>0$. Then we may select some $b_{n,P}^*\in\Lambda_\phi^\circ$ such that
\begin{align}\label{Eqn: anti concentration, aux7}
\bar\sigma_{n,P}^2\le E[\langle b_{n,P}^*,\mathbb Z_{n,P}\rangle^2]+\eta~.
\end{align}
Since $n$ and $P\in\mathbf P_0$ are being fixed and $\zeta_n\ge 1$ by Assumption \ref{Ass: critical value}(ii), it follows that $e_{b_{n,P}^*}(n,P)\ge -CM\zeta_n$ for all $M$ sufficiently large, so that $b_{n,P}^*\in \Lambda_{\phi,M}^\circ(n,P)$ and
\begin{align}\label{Eqn: anti concentration, aux8}
E[\langle b_{n,P}^*,\mathbb Z_{n,P}\rangle^2]\le  \bar\sigma_{n,P,M}^2~.
\end{align}
Combining results \eqref{Eqn: anti concentration, aux7} and \eqref{Eqn: anti concentration, aux8}, we may then conclude that, for all $M$ large,
\begin{align}\label{Eqn: anti concentration, aux9}
\bar\sigma_{n,P}^2-\eta\le \bar\sigma_{n,P,M}^2 \le \bar\sigma_{n,P}^2~.
\end{align}
Since $\eta$ is arbitrary, result \eqref{Eqn: anti concentration, aux9} implies that, as $M\to\infty$,
\begin{align}\label{Eqn: anti concentration, aux10}
\bar\sigma_{n,P,M}^2\to \bar\sigma_{n,P}^2~.
\end{align}
The claim \eqref{Eqn: anti concentration, aux6} then follows from result \eqref{Eqn: anti concentration, aux10} and the fact $\bar\sigma_{n,P}^2>0$.

With \eqref{Eqn: anti concentration, aux6} in hand, we may proceed as in \citet{FangSeo2019Shape} in view of
\begin{multline}\label{Eqn: anti concentration, aux11}
F_{n,P,M}(r)\equiv P(\sup_{b^*\in \Lambda_{\phi,M}^\circ(n,P)}\{\langle b^*,\mathbb Z_{n,P}\rangle+e_{b^*}(n,P)\}\le r) \\
= P(\sup_{b^*\in \Lambda_{\phi,M}^\circ(n,P)}\{\langle b^*,\mathbb Z_{n,P}\rangle+e_{b^*,M}(n,P)\}\le r+CM\zeta_n)~,
\end{multline}
where $e_{b^*,M}(n,P)\equiv e_{b^*}(n,P)+CM\zeta_n\ge 0$ for all $b^*\in \Lambda_{\phi,M}^\circ(n,P)$ (by the definition of $\Lambda_{\phi,M}^\circ(n,P)$). In particular, Theorem 2.2.1 in \citet{Yurinsky1995Gaussian} implies that $F_{n,P,M}$ is absolutely continuous on an interval containing $(m_{n,P,M},\infty)$, and Theorem 2.2.2 in \citet{Yurinsky1995Gaussian} in turn delivers (with $b=m_{n,P,M}+CM\zeta_n$ and $u=r+CM\zeta_n$ in Yurinsky's notation) that, for all $r>m_{n,P,M}$,
\begin{multline}\label{Eqn: anti concentration, aux12}
f_{n,P,M}(r) \equiv F_{n,P,M}'(r) \le \frac{2(r+CM\zeta_n)-(m_{n,P,M}+CM\zeta_n)}{\{(r+CM\zeta_n)-(m_{n,P,M}+CM\zeta_n)\}^2}\\
=\frac{2r-m_{n,P,M}+CM\zeta_n}{(r-m_{n,P,M})^2}~.
\end{multline}
Moreover,  \eqref{Eqn: anti concentration, aux6} and Borell's inequality (see, e.g., \citet[p.82]{Davydov1998local}) imply
\begin{align}\label{Eqn: anti concentration, aux13}
m_{n,P,M}+\bar{\sigma}_{n,P,M} z_{1-\alpha-\varpi}\ge c_{n,P,M}(1-\alpha-\varpi) ~,
\end{align}
where $z_{1-\alpha-\varpi}$ is the $(1-\alpha-\varpi)$-quantile of the standard normal distribution. Next, we note (1) $m_{n,P,M}\le c_{n,P}(0.5)$ due to $T_{n,P}\le \phi(\mathbb Z_{n,P}+\kappa_n\theta_P)$, (2) $c_{n,P}(0.5)\le C\zeta_n$ by \citet{Kwapien1994Median}, \eqref{Eqn: anti concentration, aux3} and Assumption \ref{Ass: critical value}(ii), and (3) $\bar{\sigma}_{n,P,M}\le C\{E[\|\mathbb Z_{n,P}\|_{\mathbf B}^2]\}^{1/2}$ by the definition of $\Lambda_\phi^\circ$ and result \eqref{Eqn: anti concentration, aux3} so that $\bar{\sigma}_{n,P,M}\le CE[\|\mathbb Z_{n,P}\|_{\mathbf B}]$ by Proposition A.2.4 in \citet{Vaart1996} (and that $\mathbb Z_{n,P}$ may be represented as a separable Gaussian process in $\ell^\infty(\mathbf B_1^*)$ by Assumption \ref{Ass: critical value}(i), with $\mathbf B_1^*$ the unit ball in $\mathbf B^*$; see the proof of Proposition 3.1 in \citet{FangSeo2019Shape} for formal arguments). These facts, together with \eqref{Eqn: anti concentration, aux13} and Assumption \ref{Ass: critical value}(ii), yield
\begin{align}\label{Eqn: anti concentration, aux14}
c_{n,P,M}(1-\alpha-\varpi)\le  c_{n,P}(0.5) + C\zeta_n z_{1-\alpha-\varpi} ~.
\end{align}
Note that $c_{n,P,M}(1-\alpha-\varpi)\to c_{n,P}(1-\alpha-\varpi)$ as $M\to\infty$ by $T_{n,P}\to \phi(\mathbb Z_{n,P}+\kappa_n\theta_P)$ (almost) surely and $c_{n,P}(1-\alpha-\varpi)$ being a continuity point of the cdf of $\phi(\mathbb Z_{n,P}+\kappa_n\theta_P)$ (by an application of Theorem 2.2.1 in \citet{Yurinsky1995Gaussian}). Letting $M\to\infty$, we may thus deduce from \eqref{Eqn: anti concentration, aux14} and Assumption \ref{Ass: critical value}(iii) that $\varsigma_n\le C\zeta_n z_{1-\alpha-\varpi}$. In turn, by Assumption \ref{Ass: critical value}(iv) and $\epsilon=o(c_n)$, we obtain that, as $n\to\infty$,
\begin{align}\label{Eqn: anti concentration, aux15}
\epsilon_n = o(\frac{\varsigma_n^2}{\zeta_n})= o(\varsigma_n)~.
\end{align}
Thus, $\epsilon_n\le \varsigma_n/2$ for all $n$ large, so that, whenever $x\ge m_{n,P,M}+\varsigma_n$,
\begin{align}\label{Eqn: anti concentration, aux16}
x-\epsilon_n \ge m_{n,P,M}+\varsigma_n - \epsilon_n \ge m_{n,P,M} + \frac{\varsigma_n}{2}~.
\end{align}

By the fundamental theorem of calculus, results \eqref{Eqn: anti concentration, aux12} and \eqref{Eqn: anti concentration, aux16}, and the fact that $r\mapsto(2r-m_{n,P,M}+CM\zeta_n)/(r-m_{n,P,M})^2$ is decreasing on $(m_{n,P,M},\infty)$, we may conclude that, for all $x\ge m_{n,P,M}+\varsigma_n$ and $n$ large,
\begin{multline}\label{Eqn: anti concentration, aux17}
P(|\sup_{b^*\in \Lambda_{\phi,M}^\circ(n,P)}\{\langle b^*,\mathbb Z_{n,P}\rangle+e_{b^*}(n,P)\}-x|\le \epsilon_n )\\
 = \int_{x-\epsilon_n}^{x+\epsilon_n}f_{n,P,M}(r)\,\mathrm dr
\le 2\epsilon_n \frac{2(m_{n,P,M}+\varsigma_n/2)-m_{n,P,M}+CM\zeta_n}{(\varsigma_n/2)^2}~.
\end{multline}
Since $m_{n,P,M}\le c_{n,P}(0.5)\le C\zeta_n$ and $\varsigma_n\le C\zeta_n z_{1-\alpha-\varpi}$ as argued previously, we obtain from \eqref{Eqn: anti concentration, aux17} that, for all large $n$ and all $M>0$,
\begin{align}\label{Eqn: anti concentration, aux19}
P(|\sup_{b^*\in \Lambda_{\phi,M}^\circ(n,P)}\{\langle b^*,\mathbb Z_{n,P}\rangle+e_{b^*}(n,P)\}-x|\le \epsilon_n )\le  C\epsilon_n \frac{\zeta_n+M\zeta_n}{\varsigma_n^2}~.
\end{align}
The proposition then follows by \eqref{Eqn: anti concentration, aux5}, \eqref{Eqn: anti concentration, aux19}, $\epsilon_n=o(c_n)$ and Assumption \ref{Ass: critical value}(iv). \qed

\end{appendices}


\titleformat{\section}{\normalfont\Large\bfseries}{\thesection}{1em}{}

\bibliographystyle{D:/Dropbox/Common/Latex/ecta}
\phantomsection
\addcontentsline{toc}{section}{References}
\bibliography{D:/Dropbox/Common/Latex/mybibliography}

\end{document}